\documentclass[12pt,a4paper,openright]{book}
\usepackage[T1]{fontenc}
\usepackage{fancyhdr}
\usepackage{frcursive}
\usepackage[T1]{fontenc}
\usepackage{natbib}
\usepackage{afterpage}
\usepackage{caption}
\usepackage{subcaption}
\usepackage{hyperref,color,longtable}
\usepackage{makeidx,fancybox}

\usepackage{graphicx}
\usepackage{longtable,lscape}
\usepackage{hyperref}
\usepackage{cite}
\usepackage{dcolumn}
\usepackage{amsmath}
\usepackage{amssymb}
\usepackage{bm}
\usepackage{makeidx}
\usepackage{setspace}
\usepackage{mathrsfs}
\usepackage[T1]{fontenc}
\usepackage{lmodern}

\makeindex
\begin{document}
\chapter*{}
\thispagestyle{empty}
\setlength{\textheight}{620pt}
\setlength{\oddsidemargin}{33pt}
\setlength{\evensidemargin}{0pt}
\setlength{\marginparwidth}{57pt}
\setlength{\footskip}{30pt}

\pagenumbering{gobble}
\begin{center}
{\bf \Huge Coronal Mass Ejections from \\ \vspace {0.15cm} the Sun - Propagation and \\ \vspace {0.15cm} Near Earth Effects.}\\
  \vspace{1cm}

A thesis\\
Submitted in partial fulfillment of the requirements\\
Of the degree of\\
Doctor of Philosophy\\[0.8cm]
By\\[1cm]
{\Large Arun Babu K. P. }\\
20083021 \\[1.50cm]

  \includegraphics[width=3cm]{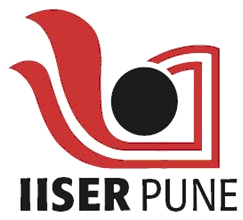}\\ 

{INDIAN INSTITUTE OF SCIENCE EDUCATION AND RESEARCH PUNE}\\[1cm]

\vfill
{July, 2014}

\end{center}

\thispagestyle{empty}
\cleardoublepage

\pagenumbering{roman}
\chapter*{Certificate}\addcontentsline{toc}{section}{\textbf{Certificate}}
Certified that the work incorporated in the thesis entitled 
``\textit{Coronal Mass Ejections from the Sun - Propagation and Near Earth Effects}'',
submitted by \textit{Arun Babu K. P. } was carried out by the candidate, under my supervision. The work presented here or any part of it has not been included in any other thesis submitted previously for the award of any degree or diploma from any other University or institution.\\[3cm]
\textit{Date} \hfill \textbf{Dr. Prasad Subramanian}
\cleardoublepage

\chapter*{Declaration}\addcontentsline{toc}{section}{\textbf{Declaration}}
{
I declare that this written submission represents my ideas in my own words and where others' ideas have been included, I have adequately cited and referenced the original sources. I also declare that I have adhered to all principles of academic honesty and integrity and have not misrepresented or fabricated or falsified any idea/data/fact/source in my submission. I understand that violation of the above will be cause for disciplinary action by the Institute and can also evoke penal action from the sources which have thus not been properly cited or from whom proper permission has not been taken when needed.\\[3cm]}
\textit{Date} \hfill \textbf{Arun Babu K. P.}\\
\begin{flushright}
Roll No.- 20083021 
\end{flushright}
\cleardoublepage
\vspace*{\fill}
\begingroup
\centering

 \Huge \textbf{\textit{To \\ my \\ Mom \& Dad}}

\endgroup
\vspace*{\fill}
\cleardoublepage

\pagestyle{fancy}
\fancyhf{}
\fancyhead[LO]{\nouppercase{\rightmark}}
\fancyhead[RE]{\nouppercase{\rightmark}}
\fancyhead[LE,RO]{\thepage}

\setlength{\parskip}{10pt}
\pagenumbering{roman}
\onehalfspacing

\tableofcontents
\listoffigures \addcontentsline{toc}{section}{\textbf{List of figures}}
\listoftables \addcontentsline{toc}{section}{\textbf{List of tables }}

\chapter*{Acknowledgements} \addcontentsline{toc}{section}{\textbf{Acknowledgements}} \chaptermark{Acknowledgements}
{

 \textit{This thesis is the fruit of my long journey in obtaining my PhD. I was not left alone in this journey, but was with support and encouragement of numerous people including, my well wishers, my friends, colleagues and collaborators. It is a pleasant task to express my thanks to all those who contributed in many ways to the success of this study and made it an unforgettable experience for me. }

 \textit{It is my great pleasure to express my gratitude to my supervisor Dr. Prasad Subramanian for his guidance, support and encouragement in pursuing this work. To be honest I was extremely impressed by his personality and his encouraging nature during my PhD interviews and the introductory talk about the research fields given by the faculties to the PhD students. There was no need of any second thought for me to join him to pursue my PhD. He was very friendly to us and was known as the `coolest guide' in IISER among my fellow students. }

 \textit{I would like to thank Dr. K. N. Ganesh, Director, IISER-Pune, for the excellent research facilities and academic environment here. I acknowledge the financial support from IISER-Pune in the form of research fellowship. Many thanks to IISER administration, library, computer, transport and house-keeping staff. }

 \textit{I acknowledge the RESPOND and CAWSES {\textsc {II}} programs administered by Indian Space Research Organisation (ISRO) for their financial support during my PhD. I acknowledge Asian Office of Aerospace Research and  Development (AOARD) for the financial support during the final year of my PhD.}

 \textit{It was an honour for me to work with the GRAPES-3 team (TIFR, Mumbai). I thank GRAPES-3 team for providing the data, which I used to study the Forbush decreases, that cover major part of my thesis. I specially thank Prof. H. M. Antia (TIFR, Mumbai) and Prof. Sunil Gupta (TIFR, Mumbai) for their help and discussions. Prof. H. M. Antia helped me in analysing the GRAPES-3 data, he was always available and never let me wait for  more than five minutes  for his help, discussion and suggestions via emails. Prof. Sunil Gupta, who was also one of my research advisory committee members helped me understand the basics of the `GRAPES-3 muon telescope', his discussions and advises have benefited me greatly during my PhD. I thank P. K. Mohanty and Atul Jain (TIFR, Mumbai) for familiarizing me  with the GRAPES-3 and I thank all other members of GRAPES-3 collaboration. It is a pleasure to thank Dr. Apratim Chatterji who was in my research advisory committee for his valuable feedback  and suggestions. I am happy to express my gratitude to Prof. Alejandro Lara S\'{a}nchez (UNAM, Mexico) for his useful discussions. }

 \textit{It is really a nice experience to be in the solar physics group of IISER with Mayur, Nishtha and Tomin. I enjoyed working with Adwiteey, Bhavesh, Ajay and Abhishikth. I have enjoyed our solar physics journal club meetings. I thank Dr. Durgesh Tripati (IUCAA, Pune) for his initiation to start our journal club. It was really nice time with you all, Prasad, Divya, Durgesh, Mayur, Nishtha, Rohit, Ajay, Tomin, Srividya,  Sargam and Girjesh. }

 \textit{It was fun and refreshing being with the fellow students. I thank Arthur, Murthy, Mayur, Kajari, Abishek, Resmi, Kanika, Padma, Somu, Ramya, Madhan, Neeraj, Shweta, Mahendra, Mandar, Vimal, Aditya, Nishtha, Shishir, Sanku, Rohit, Harshini, Kunal and Nitin. The days in common room were very cherishing with Ludo games, ice-cream parties, birthday celebrations and assignment writing. Hostel life in IISER was also so enjoyable with many hostel shiftings. It was really nice to have Arthur as my room partner, who took care of me as my brother and provided tasty food for me in our kitchen. The entertainment section was mainly managed by Mayur, who provide us movies with his passion of downloading. Kajari is a good buddy and shares my interests in photography and in food. She is always ready for any adventure. The heated unending philosophical discussions were provided by Abhishek in the company of Resmi, Kajari and Mayur. I thank Lal, Sandeep, Kajari and Anusha for their company and the bike rides we had together.  The life in IISER was not without few down and depressed moments. I thank Prasad, Arthur, Kajari, Mayur and Resmi for the comfort and assurances they gave me. }

 \textit{I thank Aaron, Abner and Adrin of `FOUR ACES', my long-time friends for their support and encouragement. Conversations with you were always cheerful and motivating.}

 \textit{At last and the most I thank my family, who support me always. I thank my mother, father and brother for their love, support and encouragement throughout my life.}
}
\begin{flushright}{\Large \textbf  {\em Arun Babu K. P.}}\end{flushright}

\chapter*{Abstract} \addcontentsline{toc}{section}{\textbf{Abstract}} \chaptermark{Abstract}

Owing to our dependance on spaceborne technology, an awareness of disturbances in the near-Earth space environment is proving to be increasingly crucial. Earth-directed Coronal mass ejections (CMEs) emanating from the Sun are the primary drivers of space weather disturbances. Studies of CMEs, their kinematics, and their near-Earth effects are therefore gaining in importance.

The effect of CMEs near the Earth is often manifested as transient decreases in galactic cosmic ray intensity, which are called Forbush decreases (FDs). In this thesis we probe the structure of CMEs and their associated shocks using FD observations by the GRAPES-3 muon telescope at Ooty. We have established that the cumulative diffusion of galactic cosmic rays into the CME is the dominant  mechanism for causing FDs (Chapter \ref{model}).

 This diffusion takes place through a turbulent sheath region between the CME and the shock. One of our main results concerns the turbulence level in this region. We have quantitatively established that cross-field diffusion aided by magnetic field turbulence accounts for the observed lag between the FD and the magnetic field enhancement of the sheath region (Chapter \ref{corrIP}).

We have also investigated the nature of the driving forces acting on CMEs in this thesis. Using CME data from the SECCHI coronagraphs aboard STEREO sapcecraft, we have found evidence for the non-force-free nature of the magnetic field configuration inside these CMEs, which is the basis for the (often-invoked) Lorentz self-force driving (Chapter \ref{fluxrope}).

Taken together the work presented in this thesis is a comprehensive attempt to characterise CME propagation from typical coronagraph fields of view to the Earth.

\chapter*{Publications}
The research work presented in this thesis has appeared in the following publications.

\textbf{\textit{ \large  Publications in international refereed journals}} \\ 
\begin{enumerate}

\item ``High-rigidity Forbush decreases: due to CMEs or Shock?'', {\bf  K. P. Arunbabu} et al., 2013, Astronomy \& Astrophysics, 555, 139

\item ``Self-similar expansion of solar coronal mass ejections: implications for Lorentz self-force driving'', P. Subramanian, {\bf  K. P. Arunbabu}, A. Vourlidas, A. Mauriya, 2014, The Astrophysical Journal, 790, 125

\item ``Relation of Forbush decreases with the Interplanetary magnetic field enhancements '', {\bf  K. P. Arunbabu} et al. in preparation

\end{enumerate}

\vspace{5mm}
\textbf{\textit{ \large Refereed conference proceedings}} \\ 
\begin{enumerate}
\item ``How are Forbush decreases related with IP magnetic field enhancements ?'', {\bf K. P. Arunbabu} et al., Proceedings of the International Symposium on Solar Terrestrial Physics, ASI Conference Series, 2013, 10, 95.
\end{enumerate}

\newpage

\textbf{\textit{ \large  Presentations in symposia/conferences }} \\ 
\begin{enumerate}
\item Meeting of the Astronomical Society of India,  2014 March 20-22 , IISER Mohali, ``How are Forbush decreases related with IP magnetic field enhancements ?'', {\bf  K. P. Arun Babu}, Prasad Subramanian, Sunil Gupta, H. M. Antia, (oral presentation)

\item International Symposium on Solar Terrestrial Physics, 2012 November 6-9 , IISER Pune, ``How are Forbush decreases related with IP magnetic field enhancements ?'', {\bf  K. P. Arun Babu}, Prasad Subramanian, Sunil Gupta, H. M. Antia, (poster presentation)

\item 39th COSPAR Scientific Assembly, 14 - 22 July 2012,Mysore, India, ``Forbush decrease observed in GRAPES-3'', {\bf  K. P. Arun Babu}, Prasad Subramanian, Sunil Gupta, H. M. Antia, GRAPES-3 team, (poster presentation)

\item Workshop on 'Physics of the Solar Transition Region and Corona' , IUCAA Pune, 2011 September 5-7 , ``Coronal Mass Ejections \& Forbush Decrease'', {\bf  K. P. Arun Babu} (oral presentation)

\item Asia Oceanic Geoscience Society International conference AOGS 2010 July 5-9, Hyderabad, `` Forbush decreases observed with GRAPES-3'', {\bf  K. P. Arun Babu}, Prasad Subramanian, Sunil Gupta, H. M. Antia, (oral presentation)

\item Workshop and Winter School on AstroParticle Physics (WAPP 2009), 2009 December 10-12, Bose Institute, Darjeeling, ``Forbush decreases observed in GRAPES-3'', {\bf  K. P. Arun Babu}, Prasad Subramanian, (oral presentation).

\end{enumerate}

\cleardoublepage
\fancyhead[LO]{\nouppercase{\leftmark}}
\fancyhead[RE]{\nouppercase{\rightmark}}
\fancyhead[LE,RO]{\thepage}
\pagenumbering{arabic}
\chapter{Introduction}
\label{ch:intro1}

\noindent\makebox[\linewidth]{\rule{\textwidth}{3pt}} 
{\textit {In this chapter we will introduce the fundamental physics and concepts that are discussed in this thesis, beginning with a short introduction to the Sun, its interior, various layers,  atmosphere, and activity. This is followed by an introduction to coronal mass ejections (CMEs) comprising a historical account of CME observations, physical properties, initiation models, and propagation models. This is followed by  an introduction to the Sun-Earth connections and the effects of the solar wind and CMEs on near-Earth space weather }  }\\
\noindent\makebox[\linewidth]{\rule{\textwidth}{3pt}}

\section{Sun }

The Sun, our nearest star and the center of our solar system, is a main sequence star of spectral type G2V. The Sun \index{Sun} has a total luminosity $L \, = \, (3.84 \pm 0.04) \times 10^{26} \, W $, mass $M \, = \, (1.9889 \pm 0.0003) \times 10^{30} \, kg$ and radius $ R \, = \, (6.959 \pm 0.007)\times  10^8 \, m$ \citep{fou04}. The Sun was born from a giant molecular cloud of approximate mass $10^4 \, - \, 10^6$ M which began to gravitationally collapse and fragment. The process of collapse and fragmentation continued until one of these fragments attained a central temperature large enough to start hydrogen fusion, about $4.6 \times 10^9$ years ago \citep{pri09}. At this point the energy produced by the hydrogen fusion was high enough to counterbalance the gravitational collapse. Currently, the Sun is in a stable configuration, on the Main Sequence, where it is in hydrostatic equilibrium $ (\nabla P = - \rho g)$. The Sun will continue to maintain this stable state for about another $5 \times 10^9$ years before entering the red giant phase. At this point the Sun \index{Sun} will expand to about 100 times its current size and begin shedding its outer layers, due to successive nuclear burning in ever more distant shells. This will ultimately leads to the total loss of the outer envelope exposing a degenerate core, in which all nuclear burning has ceased, called a white dwarf \citep{phi95}.
 
\subsection{Solar Interior}

\begin{figure*}
   \centering
      \includegraphics[width = 1.0\textwidth]{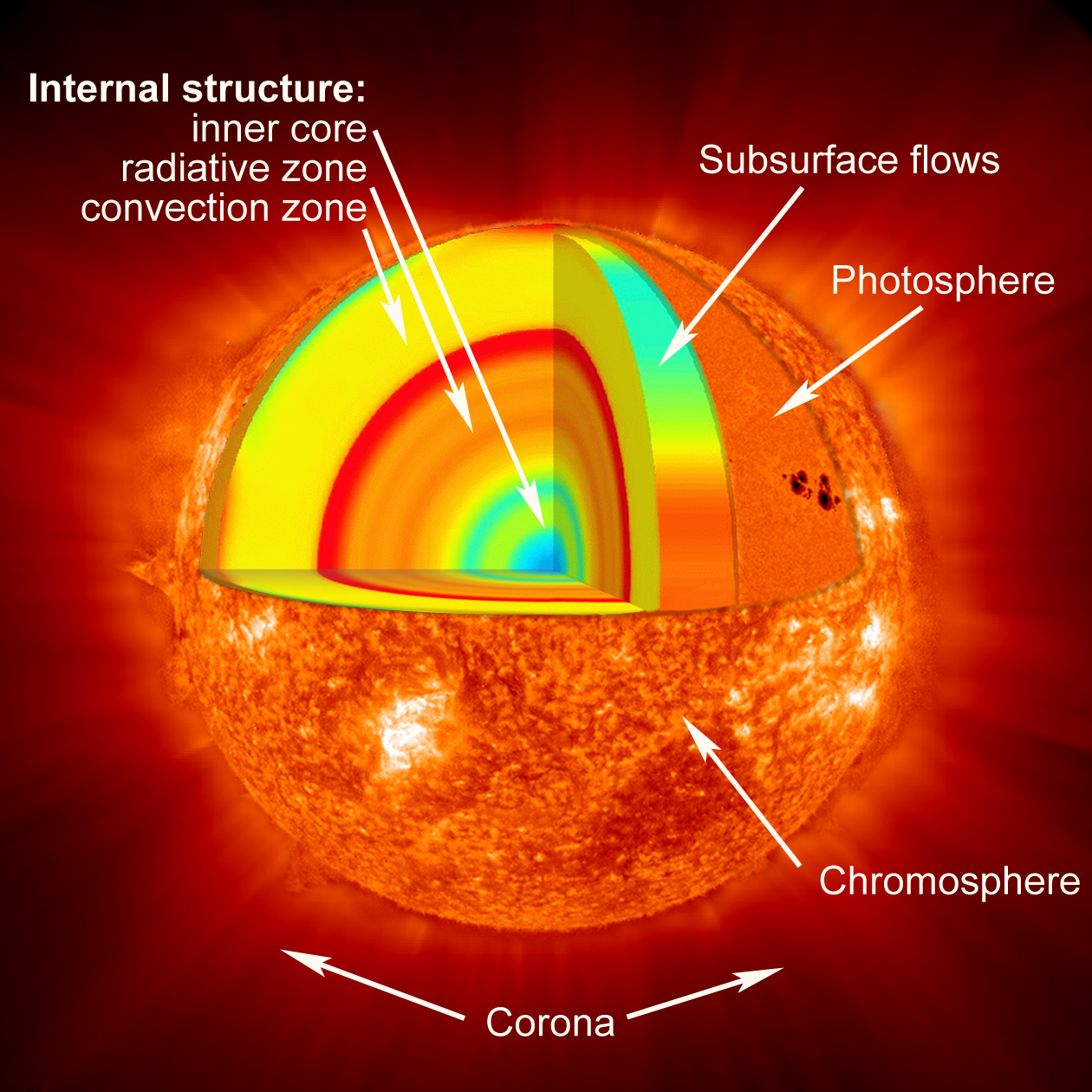}
   \caption[Showing different layers of Sun..]{Showing different layers of Sun. At the centre of the Sun is the core ($\le 0.25 \, R_{\odot}$) where temperatures reach $\sim 1.5\times 10^7 \, K$, high enough for fusion to take place. The energy generated at the core from the fusion process is transported towards surface via thermal radiation in the radiative zone ($0.25 - 0.70 R_{\odot}$). At this point the solar plasma is cool enough to from highly ionised atoms and becomes optically thick. As a result it is convectively unstable and energy is transported through mass motions in the convection zone ($0.7 - 1.0 R_{\odot}$ ). The visible surface of the Sun, the photosphere, is a thin layer in the atmosphere where the bulk of the Sun's energy is radiated, its spectra is well matched to a blackbody with peak temperature of $5600 \, K$. Above the Sun's visible surface lies the chromosphere and finally the corona where the temperature soars back up to $1 - 2 \times 10^6 \, K$. (image courtesy : \href{http://www.geyserlandobs.org/sun.html}{Geyserland observatory})}
              \label{layers}%
    \end{figure*}

The entire energy emitted by the Sun \index{Sun} is produced by `hydrogen fusion' reactions. The core is the central part of the Sun where the temperature and pressure are high enough for the fusion reaction to occur. The temperatures in the core are around $1.5 \times 10^7 \, K$ and the pressure exceeds $ 2.5 \times 10^{11} $ atmospheres. The core extends up to $0.25 \, R_{\odot}$. Outside of the core is the radiative zone ($0.25 -  0.70 R_{\odot}$), where thermal radiation is the most efficient means of transporting the intense energy generated in the core (in the form of high energy photons) outward. The temperature drops from about $7 \times 10^6 \, K$ at the bottom of the radiative zone to $2 \times 10^6 \, K$ just below the convection zone. Due to the high densities ($2 \times 10^4 - 2 \times 10^2 \, kg \, m^{-2}$ ) in the radiative zone the mean free path of the photons is very small ($\sim 9.0 \times 10^{-2} \, cm$); hence it can take tens to hundreds of thousands of years for photons to escape. The radiative zone and the convection zone are separated by a transition layer, the tachocline. The convection zone is the outer layer of the Sun ($0.7-1.0 \,  R_{\odot}$). The temperature of the convection zone is lower than that in the radiative zone and heavier atoms are not fully ionized. As a result, radiative heat transport is less effective. The density of the gases in this zone are low enough to have convective currents.  Material heated at the tachocline pick up heat and expand. This reduces the density of material and allows it to rise. Thermal convection carry the majority of the heat outward to the Sun's photosphere. The material cools off at the photosphere, which increases its density and causes it to sink to the base of the convection zone. At the convection zone it picks up more heat from the top of the radiative zone and the cycle continues. The visible surface of the Sun, the photosphere, is the layer below which the Sun becomes opaque to visible light. Above the photosphere visible sunlight is free to propagate into space, and its energy escapes the Sun entirely.

\begin{figure*}
   \centering
      \includegraphics[width = 1.0\textwidth]{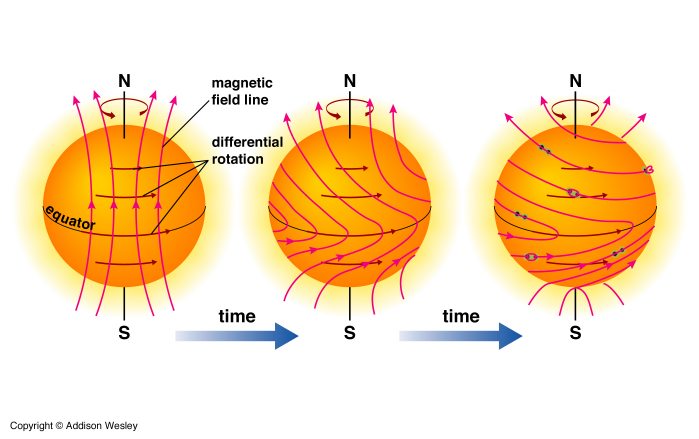}
   \caption[Differential rotation of Sun]{{\textit Left}: Shows the Sun's bipolar field. {\textit Middle}: The magnetic field is being twisted by differential rotation. {\textit Right}: Loops of magnetic field begin to break the surface forming sunspots (From The Essential Cosmic Perspective, by \citealp{bennett})}
              \label{diffrot}%
    \end{figure*}

\begin{figure*}
   \centering
      \includegraphics[width = 0.9\textwidth]{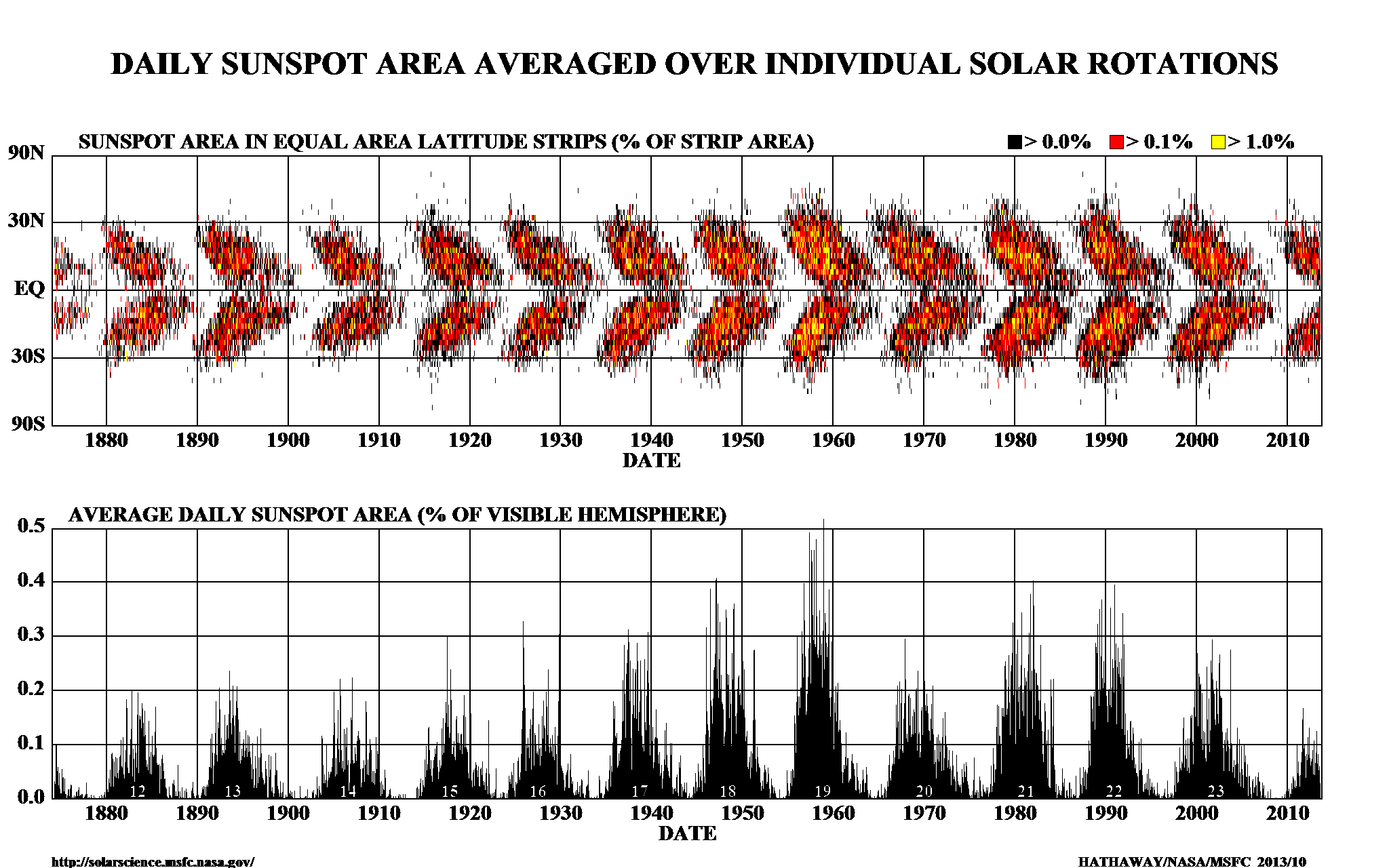}
   \caption[Butterfly diagram of sunspot cycle]{The position of the sunspots in equal area latitude strips, averaged over a solar rotation with respect to time ({\textit top}). The butterfly pattern is clear as the decrease in the upper limit of sunspot latitudes with time. ({\textit bottom}) The average sunspot area as a function of time. The 11 year modulation is clear in both of these plots. Image courtesy of \href{http://solarscience.msfc.nasa.gov/images/bfly.gif}{NASA MSFC}.}
              \label{bfly}%
    \end{figure*}

The core and the radiative zones of the Sun \index{Sun} rotate rigidly (as a solid body), where as the convection zone rotates differentially. There is a thin interface between these two regions known as the tachocline.  This region is subjected to large shear flows due to the meeting of the two bodies rotating at different rates. These flows are believed to be the mechanism which generates the Sun's large-scale magnetic field and powers the solar dynamo. The Sun's magnetic field is mainly dipolar and aligned to the rotation axis. Each hemisphere has an opposite dominant polarity (see the left panel in Figure \ref{diffrot}). The differential rotation \index{Sun!differential rotation} of the convection zone winds up this field  (see the middle panel in Figure \ref{diffrot}). This large scale twisting transforms poloidal field to toroidal field which is known as the $\Omega$-effect. As the field is twisted up the magnetic pressure increases and bundles of magnetic field lines (flux ropes) \index{flux rope} can become unstable and rise up in the from of loops. Due to solar rotation, the Coriolis effect twists these loop back towards north-south orientation, reinforcing the original poloidal field. This is known as the $\alpha$-effect (see the  right panel in Figure \ref{diffrot}) and completes the $\alpha \Omega$-dynamo. When magnetic loops become buoyant and rise up through the surface they are visible as sunspots on-disk and mark the footprints of large loops which extend into the solar atmosphere. In a given hemisphere the leading sunspot and trailing sunspot \index{sunspot} will have opposite polarities, this order is reversed in the other hemisphere (Hale's Law). The tilt angle of the sunspots pairs have a mean value of $5.6^{\circ}$ relative to the solar equator (Joy's Law). Sunspots \index{sunspot} are known to migrate from high latitudes towards the equator over an 11 year cycle (Sporer's Law; see figure \ref{bfly}). The net affect is an increase in opposite polarity field at the poles, ultimately the majority of the field will be oppositely oriented and the dipole will flip. This occurs every 11 years, thus a complete cycle takes 22 years (N to S to N). The activity of the Sun, in the form of active regions, flares, transient events, and other associated phenomenon, is modulated by this cycle (see figure \ref{bfly} lower).

\subsection{Solar Atmosphere} \label{satm}

\begin{figure}
   \centering
      \includegraphics[width = 0.9\textwidth]{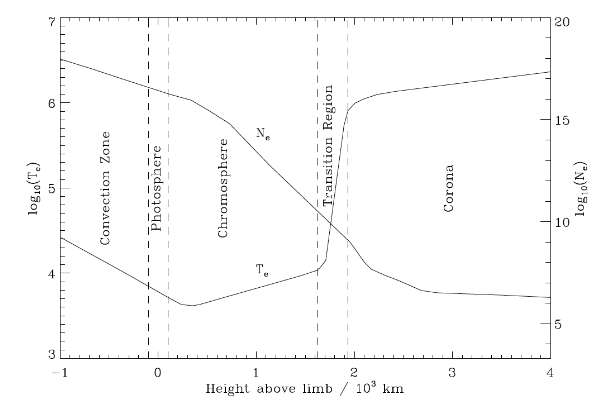}
   \caption[A 1D model of the electron density and temperature profile thought the solar atmosphere]{A 1D model of the electron density $ Ne [cm^{-3} ]$ and temperature $Te [K]$ profile through the solar atmosphere from \citet{gab82}. Neutral atoms are present in the photosphere and chromosphere but the plasma is fully ionised in the
corona due to the higher temperature}
              \label{1De}%
    \end{figure}

 The photosphere \index{photosphere} is the layer beyond which the optical depth falls below unity. The Sun's \index{Sun} atmosphere typically refers to  all the regions above the photosphere.  Based on their density, temperature, and composition the solar atmosphere is usually separated into three regions, the photosphere, chromosphere and corona  as shown in Figure \ref{1De}. However, the separation solar atmosphere is a simplification as the atmosphere is an inhomogeneous mix of different plasma properties due to up-flows, down-flows, heating, cooling and other dynamic processes. The density of the plasma generally decreases with increasing height through these regions. The temperature decreases in photosphere, reaching a minimum in the chromosphere, then slowly rises until there is a rapid increase at the transition region which continues into the corona. This rapid increase in temperature embodies the so-called `coronal heating problem'.

\begin{figure*}
   \centering
      \includegraphics[width = 0.9\textwidth]{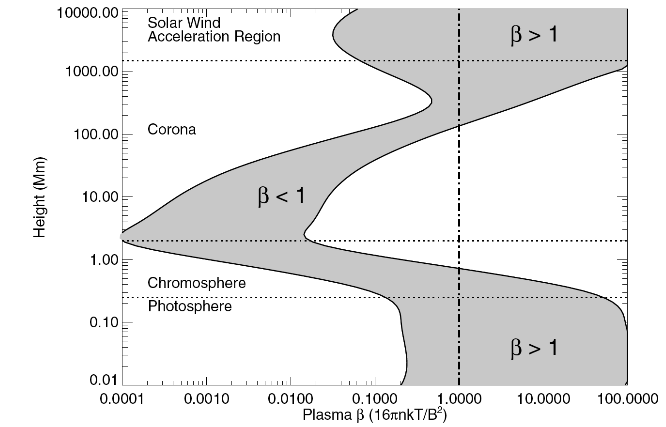}
   \caption[Plasma-$\beta$ in the solar atmosphere as a function of height]{Plasma-$\beta$ in the solar atmosphere as a function of height for two magnetic field strengths of $100 \, G$ and $2500 \, G$. The layers of the atmosphere are segregated by the dotted lines. The corona is the only region in which the magnetic pressure dominates over the thermal pressure, a low $\beta$ plasma \citep{asc06}.}
              \label{Pbeta}%
    \end{figure*}

An important parameter in describing the solar atmosphere is the plasma-$\beta$ term, the ratio of the thermal to magnetic pressures:

\begin{equation}
\beta \, = \, \frac{P_{th}}{P_{mg}} \, = \, \frac{n k_B T}{\frac{B^2}{2\mu_0}}
\end{equation}
where $n$ is the number density and $\mu_0$ the permeability of free space. In the photosphere \index{photosphere} the plasma-$\beta$ is large (Figure \ref{Pbeta}). The plasma-$\beta$ decreases farther up in the solar atmosphere, before increasing again in the outer corona. 

\subsubsection{Photosphere}
The photosphere \index{photosphere} is the lowest of the three layers of solar atmosphere. It is the visible surface of the Sun and is defined as the height where the optical depth, at visible wavelengths, equals 2/3 ($\tau_{5000} \, \approx \,  2/3, I \, = \, I_0 e^{-\tau}$ ). This is the mean optical depth at which the photospheric radiation is emitted.  The effective temperature ($T_{eff} = 5776$) and blackbody temperature of the photosphere matches here. This can be seen by substituting $B(T ) \, = \, \frac{\sigma}{\pi }T^4$ and $F \, = \, \sigma T_{eff}^4$  (F is the total radiative flux) in the general solution of the radiative transfer equation:

\begin{equation}
B(T) \, = \, \frac{3}{4} \left( \tau + \frac{2}{3}\right) \frac{F}{\pi}
\end{equation}

which gives, 

\begin{equation}
\sigma T^4 \, = \, \frac{3}{4} \left( \tau + \frac{2}{3}\right) \sigma T_{eff}^4 
\end{equation}

implying that $\tau \, = \, 2/3$ \citep{fou04}. The temperature drops from $6,400 \,  K$ at the base of the photosphere to 4,400K at the top. The spectrum of photospheric radiation is that of a blackbody with a large number of absorption features, Fraunhofer lines, due the upper layers of the atmosphere superimposed on it. The number density of photosphere ranges from $\sim 10^{19} - 10^{21} \,  m^{-3}$ over the depth of the photosphere ($500 \, km$). 

One of the main observable features in the photosphere \index{photosphere} is granulation\index{photosphere!granulation}. This represents convection currents operating just below the photosphere, which transport heat from below to the surface. Granules are small-scale features made up of brighter regions isolated by darker lanes, interpreted as the upflow of hot material to the surface which then flows horizontally and cools, flowing back down in the dark lanes. Typical granules are of the order of $1,000 \, km$ in diameter. They have lifetimes of 5-10 minutes with vertical flow velocities of hundreds to thousands $km \, s^{-1}$. There are also larger scale flow patterns known as mesogranulation and supergranulation. Mesogranules are typically $ 7000 \, km$ in diameter, and have lifetimes of hours with vertical flows of the order of tens of $m \, s^{-1}$. Supergranules are larger still at diameters of $3 \times 10^4 \, km$, and consequently have longer lifetimes of days. Supergranules have large horizontal flows and smaller vertical flows of the order of $0.5 \,  km \, s^{-1}$. Sunspots appear in the photosphere as darker regions due to their lower temperature ($4,000 \, K$). They appear dark since convection is suppressed by the strong kilogauss magnetic fields. Sunspots \index{sunspot} play an important role in the activity of the Sun \index{Sun} as they are the source of solar flares and many  CMEs\index{CME}.

\subsubsection{Chromosphere}

The chromosphere \index{chromosphere} is the middle layer of solar atmosphere, which lies above the photosphere\index{photosphere}. At this layer the temperature initially decreases to a minimum of $\sim 4,500 \, K$ before increasing to $\sim  20,000 \, K$  with increasing height. It occupies a region approximately $2,000 \, km$ thick with a density of about $10^{16} \, m^{-3}$. The  chromosphere has a  split structure between the hot bright magnetic network and the cooler darker internetwork \citep{gal99}. Spicules\index{chromosphere!spicules}, which are jets of plasma are also observed on the limb. These jet-like structures have diameters of hundreds of kilometres, attaining heights of tens of thousands of kilometres above the solar surface, with flows of the order of $30 \, km \, s^{-1}$ lasting 5 - 10 minutes.

The nature of the heating mechanism of chromosphere \index{chromosphere} is unclear. Observations imply that there must be some form of energy deposition occurring. Neither radiation nor conduction can be the source as the temperature is lower at the base of the lower chromosphere and photosphere than in chromosphere proper (and would thus violate the laws of thermodynamics). Since the chromosphere is in hydrostatic equilibrium, mass motions are neither observed nor applicable. The most likely source of the energy (heat flux) is the dissipation of compressional or sound waves \citet{sch48}.  The convective plasma motions of the photosphere launches sound waves into the chromosphere. These sound waves travel upwards with little dissipation. As the density drops, the waves steepen and form shocks which rapidly dissipate energy, thereby heating the chromosphere. This type of acoustic heating is not appropriate in the network regions.  In these regions the strong magnetic fields suppress the convective motions which drive the waves. This led to the idea of Aflv\'{e}n wave heating which is first introduced by \citet{ost61}. Aflv\'{e}n waves are magneto-hydrodynamic waves which propagate along magnetic fields. The restoring force is provided by magnetic tension and the ion mass provides the inertia. Here the magnetic field itself is responsible for transportation and deposition of the energy from the photospheric motions. This type of heating matches well with observations of plage and emerging flux regions. Both of these show strong heating which imply that the heat flux is related to the magnetic field strength.

Filaments are often seen over active regions as dark channels in on-disk $H\alpha$ observations  or as prominences when observed on the limb as bright features. The transition region lies between the chromosphere \index{chromosphere} and corona, here the temperature rapidly jumps (over 100 km) to above 1 MK. Above the transition region the magnetic field dominates and determines the structures.

\subsubsection{Corona}
The tenuous, hot, outer layer of the solar atmosphere is known as the corona\index{corona}. The electron density of the solar corona ranges from $\sim 10^{14} \, m^{-3}$ at its base, $2,500 \, km$ above the photosphere\index{photosphere}, to $ \lesssim\, 10^{12} \, m^{-3}$ for heights $\gtrsim \, 1 R_{\odot}$ \citep{asc06}. The density varies depending on the feature, such as  the open magnetic structures of coronal holes can have densities in the region of $(0.5 - 1.0) \times 10^{14}\, m^{-3}$ , streamers have densities in the region of $(3 - 5) \times 10^{14}\, m^{-3}$ while active regions have densities in the region of $2\times 10^{14} -  10^{15} \, m^{-3}$. The temperature in the corona is generally above $1\times 10^6 \, K$ but again varies across different coronal features. Coronal holes have the lowest temperature (less than $1\times 10^6 \, K$) followed by quiet Sun \index{Sun} regions at $1 -  2\times 10^6 \, K$. Active regions are the hottest at $2-6 \times 10^6 \, K$ with flaring loops reaching even higher temperatures. The high temperatures reached in the corona give rise to EUV and X-ray emission, which have highly ionised iron lines as a prominent feature. The visible corona during eclipses is due to Thomson scattering of photospheric light from free electrons in the coronal plasma. The corona \index{corona} has a number of components:

\begin{itemize}
\item{K-corona (kontinuierliches spektrum) is composed of Thomson-scattered photospheric radiation and dominates below $\sim2 R_{\odot}$.  As a result of the Thomson scattering mechanism, the scattered light is strongly polarised parallel to the solar limb. The high temperatures mean the electrons have high thermal velocities. This will wash out (due to thermal broadening) the Fraunhofer lines, producing a white-light continuum. The intensity of the K-corona \index{corona} is proportional to the density summed along the line-of-sight.}
\item{F-corona (Fraunhofer) is composed of  Rayleigh-scattered photospheric radiation by dust particles, and dominates above $\sim2 R_{\odot}$. It forms a continuous spectrum with the superimposed Fraunhofer absorption lines. The radiation has a very low degree of polarisation. The F-corona is also know as Zodiacal light, it can be seen with the naked eye at dawn or dusk under favourable conditions.}
\item{E-corona (Emission) is composed of line emission from visible to EUV due to various atoms and ions in the corona. It contain many forbidden line transitions, thus it contains many polarisation states. Some of the strongest lines are {\textrm Fe {\textsc {xiv}} 530.3 nm}  (green-line; visible), H-$\alpha$ at 656.3 nm (visible), and Lyman-$\alpha$ 121.6 nm (UV).}
\item{T-corona (Thermal) is composed of thermal radiation from heated dust particles. It is a continuous spectrum according to the temperature and colour of the dust particles.}
\end{itemize}

\section{Coronal mass Ejections (CMEs)} \label{inCME}
Every main sequence star loses mass via dynamic phenomena in its atmosphere that accelerate plasma or particles beyond the escape speed. In the case of our Sun, the outer plasma atmosphere, known as the corona\index{corona}, extends to millions of kilometres.  There are two forms of mass loss in Sun: the steady solar wind \index{solar wind} flow and episodic coronal mass ejections (CMEs\index{CME}).  CMEs \index{CME!definition}  are large scale eruptions of plasma and magnetic filed which propagate from the Sun \index{Sun} into the Heliosphere. Post-eruption, CMEs   are driven by the release of energy carried by the magnetic fields advected by the CME  \citep{vr00, sv07}. The more energetic  CMEs  are associated with flares. The solar wind outflows amount to $\sim \, 2 \times 10^{-10} \, (g \, cm^{-2} \, s^{-1})$ in coronal holes, and $ \lesssim \, 4 \times 10^{-11} \, (g \, cm^{-2} \, s^{-1})$ in active regions. The CME \index{CME!occurence} occurence rate is highly solar cycle dependent. This occurs with an average frequency of few events per day, carrying the mass in the range of $ m_{CME} \, \sim \, 10^{14} - 10^{16} g$, which corresponds to an average mass loss of $ m_{CME}/(\Delta t .4\pi R_{\odot}^2) \, \sim \, 2\times 10^{-14} - 2 \times 10^{-12} \, (g \, cm^{-2} \, s^{-1})$, which is $ \lesssim \, 1\% $ of solar wind mass loss in coronal holes, or  $ \lesssim \, 10\% $ of that in active regions. Further details regarding CME occurrence rate are mentioned below in section \S~\ref{occR}

 A typical CME \index{CME!mass} has a mass of $ 10^{13} \, - \, 10^{16} \, g $ \citep{vr02} and a  velocity between $\sim 10 \, - 2000 \, km \, s^{-1} $ some times reaching up to $3500 \, km \, s^{-1} $ close to the Sun \citep{yash04}. At 1 AU, CME \index{CME!velocity} velocities ($300 \, - 1,000 \, km \, s^{-1} $) tend to equilibrate to the solar wind speed \citep{gop06,lind99}.  The energies associated with  CMEs \index{CME!energy}  are of the order of $10^{24} -10^{25} \, J$, making  them among comparable to flares,  the most energetic events on the Sun \index{Sun} \citep{vr02}.

 In this section we give only a brief introduction to CMEs, more detailed reviews can be found in \citet{webb12} and \citet{chen11}.

\subsection{Observations of  CMEs }
Observations of the corona and CMEs \index{CME!observation} are possible only when the photosphere \index{photosphere} of the visible Sun is occulted. Such occultation occurs naturally during solar eclipses and is artificially created by spaceborne coronagraphs. A coronagraph \index{coronagraph} produces an artificial solar eclipse. It uses an occulting disk to block the Sun's bright surface, revealing the faint solar corona, stars, planets and sungrazing comets.   Coronagraphs view the outward flow of density structures emanating from the Sun by observing photons emitted from photosphere Thomson-scattered by free electrons in solar corona.

\begin{figure}
  \centering
  \includegraphics[width=1.0\textwidth]{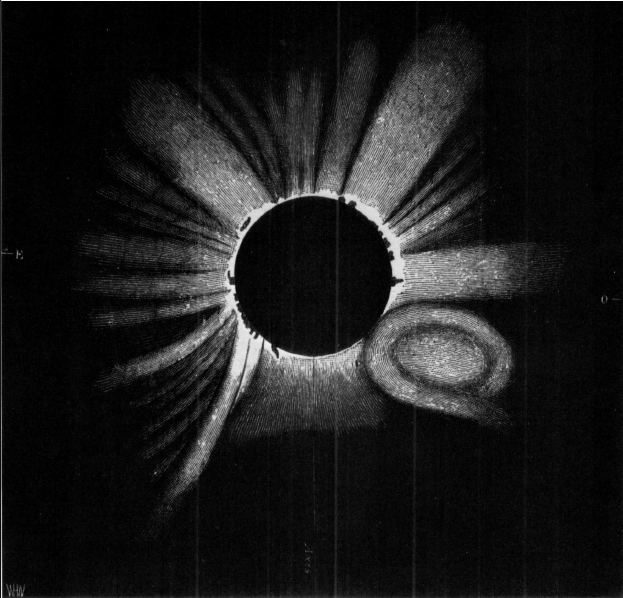}
  \caption[Drawing of corona as it appear to G. Tempel at Torreblanca]{Drawing of corona \index{corona} as it appear to G. Tempel at Torreblanca, Spain during the total solar eclipse of 18 July 1860. South is at bottom west at right. \citep{eddy74}}
  \label{1obsa}
\end{figure}

The earliest observation of a CME \index{CME!observation} probably dates back to the total solar eclipse of 18 July 1860 in a drawing recorded by Gugliemo Temple shown in figure \ref{1obsa} \citep{eddy74}. It took more than 100 years after that for the CME  to be formally discovered. The first definitive observation being made on 14 December 1971 by \citet{tou73} using the coronagraph on-board the seventh Orbiting Solar Observatory (OSO-7).

\subsubsection{Space-based coronagraphs}
A number of orbital coronagraphs \index{coronagraph} were flown in space which provided  better data and longer periods of CME \index{CME!observation} observations.  Spaceborne coronagraphs which observed CMEs include  OSO-7 coronagraph in the early 1970s \citep{tou73},  Skylab (1973 - 1974; \citealp{mq80}), P78-1 (Solwind) (1979 - 1985; \citealp{shee80}), and Solar Maximum Mission (SMM) (1980; 1984 - 1989; \citealp{hund99}). In late 1995, the Solar and Heliospheric observatory (SOHO) was launched.  The SOHO spacecraft carries a set of three Large Angle and Spectrometric Coronagraph (LASCO) \index{coronagraph!LASCO} coronagraphs with nested field of view which image the Sun from $1.1 - 32 R_{\odot}$ \citep{bruk95}. Of the three coronagraphs, the first one, C1 was an internally occulted coronagraph designed to image the innermost corona from  $1.1 - 3 R_{\odot}$ in emission lines. C1 ceased operation in June 1998. The other two (C2 and C3) are externally occulted and designed to image the outer corona. The field of view of C2 and C3 are   $1.5 - 6 R_{\odot}$ and  $3- 32 R_{\odot}$ respectively. Two of the three LASCO \index{coronagraph!LASCO} coronagraphs (C2,C3) are still operating.  In October 2006 two sets of identical coronagraph packages called, the Sun Earth Connection Coronal Heliospheric Investigations (SECCHI) \index{coronagraph!SECCHI} were flown on two spacecrafts called Solar TErrestrial RElations Observatory (STEREO) A and B. The SECCHI package consists of coronagraphs having unprecedented field of view ranging from $1.4 - 4 R_{\odot}$ (COR1), $2 - 15 R_{\odot}$ (COR2), $12 - 84 R_{\odot}$ (HI1) and $66 - 318 R_{\odot}$ (HI2) in white light. These coronagraphs are designed to provide 3-D view of  CMEs \index{CME} in the inner and outer corona.

\subsubsection{Ground-based coronagraphs}

Ground-based instruments for imaging corona are complementary to the space-based ones because they can achieve a better temporal resolution and are not limited by the telemetry rate. However, they are limited by the intensity and temporal variability of the sky. The main operating ground based coronagraphs today  include  the Mauna Loa Solar Observatory (MLSO) K-coronameter ($1.2 - 2.9 \, R_{\odot}$) \citep{fish81,koom74}  and the green line  coronagraphs \index{coronagraph} at Sacramento Peak, New Mexico \citep{Dem73} and Norikura, Japan \citep{hir74}.

\subsection{Physical properties of CMEs}

CME \index{CME!observation}  observations have not only led to studies of their morphological properties, but also to statistical analyses of their physical and kinematical properties.

\subsubsection{Morphology}

\begin{figure*}
   \centering
      \includegraphics[width = 1.0\textwidth]{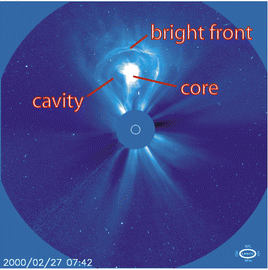}
   \caption[Example of a classic three-part CME ]{Example of a classic three-part CME \index{CME!structure} with the core, cavity, and bright front marked. Adapted from \citet{Riley}}
              \label{3ps}%
\end{figure*}

A classical picture of a CME \index{CME!structure} observed in white light comprises a three part structure (figure \ref{3ps}). It displays a bright leading edge which contains the material swept by the CME. The leading edge is followed by a darker cavity which is due to its low density but has a high magnetic field. The cavity is followed by a bright knot or core, which corresponds to the erupting filament \citep{hou81}. CMEs may also exhibit more complex structures \citep{pick06}. In fact less than $\sim30\%$ of CME \index{CME!structure} events possess all the three parts \citep{webb87}.  CMEs   come in many different shapes, and much of the variety is believed simply due to the projection effects \citep{schw06}. Fundamental difference can be found between narrow  CMEs  and  normal  CMEs. The narrow  CMEs \index{CME!narrow CME}  show jet-like motions probably along open magnetic field, on the other hand  normal  CMEs   are characterized by a closed frontal loop. There are also CMEs without a bright core. Some are due to the fact that the filament material drained down to the solar surface along the stretched magnetic field, some due to the fact that thermal instability had not started to form a filament in the pre-eruption structure, and others might not be related to filament or filament-supporting structure at all.

\subsubsection{Angular width}

\begin{figure*}
   \centering
      \includegraphics[width = 1.0\textwidth]{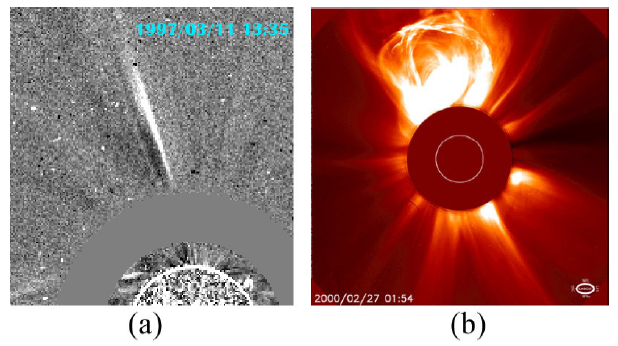}
   \caption[White-light images of narrow CME and normal CME]{White-light images of two types of typical  CMEs  (from SOHO/LASCO \index{coronagraph!LASCO} database). (a) A narrow CME \index{CME!narrow CME} on 1997 March 11 (b) a normal CME  on 2000 February 27 with a three-part structure, i.e., a frontal loop, a cavity, and a bright core, where the white circle marks the solar limb. (Adapted from \citet{chen11})}
              \label{nnCME}%
\end{figure*}

 The angular width of  CMEs \index{CME!angular width}  projected in the plane of the sky ranges from $\sim 2^{\circ} \,  to \,  360^{\circ}$ \citep{yash04}, with a significant fraction in the low end (e.g., $< 20^{\circ}$) and a small fraction in the high end (e.g., $> 120^{\circ}$). CMEs  with  angular widths less than $\sim 10^{\circ}$ can be called narrow  CMEs \index{CME!narrow CME}  \citep{wang98}, and others are sometimes called normal  CMEs  (\citealp{yash04}, see Figure \ref{nnCME}). The average width of LASCO/SOHO \index{coronagraph!LASCO} CMEs is found to be $40^{\circ}$ for limb CMEs. The average width is relatively smaller ($47^{\circ}$) during solar activity minimum compared to that during solar maximum ($61^{\circ}$) \citep{yash04}. Halo  CMEs \index{CME!Halo} , with an apparent angular width of or close to $360^{\circ}$, are simply due to the fact that the  CMEs (probably with an angular width of tens of degrees) propagate near the Sun-Earth line, either toward or away from the Earth.

\subsubsection{Frequency of occurence} \label{occR}

\begin{figure*}
   \centering
      \includegraphics[width = 1.0\textwidth]{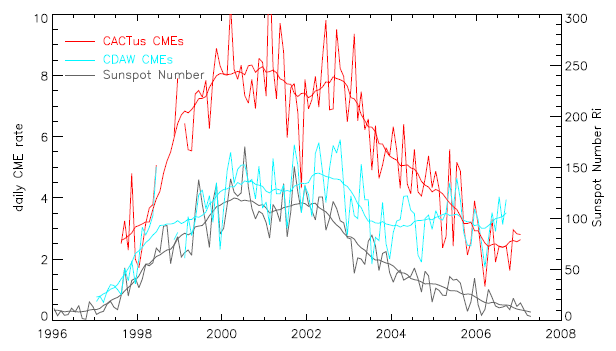}
   \caption[The CME daily occurrence rate]{The CME \index{CME!occurence} daily occurrence rate detected by the CACTus archive (red) and the CDAW archive (blue) compared with the daily sunspot number (gray) during solar cycle 23. Thin curves: smoothed per month, thick curves: smoothed over 13 months (from \citealp{rob09})}
              \label{CMEocc}%
\end{figure*}

During  solar cycle 23, LASCO \index{coronagraph!LASCO} observations provided unprecedented observations of  CMEs \index{CME!observation}. The occurrence rate of  CMEs \index{CME!occurence}  was found to largely track the solar activity cycle,  but with a delay of 6- 12 months \citep{ray05,rob09} between the solar cycle maximum and maximum occurence rate of CMEs.  Before the launch of SOHO, the average occurrence rate was found to increase from 0.2 per day at solar minimum to 3.5 per day at solar maximum \citep{web94}. With the increased sensitivity and wider field of view, the SOHO/LASCO coronagraphs \index{coronagraph!LASCO}detected CMEs   more frequently. The CDAW catalog \footnote{$http://cdaw.gsfc.nasa.gov/CME \_ list /$} lists around 13,000 CMEs identified visually. The results from this catalog suggest that the CME occurence rate increases from $\sim 0.5$ per day near solar minimum to $\sim 6$ near solar maximum \citep{gop03, yash04}. However, the automated software, CACTus3, identified many more events for the same period, with the occurrence rate increasing from $< 2$ per day near solar minimum to $\sim 8$ per day near solar maximum \citep{rob09}.  The CME \index{CME!occurence} daily occurrence rate detected by the two methods, along with the sunspot number are shown in figure \ref{CMEocc} for the comparison.

\subsubsection{CME speeds}

Estimates of the apparent speeds of the leading edges of  CMEs \index{CME!velocity}  range from about $20  \, to \, > \, 2500 \, km \, s^{-1}$. The annual average speeds of Solwind and SMM  CMEs  varied over the solar cycle from about $ 150 - 475  \, km \, s^{-1}$. Even though their relationship to sunspot \index{sunspot} number was unclear \citep{how86, hund94}. On the other hand LASCO \index{coronagraph!LASCO} CME \index{CME!velocity} speeds did generally track sunspot number in Solar Cycle 23 \citep{yash04, gop10}, from $ 280 \,  to \, \sim 550  \, km \, s^{-1}$. For a  typical  CMEs, above a height of about $2R_{\odot}$ the speeds  are relatively constant in the field of view of coronagraphs\index{coronagraph}. The slowest  CMEs   tend to show acceleration while the fastest tend to decelerate \citep{st00, yash04, gop06b}. This can be expected, considering that  CMEs   must push through the surrounding solar wind\index{solar wind}, believed to have a speed of around $400 \, km \, s^{-1}$ in the outer corona.

 CMEs   typically  accelerate fast low in the corona \index{corona} until gravity and other drag forces slow them further out. This process continues into the interplanetary medium. The early acceleration for most  CMEs \index{CME!acceleration} occurs low in the solar corona ($< 2R_{\odot}$). Only 17\% of all LASCO  CMEs   experience acceleration out to $30R_{\odot}$ \citep{st00}.  \citet{zha01, zha04} used the observations of flare-associated  CMEs  close to the limb in the LASCO C1 field of view ($1.1 - 3.0R_{\odot}$) and found a three-phase kinematic profile. The first phase is  a slow rise ($< 80 \, km \, s^{-1}$) over tens of minutes. The second phase shows rapid acceleration of $100 - 500 \, m \, s^{-2}$ in the height range $1.4 - 4.5R_{\odot}$ during the flare rise phase, and a final phase exhibits propagation at a constant or declining speed. \citet{gal03} and others have identified the strong acceleration region of impulsive  CMEs   to $\sim 1.5 - 3.0 R_{\odot}$. The studies of \citet{shee99} and \citet{sri99} using LASCO \index{coronagraph!LASCO} data found that gradually accelerating  CMEs \index{CME!acceleration}  looked balloon-like in coronagraph \index{coronagraph} images, but fast  CMEs   moved at constant speed even as far out as $30 R_{\odot}$.  \citet{shee99} found that when viewed well out of the sky plane, gradual  CMEs looked like smooth halos which accelerated to a limiting value then faded, while fast  CMEs \index{CME}  had ragged structure and decelerate. In the LASCO field of view slow CMEs   tend to accelerate and fast  CMEs   decelerate,  while  those around the solar wind \index{solar wind} speed having constant speeds \citep{yash04}. 

The aerodynamic drag experienced by  CMEs \index{CME!aerodynamic drag}  while they travel through the interplanetary medium between the Sun and Earth is generally thought to arise due to the coupling of the  CMEs \index{CME}  to the ambient solar wind.  The solar wind strongly mediates CME  propagation \citep{gpl00, mano06}  in the interplanetary medium,  CMEs   which start out slow (with respect to the solar wind speed) near the Sun seem to accelerate en route to the Earth, while fast  CMEs \index{CME!acceleration}  are decelerated. This fact has been invoked in several studies that derive a heuristic aerodynamic drag coefficient for  CMEs to investigate CME slowdown using a simple 1D hydrodynamical model that lends itself to analytical solutions \citep{bor09, cgl04, byr10, mal10, vrs10, vrs13}. \citet{slb12} used the viscous drag by the solar wind \index{solar wind} on the CME \index{CME!velocity} to study the velocity of CME  in the interplanetary medium.

\subsubsection{CME mass \& energy estimates }

Masses of CMEs \index{CME!mass} are derived from the white light images obtained by coronagraphs. The mass estimates are based on the fact that the white light emission from the corona is mainly due to Thomson scattering of photospheric light by the electrons in the corona\index{corona}. This therefore requires estimation of the coronal density from a series of time-lapse images of a CME by the coronagraph. This involves measurement of electron density, which is computed from the excess brightness after removing the pre-CME brightness. The method is based on the assumption that a single electron at a certain point in the atmosphere will scatter a known amount of solar disk intensity. Thus by measuring the intensity and assuming that all of the mass is in a single volume element, the number of electrons can be computed. The CME mass can be estimated assuming charge neutrality. 

Masses and energy calculations of  CMEs \index{CME!energy} require difficult instrument calibrations and often suffer from significant uncertainties. The older coronagraph \index{coronagraph} data (Skylab, SMM and Solwind) derived the average mass of  CMEs \index{CME!mass} to be a few times $10^{15} \, g$. Calculations using LASCO \index{coronagraph!LASCO} observations indicate a slightly lower average CME  mass, $1.6 \times 10^{15} \, g$. This is  because LASCO can measure smaller masses down to the order of $10^{13} \, g$ \citep{vr02, vr10, vr11, kah06}. Studies using Helios \citep{webb96} and LASCO \citep{vr00, vr10,vr11} data suggest that the older CME \index{CME!mass} masses may have been underestimated. This is  because mass outflow may continue well after the CME's leading edge leaves the instrument field of view. The true mass calculations using the STEREO data \citep{col09} states that the CME mass increases with time and height then reaches a constant value above about $10 \, R_{\odot}$.

\begin{figure*}
   \centering
      \includegraphics[width = 1.0\textwidth]{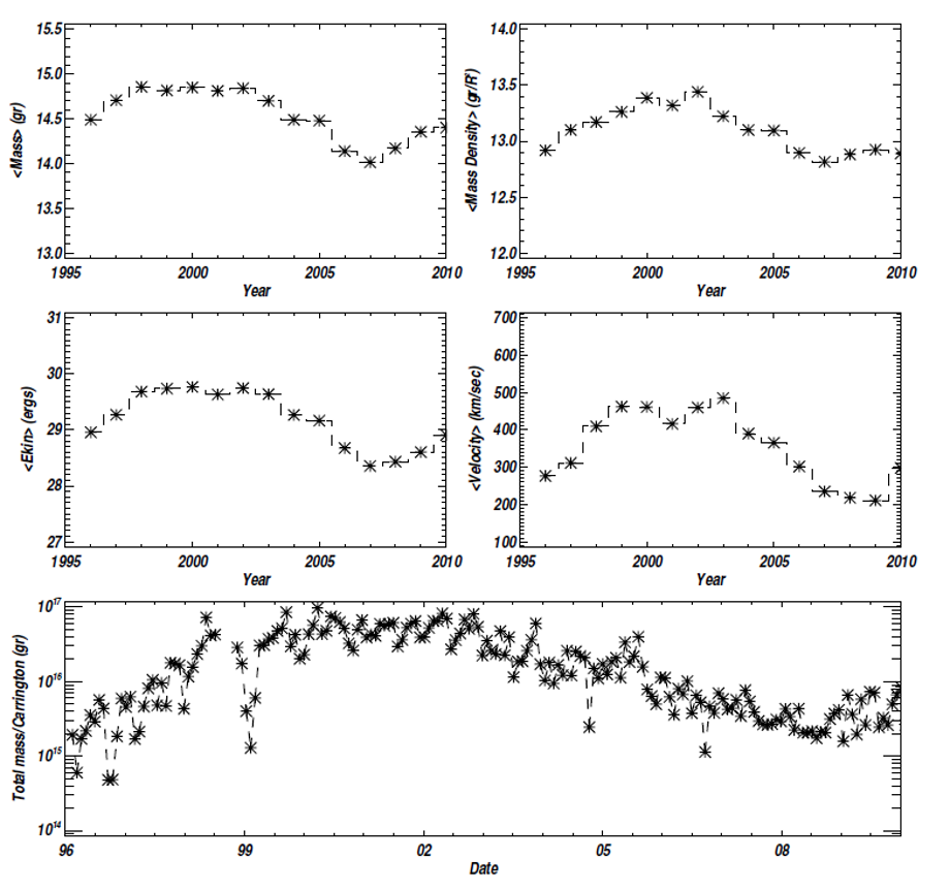}
   \caption[Solar cycle dependence of the CME mass and kinetic energy]{Solar cycle dependence of the CME \index{CME!mass} mass and kinetic energy. {\textit {Top left}}: log CME  mass. {\textit {Top right}}: log CME  mass density in g $R^{-2}$. {\textit {Middle left}}: log CME \index{CME!energy} kinetic energy. {\textit {Middle right}}: CME \index{CME!velocity} speed. All four plots show annual averages. {\textit {Bottom panel}}: total CME \index{CME!mass} mass per Carrington rotation. The data gaps in 1998 and the drop in 1999 are due to spacecraft emergencies. The plot is an update of Figures 14 and 1 in \citet{vr10,vr11} to include events to July 31, 2010, Adapted from \citet{webb12}}
              \label{CMEmass}%
\end{figure*}

\begin{figure*}
   \centering
      \includegraphics[width = 0.8\textwidth]{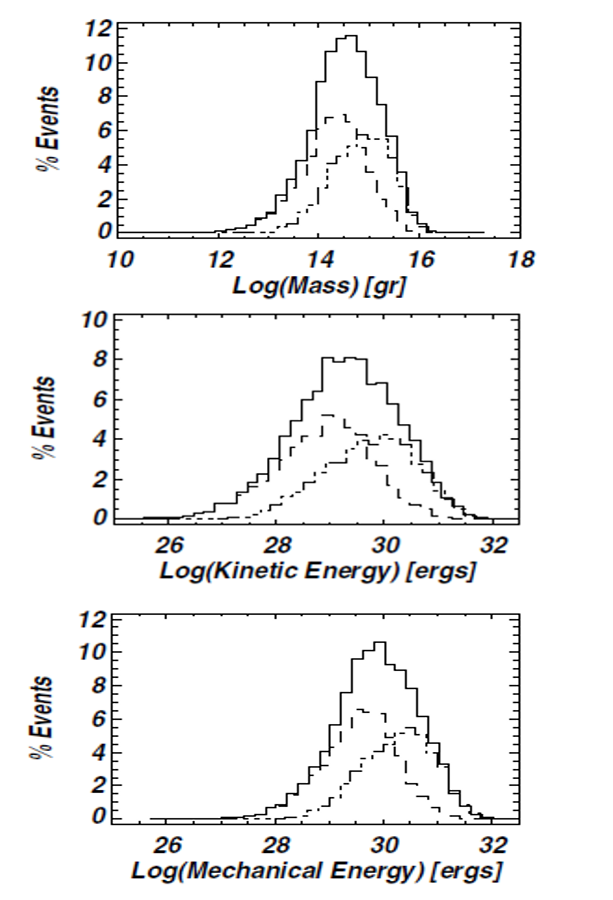}
   \caption[Histograms of LASCO CME  mass distribution]{Histograms of LASCO CME \index{CME!mass} mass distribution ({\textit {upper left}}), kinetic energy \index{CME!energy} ({\textit {upper right}}), and total mechanical energy ({\textit{ bottom left}}) for 7668 events. Also shown are the histograms for events reaching maximum mass  $< \, 7 R_{\odot}$ (dashed lines) and events reaching maximum mass $7 R_{\odot}$ (dash-double dot). Not all detected  CMEs   have been included because mass measurements require: (i) a good background image, (ii) three consecutive frames with  CMEs  , and (iii)  CMEs   well separated from preceding  CMEs. Adapted from \citet{vr10,vr11}.}
              \label{CMEen}%
\end{figure*}

Average CME \index{CME!energy} kinetic energies measured by LASCO \index{coronagraph!LASCO} are less than previous measurements, $2.0 \times 10^{30} \, erg$ (\citealp{vr10} - Figure \ref{CMEen}).  The kinetic energy distribution of CME events appears to have a power law index of -1 \citep{vr02}.

 The solar-cycle dependence of the LASCO CME \index{CME!mass} mass and kinetic energy \index{CME!energy} \citep{vr10, vr11} are shown in figure \ref{CMEmass}. The bottom panel shows the total CME mass per Carrington rotation. The mass, mass density, and kinetic energy all have minima in 2007.  These are 2 - 4 times below the 1996 minimum and  reflect the unusual extended activity in Solar Cycle 23. The total mass reaches a minimum in 2009 and is roughly equivalent to the  minimum in 1996. According to \citet{mq01}, the mass density variation between Solar Cycle 22 minimum and maximum varied by a factor 4 even in the background corona\index{corona}.

It is a difficult task to measure  CME \index{CME!mass} \index{CME!energy} masses and energies using white light images farther from the Sun \index{Sun}. This is because of  the lack of calibration information and the uncertainties imposed by the faintness of the  CMEs  compared to the background noise. The 3-D density reconstructions of a few  CMEs \index{CME}  observed in the heliosphere by SMEI \citep{jack08, jack10} also give the mass and energy estimates. These mass estimates generally agree with the mass of the same  CMEs \index{CME}  as derived from LASCO \index{coronagraph!LASCO} data. The  kinematical properties, mass calculations are based on coronagraph \index{coronagraph} images and, therefore, subject to the same problems of projection and perspective. More recent work making use of the stereoscopic capabilities of STEREO have provided more accurate measurements \citep{col09}.

\subsection{CME initiation models}

A variety of observations of  CMEs \index{CME!initiation}  from different space-based and ground-based missions prompted theoreticians to come up with several models or mechanisms to explain the initiation and eruption of CME. It is generally accepted that the required energy for powering a CME comes from the coronal magnetic field. The pre-eruption state of the coronal magnetic field is stressed beyond its minimum energy configuration. Several triggering mechanisms have been proposed either conceptually or through MHD analysis and/or simulations which are described below.
 
\subsubsection{Magnetic breakout model}

\begin{figure*}
   \centering
      \includegraphics[width = 0.9\textwidth]{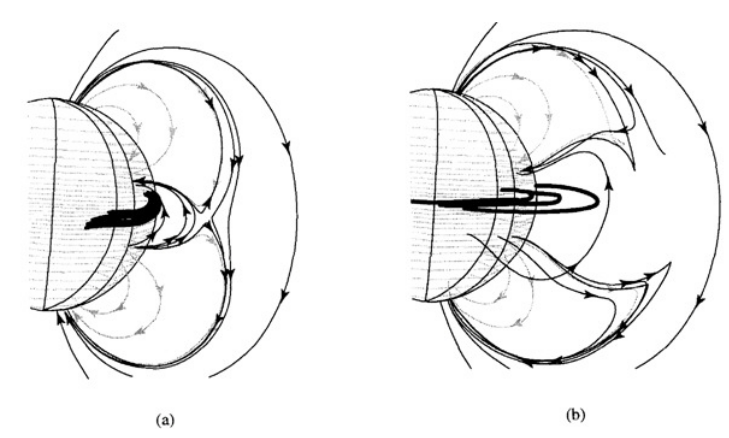}
   \caption[The evolution of the magnetic field in the breakout model]{The evolution of the magnetic field in the breakout model, showing the reconnection above the central flux system removes the constraint over the core field (thick lines), and results in the final eruption (adapted from \citet{ant99})}
              \label{breakout}%
\end{figure*}

The Magnetic break-out model was initially proposed by Antiochos \citep{ant99}. The initiation of a CME \index{CME!initiation!Breakout model} occurs in multipolar topological configurations where reconnection between a sheared arcade and neighboring flux systems triggers the eruption. The term breakout refers to the process of reconnection which removes the unsheared field above the low-lying sheared core flux, allowing it to burst open. A schematic is shown in figure \ref{breakout}.  This model strongly supports the idea that the eruption is solely driven by magnetic energy stored in a closed sheared arcade. 

\subsubsection{Tether cutting or flux cancellation  model}

\citet{moor80} analyzed the filament eruption event on 1973 July 29, and found that: 
\begin{itemize}
\item the magnetic field is strongly sheared near the magnetic neutral line
\item the filament eruption and the two-ribbon flare were preceded by precursor activities in the form of small H$\alpha$ brightening and mass motion along the neutral line
\item H$\alpha$ precursor brightening and the initial brightening of the flare are both located in the vicinity of the steepest magnetic field gradient
\end{itemize}
Piecing these features together, they proposed the tether-cutting mechanism\index{CME!initiation!Tether cutting model}.

This model is based on reconnection which occurs in initially sheared bipolar arcades, leading to formation of a magnetic island or plasmoid, which is then ejected. A set of largely closed post-flare loops are formed subsequently underneath the erupting flux-rope, which is an important observational signature for this model. These post flare loops form a new arcade that grows with time and sustained reconnection. In this model reconnection process is considered as an essential for the onset of magnetic explosion.

 \citet{van89} proposed a similar mechanism. They pointed out that the magnetic flux cancellation near the neutral line of a sheared magnetic arcade would produce helical magnetic field lines (flux rope)\index{flux rope}. These can support a filament, and on further cancellation can result in the eruption of the previously-formed filament. This model was  numerically simulated by \citet{amari03}.

 Tether-cutting model and  flux cancellation model are similar in nature. The flux cancellation model might emphasize a more gradual evolution of magnetic reconnection in the photosphere\index{photosphere}, whereas  tether-cutting is a relatively more impulsive process occurring in the low corona\index{corona}.

\begin{figure*}
   \centering
      \includegraphics[width = 1.0\textwidth]{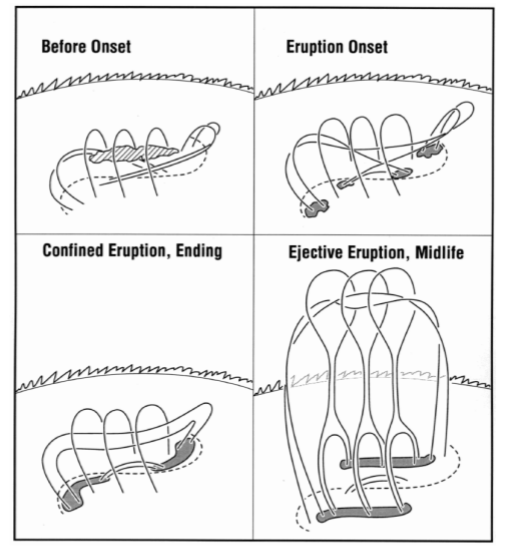}
   \caption[Standard model for the magnetic field explosion in single-bipole eruptive solar events]{Standard model for the magnetic field explosion in single-bipole eruptive solar events ( Adapted from \citet{moor01}) The dashed curve is the photospheric neutral line, bright patches are ribbons of flare in the chromosphere at the feet of reconnected field lines. The diagonally lined feature above the neutral line in the top left panel is the filament of chromospheric temperature plasma.}
              \label{tether}%
\end{figure*}

\subsubsection{Flux rope model}
In this model it is assumed that the initiation of the CME \index{CME!initiation!Flux rope model} consists of two phases, photospheric shearing and flux emergence. This leads to formation of a twisted flux tube. The pre-eruption configuration consists of an infinitely long flux rope and an overlying arcade, which  starts to rise in the initial phase, sets of magnetic field lines then form an island through which runs the twisted flux rope \index{flux rope} closing down below with field lines reconnecting region and finally a set of arcades close to the boundary that reforms with sustained reconnection \citep{chn96}.

\subsubsection{Flux-injection model}

The magnetic configuration of a CME \index{CME!initiation!Flux injection model} is that of a flux rope \index{flux rope} with footpoints anchored below the photosphere\index{photosphere}. The eruption of such a configuration can be brought by `` flux injection'' process or a rapid increase in poloidal flux. This mechanism is quite successful in reproducing not only the observed features close to the Sun \index{Sun} but in the interplanetary medium of a CME  \citep{kcs00}.

\section{Sun-Earth Connection (Space weather)} \label{SW}

The term space weather \index{space weather!definition} refers to conditions on the Sun \index{Sun} and in the solar wind\index{solar wind}, magnetosphere, ionosphere, and thermosphere that can influence the performance and reliability of space-borne and ground-based technological systems and that can affect human life and health (definition used by the U.S. National Space Weather Plan). Our society has become increasingly vulnerable to disturbances in near-Earth space weather, in particular to those initiated by explosive events on the Sun like solar flares, solar energetic particles (SEPs) and coronal mass ejections (CMEs)\index{CME}. 

Solar flares release flashes of radiation covering wavelength ranging from radio waves to Gamma-rays, that can often heat up the terrestrial atmosphere within minutes such that satellites drop into lower orbits. SEPs accelerated to near-relativistic energies during major solar storms arrive at the Earth's orbit within minutes and may, among other things,  severely endanger astronauts travelling through interplanetary space, i.e., outside the Earth's protective magnetosphere\index{magnetosphere}. Earth directed CMEs hit the Earth's magnetosphere and cause (among other effects) geomagnetic storms\index{geomagnetic storms}.

A fleet of spacecraft (ULYSSES, SOHO, YOKHOH, WIND, ACE, TRACE, RHESSI, Hinode, SDO) has enabled us to advance our understanding of the processes involved near the Sun, in interplanetary space, and in the near-Earth environment, thus sharpening our understanding of the Sun, the heliosphere, and the solar-terrestrial relationships. It is useful to mention the famous ``Halloween events''  that occurred during several days in late October/early November 2003 to understand the importance of space weather study. A few very active regions moved across the Earth-facing side of the Sun and produced several bright flares and massive eruptions. Some of them resulted in powerful CMEs   that were pointed towards the Earth and caused major geomagnetic storms. Intense fluxes of SEPs with relativistic energies were also generated, capable enough to penetrate the skins of spacecraft and instruments and even damage some. Fortunately, the CCD cameras in these telescopes recovered after few hours. When it was  realized how high  the radiation dose from such giant events can actually be, this issue became a primary concern in space exploration. Adequate protective measures must be found to ensure the astronauts' safety on their future journeys.

\begin{figure*}
   \centering
      \includegraphics[width = 1.0\textwidth]{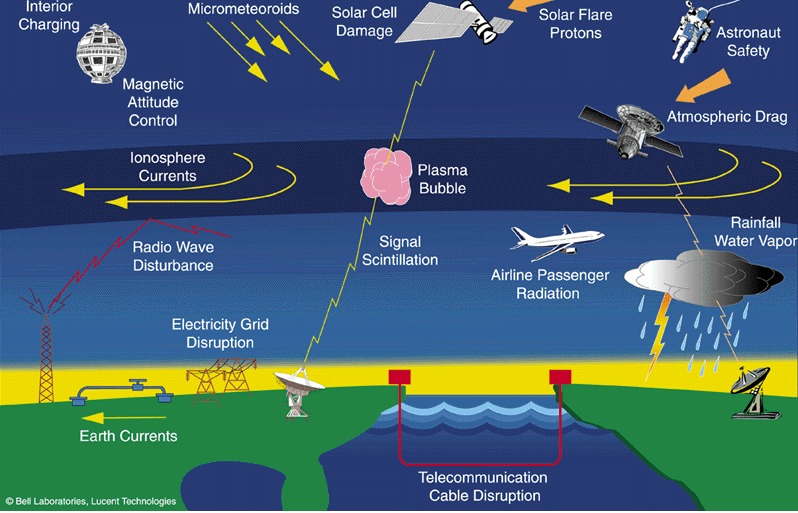}
   \caption[Effects of solar disturbances in human society]{This cartoons shows how the solar disturbances can effect the human society. It can damage things from the satellites to the underground communications. }
              \label{spw}%
\end{figure*}

Figure \ref{spw} shows the different regions where all the solar disturbances can effect our human society. It can impact the satellites and damage the electronics and solar panels. It can effect the GPS systems and air lines, the communications system and electrical grids. The ``Halloween events'' of Oct/Nov 2003 are an example of these. 

\subsection{Solar wind and space weather}\index{solar wind}

The space between the Sun and its planets is filled by the solar wind, a tenuous magnetized plasma, which is a mixture of ions and electrons flowing away from the Sun. The Sun's outer atmosphere is so hot that the Sun's gravity cannot prevent it from continuously flowing outward. The escaping plasma carries the solar magnetic field along with it  out to the border of the heliosphere where its dominance finally ends.

The solar wind \index{solar wind} and the Interplanetary magnetic field (IMF) \index{Interplanetary magnetic field} carried with it are a key link between the solar atmosphere and the Earth system. Although the energy transferred by the solar wind is extremely small compared to both sunlight and those energies involved in Earth's atmosphere, the solar wind is capable of imparting small impulses to the Earth system, which eventually may react in a highly non-linear way. There are indications of these effects reaching down as far as the troposphere.

\begin{figure*}
   \centering
      \includegraphics[width = 0.8\textwidth]{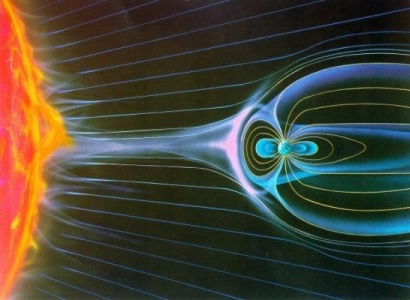}
   \caption[Earth's magnetosphere]{An artistic view of solar wind and Earth's magnetosphere\index{magnetosphere}. showing how the solar wind \index{solar wind}  rearranges the Earth's magnetosphere. it is compressing the magnetic field in the side facing sun and elongates the same in other end.}
              \label{magsphr}%
\end{figure*}

The solar wind rearranges the Earth's magnetosphere. The dynamic pressure (of) and magnetic fields (carried by) the solar wind compresses the magnetic fields of Earth's magnetosphere in the Sun facing side and stretches out the magnetic fields to extend as far as the night side of Earth. An artistic view of this is shown in the figure \ref{magsphr}. Generally, the solar wind \index{solar wind} flow is diverted around Earth by its magnetosphere which is maintained by the Earth's intrinsic magnetic field. Solar wind particles cannot enter into the magnetosphere, unless there is magnetic reconnection between interplanetary and planetary magnetic field lines. This may happen when the northward pointing Earth field on the front of the magnetosphere \index{magnetosphere} is hit by solar wind, which  carry a southward pointing interplanetary magnetic field. In such cases, significant geomagnetic disturbances of various kinds will be initiated. 

\subsection{CMEs and space weather}

Earth-directed halo CMEs \index{CME!Halo} can cause major space weather disturbances. CMEs \index{CME} with their enhanced magnetic fields can cause major deformations in Earth's magnetosphere\index{magnetosphere}. The famous Halloween events were associated with very fast CMEs. 

The study of terrestrial consequences of earthward-directed CMEs are important for  space weather predictions. Broadly this comprises 

\begin{itemize}
\item identification of solar sources or origins, the coronal mass ejections.
\item understanding the propagation of CMEs in the interplanetary medium.
\item identifying the key interplanetary parameters such as solar wind \index{solar wind} velocity, solar wind density, total interplanetary magnetic field and its southward component.
\item understanding the physical relationship of interplanetary parameters with solar parameters
\item understanding the relationship with the  strength of geomagnetic storm \index{geomagnetic storms}with the interplanetary parameters.
\item study the effect of geomagnetic storms and the interplanetary parameters on the navigations systems, satellite communications, etc. 
\end{itemize}
Several attempts have been made to address these aspects. Even so, the prediction schemes are not yet very reliable. 

\section{Summary}

We have given a brief introduction to the fundamental physics and concepts related to the Sun, CMEs and space weather. We started with an outline of the structure of Sun and its various layers. 
Section \S\ref{inCME} gives a basic introduction about CMEs. Observations of  CMEs are outlined in this section starting from historical observations to current ground and space based observations. Physical properties of CMEs such as morphology, angular width, frequency of occurence and CME mass \& energy estimates are also outlined in this section. A flavour of different initiation models and kinematic models are also mentioned. 
An short introduction to space weather and its relation with the solar wind and CMEs are included in  section \S\ref{SW}.

\chapter{Forbush decreases observed in  GRAPES-3 }
\label{GRP3}

\noindent\makebox[\linewidth]{\rule{\textwidth}{3pt}} 
{\textit { In this chapter we give an introduction to Forbush decreases, specifically as a probe of the near-Earth structure of CMEs and their associated shocks. We then briefly introduce the GRAPES-3 muon telescope at Ooty. Thereafter we explain data analysis methods used to identify Forbush decreases in GRAPES-3 data. }  }\\
\noindent\makebox[\linewidth]{\rule{\textwidth}{3pt}}

Cosmic rays \index{cosmic rays} are broadly defined as massive, energetic particles which reach Earth from anywhere beyond its atmosphere. These cosmic rays carry information on the composition of astrophysical sources in our immediate neighbourhood as well as sources far away from our galaxy. They also provide information on acceleration processes operative therein. Low energy cosmic rays can originate in the solar corona, as a result of transient eruptions, while high energy cosmic rays are typically of galactic origin (galactic cosmic rays), from outside our solar system. Galactic cosmic rays \index{cosmic rays} provide us with some of the few direct samples of matter from outside our solar system. They are mostly atomic nuclei whose  electrons have stripped out during  their passage through the galaxy at relativistic speeds. The  galactic cosmic rays  are isotropic when detected at the earth because the intervening turbulent magnetic fields in our galaxy  ``scramble'' the directions of these charged particles. Shock \index{shock} waves driven by supernova explosions are believed to be responsible for the acceleration of galactic cosmic rays \index{cosmic rays}. However, there are several observations of very high energy cosmic rays (of the order of $10^{20}$ eV) for which this explanation is inadequate. It is often conjectured that these ultra-high energy cosmic rays could be coming from outside the galaxy, from active galactic nuclei and/or  gamma ray bursts. They could also represent signatures of topological defects in the structure of the universe, or of exotic mechanisms such as strongly interacting neutrinos.

\begin{figure}[h]
\centering
\includegraphics[width = \textwidth]{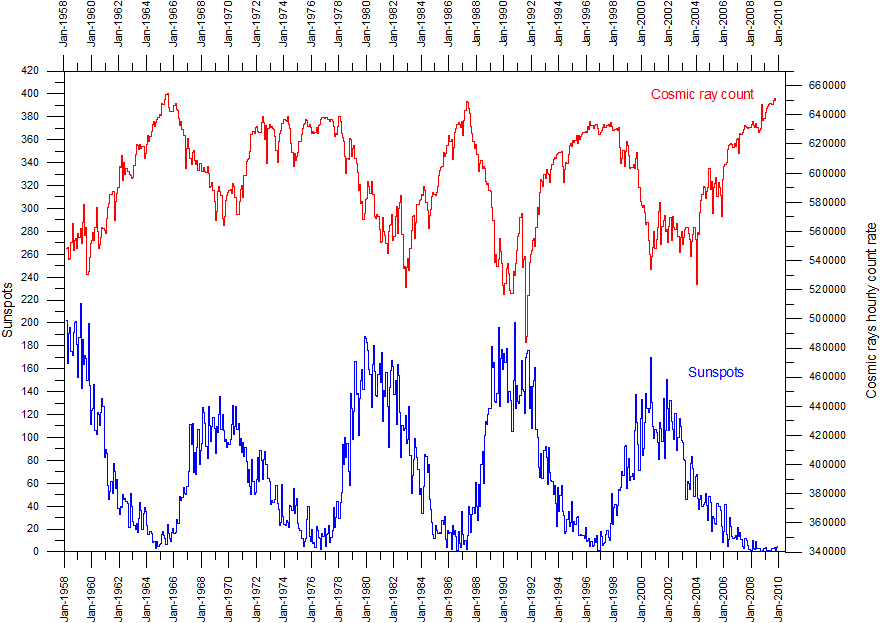}
\caption[Anticorrelation of cosmic ray intensity with solar cycle]{ Variation of cosmic ray intensity and monthly sunspot activity since 1958. The red line shows the cosmic ray intensity and the blue line shows the monthly sunspot activity. The cosmic ray intensity is in anticorrelation with the sunspot activity. (image courtesy : \href{http://www.climate4you.com/Sun.htm}{www.climate4you.com})}\label{CRSS}
\end{figure}

The sun itself is a source of low-to-medium energy cosmic rays \index{cosmic rays}. Solar activities such as flares and coronal mass ejections (CMEs)\index{CME}, which occur frequently during its active phase, accelerate nuclei and electrons to energies ranging from 10-100 MeV. There are rare instances where  solar cosmic rays will reach energies as high as 10 GeV.  The cosmic rays  can  be affected by turbulent magnetic fields carried by the solar wind\index{solar wind}. It is well established that the intensity of cosmic rays \index{cosmic rays} arriving at the Earth is anticorrelated with the solar cycle (Figure \ref{CRSS}). 

\section{Forbush decrease} \label{fode}
Decreases in the intensity of cosmic ray \index{cosmic rays} which last typically for about a week, were first observed by \citet{forb37, forb38} and \citet{hess37} using ionisation chambers. The transient decrease in the observed galactic cosmic ray intensity observed at the Earth are  called Forbush decreases (FDs) \index{Forbush decreases!definition} which are named after their discoverer. 

These shorter term decreases in the cosmic ray intensity exhibit a sudden decrease and a gradual recovery.  Initially it was assumed that these variations were produced, either directly or indirectly, by geomagnetic disturbances such as perturbation of geomagnetic field during geomagnetic storms \index{geomagnetic storms} and these variations were thought to be of terrestrial origin. It was in the early 1950s, when  work of Simpson using neutron monitors \citep{sim54}  showed that the origin of these FDs was in the interplanetary medium. They concluded that these variations are not related with the terrestrial activity but with solar activity. FDs are generally correlated with co-rotating interaction regions (CIRs) or with Earth-directed CMEs from the sun. In this chapter we give a brief introduction to FDs. Comprehensive reviews of this can be found in \citet{cane00} and \citet{loc71}.

There are two broad types of FDs \index{Forbush decreases}: the first one is  `Recurrent decreases' \index{Forbush decreases!Recurrent decreases} \citep{loc71} which have a more gradual onset, and are more symmetric in profile.  These are well associated with corotating high speed solar wind \index{solar wind}streams (e.g., \citealt{iuc79}). The second types are  `Non-recurrent decreases'  \index{Forbush decreases!non-recurrent decreases} which are caused by transient interplanetary events which are related to coronal mass ejections from the Sun. These have the characteristics of  a sudden onset, reach maximum depression within about a day and have a more gradual recovery.  Historically, all short term decreases have been called `Forbush decreases' \index{Forbush decreases}. In this  thesis we are using the term Forbush decrease   more selectively to apply to only those with a sudden onset and a gradual recovery.

 FDs \index{Forbush decreases} are variously thought to be due to
\begin{itemize}
\item The CME-driven shock, which acts as a propagating diffusive barrier, shielding cosmic rays.  The cosmic ray flux behind the shock is lower in comparison to that ahead of the shock\index{shock}. This is manifested as a FD as the shock sweeps past the observer. 
\item The magnetic cloud\index{magnetic cloud}; there is a deficit of high energy cosmic rays inside the CME\index{CME}/ magnetic cloud (we will discuss this further in chapter \ref{model}). The FD is a manifestation of the low-density cavity (magnetic cloud) engulfing  the earth. 
\item There are  "Two-step" FDs in which the first step is due to the shock (propagating diffusive barrier) and the second step due to near-earth CME/magnetic cloud.

\end{itemize}

\begin{figure}[h]
\includegraphics[width = \textwidth]{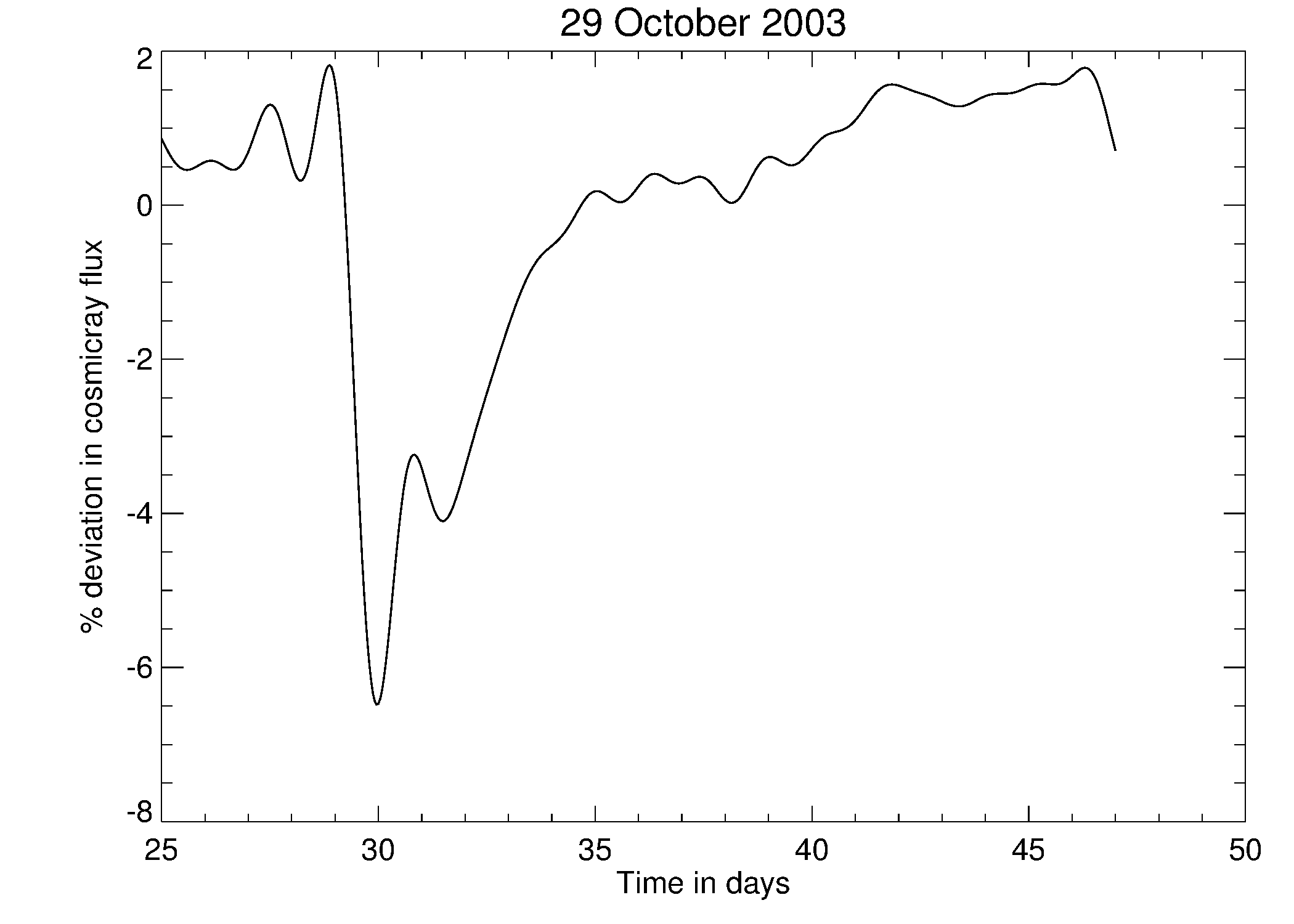}
\caption[FD on 2003 October 29 ]{FD \index{Forbush decreases} observed in GRAPES-3 \index{GRAPES-3} on 29 October 2003, This FD was associated with the Halloween events in 2003. The `X' axis in the time in days starting from the date 1 October 2003 and the percentage deviation of the cosmic ray intensity from the pre-event background is plotted on the `Y'-axis.  }\label{oct29}
\end{figure} 

Figure \ref{oct29} is an example of a classic FD \index{Forbush decreases}. This event was observed using the GRAPES-3 muon telescope\index{GRAPES-3!muon telescope}. The graph shows the percentage deviation of the cosmic rays \index{cosmic rays} from an average cosmic ray intensity which is passed through a low-pass filter to eliminate the fluctuations having frequency more than 1/day. The details of the filter will be discussed later in this chapter. The average value is calculated using a suitable 28 day period; for this event it is from 20 October to 16 November 2003. 

\begin{figure}
\includegraphics[width=0.9\textwidth]{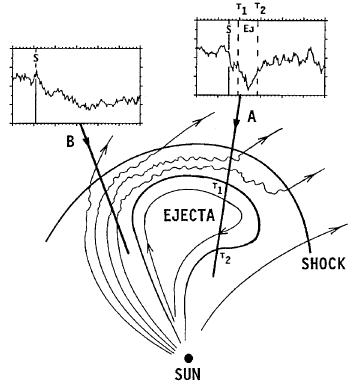}
\caption[The large-scale structure of a fast ejecta and associated shock.....]{ Adapted from \citet{cane00}. The large-scale structure of a fast ejecta and associated shock\index{shock}. The upstream solar wind \index{solar wind} is draped around the ejecta and heated and compressed at the front of the ejecta. Two paths through the ensemble are indicated with differing resultant cosmic ray profiles. The time of shock passage is indicated by a vertical line marked S and the start and end times of ejecta passage are marked T1 and T2. Only if the ejecta is intercepted is a two-step decrease be observed.} \label{CAN00}
\end{figure}

Figure \ref{CAN00} (adapted from \citealp{cane00}) illustrates the large scale structure of the CME (often called  ``ejecta'') and its associated shock and how the cosmic ray \index{cosmic rays} response is related to the path through the CME shock system. If an observer encounters the shock and its associated ejecta as shown in path A, a two-step FD \index{Forbush decreases!two-step} is observed. A less energetic ejecta which does not create a shock \index{shock} will cause only a short-duration one component/step decrease as it passes by. Since shocks have a greater longitudinal extent than ejecta,  there is a   a possibility of intercepting the shock but not the ejecta as shown by path B. In this case, only the effect due to the shock is evident. 
The relative contribution of CME \index{CME} and shock for causing  a  FD is a matter of debate in the scientific community. We will be addressing this problem in  Chapter \ref{model}.

Very energetic CMEs \index{CME} create shocks which are strong enough to cause significant cosmic ray \index{cosmic rays} decreases for observers who detect the shocks beyond the azimuthal extent of the CMEs. In such cases the shocks will also produce major solar energetic particle increases with profiles characteristic of events \citep{can88}. These solar energetic particles allow one to be sure that the cosmic ray decrease was caused by a CME-driven shock \index{shock} and not by a co-rotating stream.

\subsection{Observations of FDs} \index{Forbush decreases!observation}
It was \citet{forb37, forb38} who first identified the sudden decrease in the intensity of ionization chambers located at Earth, a day or two after large solar flares and almost simultaneously with large geomagnetic storms\index{geomagnetic storms}. \citet{fan60a, fan60b} made the first experimental observation using the cosmic ray detector carried by the Explorer {\textsc {vi}} satellite,  which  clearly showed that these decreases took place at large distances from Earth.  FDs were studied using data from the satellite based, balloon-borne and ground based cosmic ray detectors. Currently FDs are extensively studied using neutron monitors and muon detectors. In this thesis we use data from the GRAPES-3 muon telescope\index{GRAPES-3!muon telescope}, which is described in section \S \ref{grapes}.

\subsection{FD magnitude} \index{Forbush decreases!magnitude}
In general the cosmic ray intensity is represented as a percentage variation from an average value. The magnitude of FD is the difference between the pre-event intensity of cosmic rays and the intensity at the minimum. The largest FDs observed in neutron monitors have magnitudes of 10-25 \%. The FD magnitudes reported by different neutron monitors may vary because of the anisotropies present in neutron monitor data. The magnitude will be smaller if daily averages are used rather than hourly averages. FD magnitudes are typically higher for lower rigidities. \citet{loc86} and \citet{can93} found that the ratio of FD magnitudes \index{Forbush decreases!magnitude} found by IMP 8 (median rigidity \index{rigidity!median rigidity}$\sim 2 \, GV $) relative to Mt.Wellington/Mt. Washington (median rigidity $\sim 8 \, GV $) was about 2 for those events in which there were no accelerated particles.

\subsection{Rigidity dependence} \index{rigidity}

The rigidity of a high energy proton is its momentum per unit charge; it may be represented as 
\begin{equation}
Rg (volts) \,  = \, \frac{Pc}{Ze} \, = \, 300 B (gauss) R_L (cm) 
\end{equation}
where $P$ is the proton momentum, $c$ the speed of light, $Z$ the charge state (= 1 for a proton), $e$ the charge of an electron, $B$ the magnetic field in Gauss, and $R_L$ the proton gyro-radius in $cm$.

The rigidity dependence of the magnitude of FDs \index{Forbush decreases!magnitude} is approximately equal to $Rg^{-\gamma}$ where $\gamma$ ranges from about 0.4-1.2 \citep{cane00}. A number of researchers have examined whether the rigidity dependence of FDs \index{Forbush decreases} varies with the Sun's polarity.  All these  groups have concluded that the rigidity dependence does not vary with the polarity of the Sun's magnetic field (see, e.g., \citealp{mor90}). \citet{koj13} found that for neutron monitors of median rigidity 10.0$\sim $31.6 GV the power law index $\gamma$ of rigidity dependence is $0.65\pm 0.05$, where as for the muon telescopes of median rigidity \index{rigidity!median rigidity}64.4$\sim $92.0 GV the power law index is $1.26\pm0.08$.

\subsection{Recovery characteristics}\index{Forbush decreases!Recovery}

The cosmic ray intensity typically recovers gradually in the FD profile. For FDs with relatively clean profiles, the recovery can be exponential with an average recovery time of $\sim$5 days but ranging from $\sim$3 to $\sim$10 days \citep{loc86}. The recovery time of a FD is dependent on the longitude of the solar source region \citep{bar73, iuc79b, can94}. \citet{loc86} found that the recovery time was independent of rigidity and there is  no dependence on solar polarity or time in the solar cycle. \citet{kumar14} state that the characteristic recovery time of galactic cosmic ray depression increases with galactic cosmic ray effectiveness; it is larger for shock- associated CMEs \index{CME} as compared to those not associated with shocks\index{shock}.

\subsection{FD and CR anisotropies} \index{cosmic rays!anisotropy} 

Anisotropies are often observed preceding and during the FDs\index{Forbush decreases}. FDs display anisotropies both in, and perpendicular to, the ecliptic plane. These anisotropies are related to the structure of the associated solar wind\index{solar wind}. Anisotropies are mostly observed near shock passage and inside ejecta \citep{cane00}. Although the anisotropy of CR changes by magnitude and direction during the FD, the fastest variations typically occur near the interplanetary shock \index{shock} and close to the FD minimum \index{Forbush decreases!minimum} \citep{bel09}. Details of the structure of the interplanetary disturbance are reflected by the cosmic ray  anisotropy. In particular, the boundaries of magnetic cloud \index{magnetic cloud} are normally clearly seen in the behaviour of the anisotropy \citep{bel09}. There are also periods of enhanced diurnal waves during the recovery of FDs\index{Forbush decreases!Recovery}. For a summary of early work see \citet{dug78}. 

\subsection{FD precursors}
Many FDs are associated with precursors\index{Forbush decreases!precursors}. Anomalies in the cosmic ray intensity distribution such as pre-increases or pre-decreases along with the anisotropy are often observed. Precursor increases can be due to  reflection of particles from the shock or acceleration at the shock. The Precursor decreases are due to the collimated outflow of the low-density cosmic rays across the shock from the inside of FD \citep{nag92}. The collimation of the flow is due to the bottle-neck nature of the interplanetary magnetic field \index{Interplanetary magnetic field} infront of the shock\index{shock}.  \citet{mun00} showed that FD precursors can be observed prior to the geomagnetic disturbances. These changes are observed from one to 24 hours before the arrival of the shock \citep{pap12} and can be used to forecast the intensity of the impending geomagnetic storms\index{geomagnetic storms}.

\subsection{Solar associations}

Large FDs \index{Forbush decreases} are caused by fast CMEs \index{CME} and  interplanetary shocks \index{shock} associated with them,  which can be associated with specific solar flares. \citet{dug77} suggested that flares could not be the causes of FDs based on a superposed epoch analysis between flares and cosmic ray variations. \citet{can96} studied FDs with magnitude \index{Forbush decreases!magnitude} $\ge \, 4 \%$ and determined which are flare related based on the presence of associated energetic particle events. The flare-associated FDs are in general caused by more energetic CMEs.

FDs are generally associated with fast CMEs. 
\citet{bel14} found that the  mean speed for FD associated CMEs is $(727 \pm 24) \,  km \, s^{-1}$, compared with $(402 \pm 2) \,  km \, s^{-1}$, for the general population. The average speed of CMEs, which are associated with FDs is close to that of CMEs associated with magnetic clouds \index{magnetic cloud} \citep{gopal10}. 

FDs are caused by the Earth-directed halo or partial halo CMEs. \citet{bel14} found that the mean width of FD  associated CMEs is $220 \pm 6^{\circ} $ , compared with $58 \pm 1 ^{\circ}$ for the general population. The wider the CME, the higher the possibility of FD occurrence. 
 
\subsection{Associations with interplanetary disturbances}

Faster the propagation of interplanetary disturbance and stronger its magnetic field the stronger will be the FD and faster will be decrease of CR density. \citet{bel09} states that the magnitude of FD \index{Forbush decreases!magnitude} is proportional to the the product of solar wind velocity and its magnetic field. \citet{bel01} calculated the correlation \index{correlation} of FD magnitude with the quantity $V_{max} H_{max}$, where $V_{max}$ is the ratio of maximum solar wind \index{solar wind} speed to the average ambient solar wind speed ($400 \, km \, s^{-1}$) and $H_{max}$ is the ratio of maximum magnetic field strength to ambient magnetic field \index{Interplanetary magnetic field} strength ($ 5 \, nT$). They found a correlation of 70 \% between the FD magnitude and $V_{max} H_{max}$, where as the correlation of FD magnitude with $H_{max}$ is 66 \% and with $V_{max}$ is 22 \%. For the powerful disturbances ($ V_{max} H_{max} \, \ge \, 8.4$) the correlation was 99 \%. 

\citet{dum12} studied the correlation of FD parameters (FD magnitude $|FD|$, duration of FD $t_{FD}$ ) with the solar wind parameters (amplitude of magnetic field enhancement \index{Interplanetary magnetic field!enhancement} $B$, amplitude of the magnetic field fluctuations $\delta B$, maximum solar wind speed associated with the disturbance $v$, duration of the disturbance $t_B$). They have also used products of these parameters such as $Bt_B$ as the proxy for the time integral of the IMF perturbation, the $Bv$ product as the proxy for maximum (convective) electric field, $Bvt_B$  as the proxy for the magnetic flux (per unit-width of the disturbance cross-section), and $|FD|t_{FD}$ as a measure of the total amount of CRs reduced by the passage of the disturbance. They found a correlation \index{correlation} with the FD parameters with the solar wind parameters for the ICME associated events, where as there was a lack of correlation for the CIR associated events. Similarly, they found good correlation for the shock \index{shock} associated events and no correlation for events lacking shocks.




\subsection{FDs and cosmic ray diffusion} \index{diffusion}

It is well known that charged particles cannot freely move ``across'' magnetic field lines. A typical charged particle is caught by a magnetic field line in a circular trajectory, whose radius (called the gyroradius or Larmor radius) is directly proportional to the speed of the particle and inversely proportional to the strength of the magnetic field. The CME \index{CME} is composed of coronal plasma and structured magnetic fields and the CME-driven shock compresses plasma ahead of it. The magnetic field has an ordered, as well as a turbulent component. Cosmic rays \index{cosmic rays} generally diffuse through the tangled, turbulent magnetic field; this diffusion process is inhibited in the vicinity of the CME and shock\index{shock}, which has a strong ordered magnetic field component. It is the inability of the cosmic ray particles to diffuse efficiently across the magnetic field structures that causes the observed FD\index{Forbush decreases}.

\section{GRAPES-3} \label{grapes}

\begin{figure}
\centering
\includegraphics[width=\textwidth]{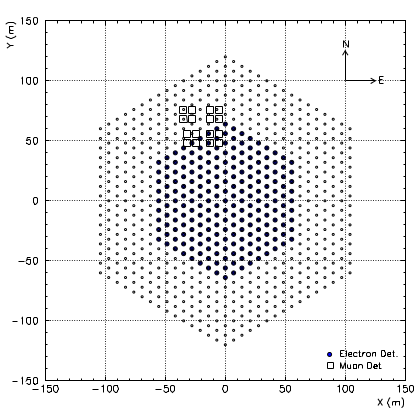}
\caption[A schematic layout of  GRAPES-3  shower array]{(Adapted from \citealp{non03}) A schematic layout for the 721 detector (open circles) GRAPES-3 \index{GRAPES-3} shower array of which 217 detectors (filled circles) used at present are shown. Each of the 16 squares represents a $35 \, m^2$ area muon tracking detector with $E_{\mu}$ $\ge$ 1 GeV.} \label{GR3}

\end{figure}

The  GRAPES-3 \index{GRAPES-3}  (Gamma Ray Astronomy at PeV EnergieS- phase 3) experiment is located at Ooty ($11.4^{\circ}$ N latitude, $76.7^{\circ}$ E longitude, and 2200 m altitude), a popular mountain resort town in southern India. The GRAPES-3 \index{GRAPES-3}  is a high density extensive air shower (EAS) array designed for precision study of the cosmic ray (CR) \index{cosmic rays} energy spectrum and its nuclear composition using the muon multiplicity distribution (MMD) in the energy range from $3\times 10^{13} \, eV$ to $3\times 10^{16} \, eV$. The  GRAPES-3 \index{GRAPES-3}  air shower experiment has been designed to have one of the most compact configurations of the conventional type arrays with a separation of only 8 m between the adjacent detectors which are deployed in a symmetric hexagonal geometry. A schematic layout of the  GRAPES-3   array is shown in figure \ref{GR3}. Observations were started in early 2000 with 217 detectors, located within the inner 8 rings which are shown as filled circles in figure \ref{GR3}. The array also contains a large area (560 $m^2$) tracking muon telescope \index{GRAPES-3!muon telescope} to measure the muon component and obtain the MMD of the EAS.

A very large-area tracking muon telescope \index{GRAPES-3!muon telescope} operating is a part of the  GRAPES-3 \index{GRAPES-3}  experiment (\citealp{gupta05, hay05}). It is capable of providing a high-statistics, directional study of muons. The  GRAPES-3   muon telescope covers an area of 560 $m^2$, consisting of a total 16 modules, each 35 $m^2$ in area. The square boxes in the Fig \ref{GR3} represents these modules, which are located close to each other as shown in the figure. A cluster of four neighbouring modules of 35 $m^2$ area, located inside a common hall, constitutes one supermodule with a total area of 140 $m^2$. The energy threshold of the telescope is 1 $GeV$ for the muons, which are  arriving along the vertical direction. The cut-off rigidity \index{rigidity!cut-off rigidity} due to the magnetic field of the Earth at Ooty is $17  \, GV $ in the vertical direction and  across the field of view of the telescope in different directions the cut-off rigidity  varies from $12 \, to \, 42 \, GV$  as shown  in figure \ref{CRig}.

\begin{figure}
\includegraphics[width=\textwidth]{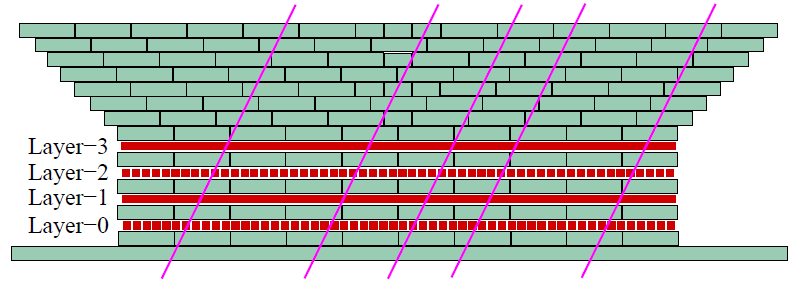}
\caption[ A schematic display of the 4-layer tracking muon telescope \index{GRAPES-3!muon telescope} module with 58 PRCs per layer]{ (Adapted from \citealp{sub09}) A schematic display of the 4-layer tracking muon telescope module with 58 PRCs per layer. The four layers of the PRCs labelled Layer-0, Layer-1, etc. are embedded in concrete blocks. Inclined lines represent a set of parallel muon tracks.} \label{muontel}

\end{figure}

A rugged proportional counter (PRC) is the basic detector element of the GRAPES-3 muon telescope. This PRC is a $600 \, cm $ long mild steel square pipe with a square cross-sectional area of $10 \, cm  \times 10 \, cm$, and a wall thickness of $2.3 \, mm $. A muon telescope \index{GRAPES-3!muon telescope} module with a sensitive area of 35 $m^2$ consists of a total of 232 PRCs arranged in four layers, each layer is of 58 PRCs. The  alternate layers are placed in orthogonal directions. Two successive layers of the PRCs are separated by a 15 $cm$ thick concrete layer, consisting of $60\times 60\times 15 \, cm^3$ blocks as shown in figure \ref{muontel}. The four-layer PRC configuration of the muon modules allows a 3-D reconstruction of the muon track direction to an accuracy of $\sim$ $6^{\circ}$. The accuracy gradually increases with increasing zenith angle due to the greater separation of the triggered PRCs. 

\begin{figure}
\includegraphics[width=\textwidth]{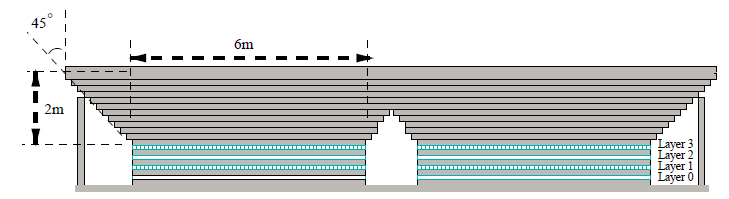}
\caption[Schematic of a muon detector module showing 4 layers of proportional counters embedded within concrete blocks.]{ (Adapted from \citealp{hay05}) Schematic of a muon detector module showing 4 layers of proportional counters embedded within concrete blocks.} \label{muontel1}

\end{figure}

\begin{figure}[h]
\includegraphics[width=\textwidth]{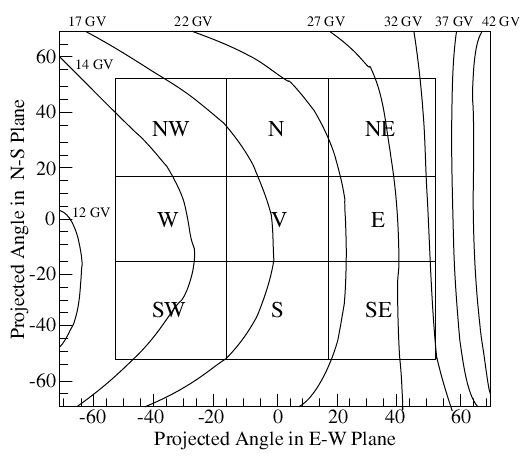}
\caption[The 9 coarse solid-angle bins are shown along with the contours of constant geomagnetic cut-off rigidity  \index{rigidity!cut-off rigidity} in the FOC of GRAPES-3]{(Adapted from \citealp{sub09}) The 9 coarse solid-angle bins are shown along with the contours of constant geomagnetic cut-off rigidity in the field of view (FOV). Cut-off rigidity varies from 12 to 42 GV in the FOV of  GRAPES-3 \index{GRAPES-3}.} \label{CRig}

\end{figure}

To achieve the  1 $GeV$  energy threshold for vertical muons, an absorber of total thickness $\sim \, 550 \, g \, cm^{-2}$ in the form of concrete blocks is employed. This is achieved by placing 15 layers of concrete blocks above Layer-1, as shown schematically in figure \ref{muontel} and \ref{muontel1}. The robust structure of PRCs permits it to support the huge load of 2.4 $m$ thick of concrete absorber in a self-supporting manner. The concrete blocks are arranged in the shape of an inverted pyramid to shield the PRCs, with coverage up to $45^{\circ}$ around the vertical direction for the incident muons. {The threshold energy for muons arriving  other than vertical direction will increase depend up on their incident angel. The threshold energy changes to $sec \theta$ GeV ($\frac{1GeV}{cos \theta}$) for the muons incident at a zenith angle of $\theta$}. The cross section of a muon telescope \index{GRAPES-3!muon telescope} module is shown schematically in figure \ref{muontel}. A cluster of four such modules, which are separated by a horizontal distance of 130 $cm$ at the base constitutes one supermodule. The  GRAPES-3 \index{GRAPES-3}  muon telescope contains a total of four supermodules \citep{hay05}. A cross-section of a supermodule is shown in figure \ref{muontel1}.

The top 7 layers of the concrete blocks are shared as an absorber by all four modules in a super-module. Therefore, the absorber above the $10^{th}$ layer from the base, is essentially a full-size layer, which  cover all four modules \citep{hay05} as seen in figure \ref{muontel1}. Finally, a 30 $cm$ thick and $18 m \times 18m$ concrete slab was cast as the top layer of the absorber in order to make it weather-proof. Similarly at the bottom, a single 30 $cm$ thick, $16m \times 16m$ concrete block cast on the ground, serves as the floor for the super-module. This has been done to uniformly distribute the 1200t load of the absorber  to the soil below.

The PRC absorber assembly of each super-module has been enclosed within a large hall. The hall is provided with suitable doors and windows for access and illumination, and to control the humidity for safe operation of the PRCs. Two heavy duty de-humidifiers are operated round the clock, inside each of the four halls to maintain a low level of humidity ($ < 50\%$) , in view of nearly eight month long rainy season at Ooty. 

The PRCs are sensitive to the low-energy $\gamma$ rays from the radioactivity present in the concrete absorber. Because of this individual PRCs display sizable counting rates of $\sim \, 200 \, Hz$.  When any one of the 58 PRCs produces a signal an output will be generated. A logical `OR' of outputs from all 58 PRCs in a layer is generated, after suitable amplification and shaping to form the layer OR output. A coincidence of the four OR outputs from the four layers in a module is used to generate the 4-layer trigger.  Despite the high counting rates of the individual PRCs due to the radioactivity that is present in the surrounding absorber, the 4-layer coincidence triggers are relatively free from this background and it is caused only by the passage of a muon. The observed 4-layer muon counting rate of $\sim \, 3200 \, Hz$ per module yields a total counting rate $\sim \, 3 \times 10^6 \, min^{-1}$ for all 16 modules. This high rate permits detection of small changes of $\leq$ 0.1 \% in the muon flux over a time-scale of $\sim$ 5 min, after application of appropriate correction to the variation in the atmospheric pressure with time \citep{sub09, non03}.

Most of the detected muons are generated by  galactic cosmic rays \index{cosmic rays} of energy $\gtrsim$ 20 $GeV$ and form a stable and dominant background to the variation in their flux produced by the  CME \index{CME} /solar flare. The muon data is grouped online every 10 sec, into solid angle bins of $\sim \, 0.05 \, sr$, consistent with the angular resolution of the muon telescope \index{GRAPES-3!muon telescope} as described in section \S\ref{DA}. These observations can be used to probe the effect of the Sun on cosmic rays \index{cosmic rays}, since > 1 GeV muons are secondaries produced by the primary protons of energy $\gtrsim$ 20 GeV in the atmosphere.

\section{Data Analysis}
\label{DA}
\begin{figure}[h]
\includegraphics[width=\textwidth]{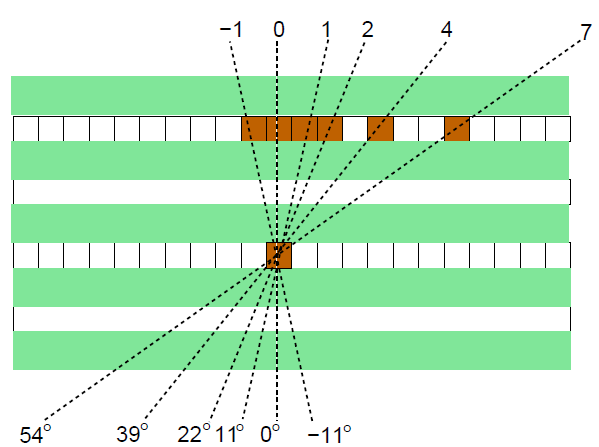}
\caption[A schematic view of muon arrival angle selection ]{(Adapted from \citealp{sub09}) A schematic view of muon arrival angle selection based on the PRC triggered in the lower and 15 PRCs in the upper layer. The triggered PRCs are shown as filled squares.} \label{ang}

\end{figure}

The direction of muons is recorded into 225 solid-angle bins in all four supermodules of the muon telescope\index{GRAPES-3!muon telescope}. A dedicated direction-sensitive trigger with an independent data acquisition system for each of the four supermodules is used to get the 225 solid-angle map of muon directions. The muon angle is determined for each PRC in the lower layer and binned into 15 angular bins based on the specific location of the PRC triggered in the upper layer from among the 15 PRCs , one directly above (central PRC) and 7 each on either side of the central PRC (see figure \ref{ang}). This angular binning is carried out in each of the two orthogonal projections (XZ and YZ; Z is the vertical direction). This generates a 2-dimensional $15 \, \times \, 15 \, = \, 225$ solid-angle map of muon directions. The contents of the 225 solid-angle bins are recorded, once every 10 $sec$, which provide a continuous monitoring of the directional flux of muons in the sky.

\begin{figure}
\includegraphics[width=\textwidth]{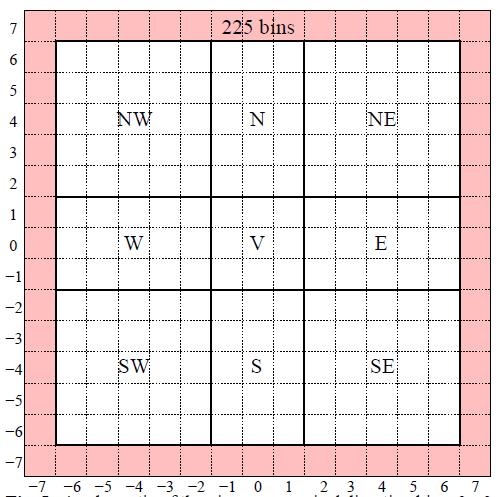}
\caption[A schematic of the nine muon arrival direction bins]{(Adapted from \citealp{sub09}) A schematic of the nine muon arrival direction bins; $3 \, \times \, 3 $
vertical bin V, and four $3 \, \times \, 5 $ central bins N, E, W, S, and four
$5 \, \times \, 5 $ outer bins NE, SE, SW, NW.} \label{G225}

\end{figure}

The muon rate variations can be studied in any of the 225 solid-angle bins. It is expected that the influence of a solar flare and/or  CME \index{CME}  would be spread over several bins. { This directional spread can be due to the influence of the terrestrial, solar, and interplanetary magnetic fields\index{Interplanetary magnetic field}, etc.} Because of this directional spreading  the detected muons have been regrouped into $3 \, \times \, 3 \, = \, 9$ coarse solid-angle bins, as shown schematically in figure \ref{G225}. In this regrouping we eliminated the outer bins hence a total of 169 out of 225 is used, in the figure \ref{G225} the white bins show the chosen 169 bins and the pink ones are the excluded. This regrouping of the data was done by combining either a set of $3 \, \times \, 5 $ or $5 \, \times \, 5 $ fine solid-angle bins. The exception being the vertical direction where $3 \, \times \, 3 $ bins have been combined. The muon flux is comparatively larger for the near central directions (N, E, W, S) than for the outer directions (NE, SE, NW, SW). The choice of angular segmentation was dictated by this fact. This choice results in a relatively uniform solid-angle coverage for the nine coarse bins. The solid-angle of acceptance includes only $13 \, \times \, 13 \, = \, 169 $ out of  225 total bins. This eventually  restricts the maximum zenith angle to $50^{\circ}$. For the PRCs at the outer edge this zenith angle  exceeds the shielding coverage of $45^{\circ}$, but such events constitute <1\% of the data. This regrouping also results in muon statistics for various bins that are almost similar.

The energies of interest here are  $\gtrsim \, 20 \, GeV$. At these energies the propagation of charged particles near the Earth (< 20 $R_E$, where $R_E$ is the radius of the Earth) is strongly influenced by the geomagnetic field\index{geomagnetic field}. The access by a charged particle to a given geographical location depends on its rigidity, the momentum per unit charge of the particle. The threshold value of the rigidity is termed ``geomagnetic cut-off rigidity ". This cut-off rigidity \index{rigidity!cut-off rigidity} depends on the geographical location on the Earth and the direction of the arriving particle. The geomagnetic cut-off rigidity can be calculated using a detailed model of the geomagnetic field \citep{cooke}. The geomagnetic cut-off rigidity for the field of view (FOV) of the  GRAPES-3 \index{GRAPES-3}  muon telescope \index{GRAPES-3!muon telescope} varies significantly for the nine coarse, solid-angle bins. We used the International Geomagnetic Reference Field 2000 (IGRF2000) geomagnetic field model \citep{mandea} in order to calculate the cut-off rigidity for the centre of each of the 169 fine solid-angle bins, which constitute the 9 elements of the FOV. Subsequently, a weighted mean of the cut-off rigidities of the fine bins (which constitute a coarse bin) is calculated for each of the 9 coarse bins. These weights are the muon counting rates for a given fine bin. It need to  be emphasised here that the knowledge of the geomagnetic field at any particular moment in time is imperfect. It is virtually impossible to determine the cut-off to a high degree of accuracy. Even though the calculated values of the geomagnetic cut-off represent a very useful approximation to the true values at the time of the observations.  In figure \ref{CRig} the contours of constant geomagnetic cut-off rigidity \index{rigidity!cut-off rigidity} in the FOV are superimposed over a schematic of the 9 solid-angle bins of muon arrival directions. Within the FOV of the  GRAPES-3 \index{GRAPES-3}  tracking muon telescope\index{GRAPES-3!muon telescope} , the geomagnetic cut-off rigidity varies from 12 $GV$ in the west to 42 $GV$ in the east

\begin{figure}
\includegraphics[width=\textwidth]{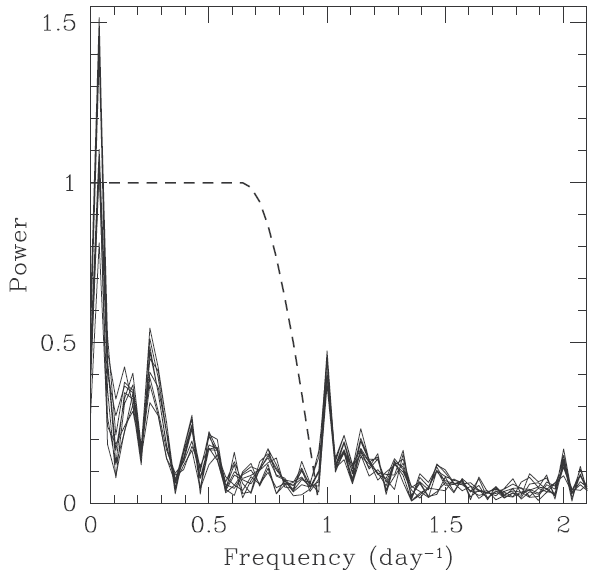}
\caption[The power spectrum of 28 day data]{(Adapted from \citealp{sub09}) The power spectrum of 28 day data covering the period from 26March to 22 April 2001. The solid lines show the results for all 9 directions used in the study while the dashed line shows the function that is used for filtering \index{GRAPES-3!filter} out the high-frequency components. The Fourier transform is multiplied by this function before taking the inverse Fourier transform to get filtered time series.} \label{filt}

\end{figure}

For the studies we carried out in this thesis, we used the  GRAPES-3 \index{GRAPES-3}  data summed over a time interval of one hour for each of the nine bins. These bins are identified as NW, N, NE, W, V, E, SW, S, and SE. The calculated mean cut-off rigidities for these nine bins are listed in the table \ref{cuto}, and depicted in figure \ref{CRig}. The summing over an interval of 1 hour improves the signal-to-noise ratio. The diurnal variations in the muon flux are present even after this summing over an interval of 1 hour. We adapted a low-pass filter\index{GRAPES-3!filter}, removing all frequencies higher than 1 $day^{-1}$ in order to filter out the oscillations due to diurnal variations. In figure \ref{filt} we show the Fourier transform of the data covering a period of 28 days that includes the FD event of 2001 April 11. The peaks corresponding to diurnal variation and its first harmonic are visible. To remove the high-frequency components,  the Fourier transform is multiplied by the function shown by a dashed line in figure \ref{filt}. This filter is found to be effective in removing high-frequency oscillations, such as  the diurnal variations and their harmonics. The FD \index{Forbush decreases} events are also clearly identifiable   in the filtered data.

\begin{table}
\caption[Cut-off rigidities for 9 bins]{ Mean cut-off rigidities \index{rigidity!cut-off rigidity} of GRAPES-3 muon telescope \index{GRAPES-3!muon telescope} for different bins are shown here. \label{cuto}}
\centering
\begin{tabular}{lc} \hline \hline
Bin & Cut-off Rigidity (GV)\\ \hline
 
	{NE} & {24.0} \\
	{N } & {18.7 }\\
	{NW} & {15.7} \\
	{E } &  {22.4} \\
	{V } &  {17.2} \\
	{W } &  {14.3} \\
	{SE } &  {22.4} \\
	{S } &  {17.6} \\
	{SW } &  {14.4} \\ \hline \hline
	\end{tabular}
\end{table}

\begin{figure}
\includegraphics[width=\textwidth]{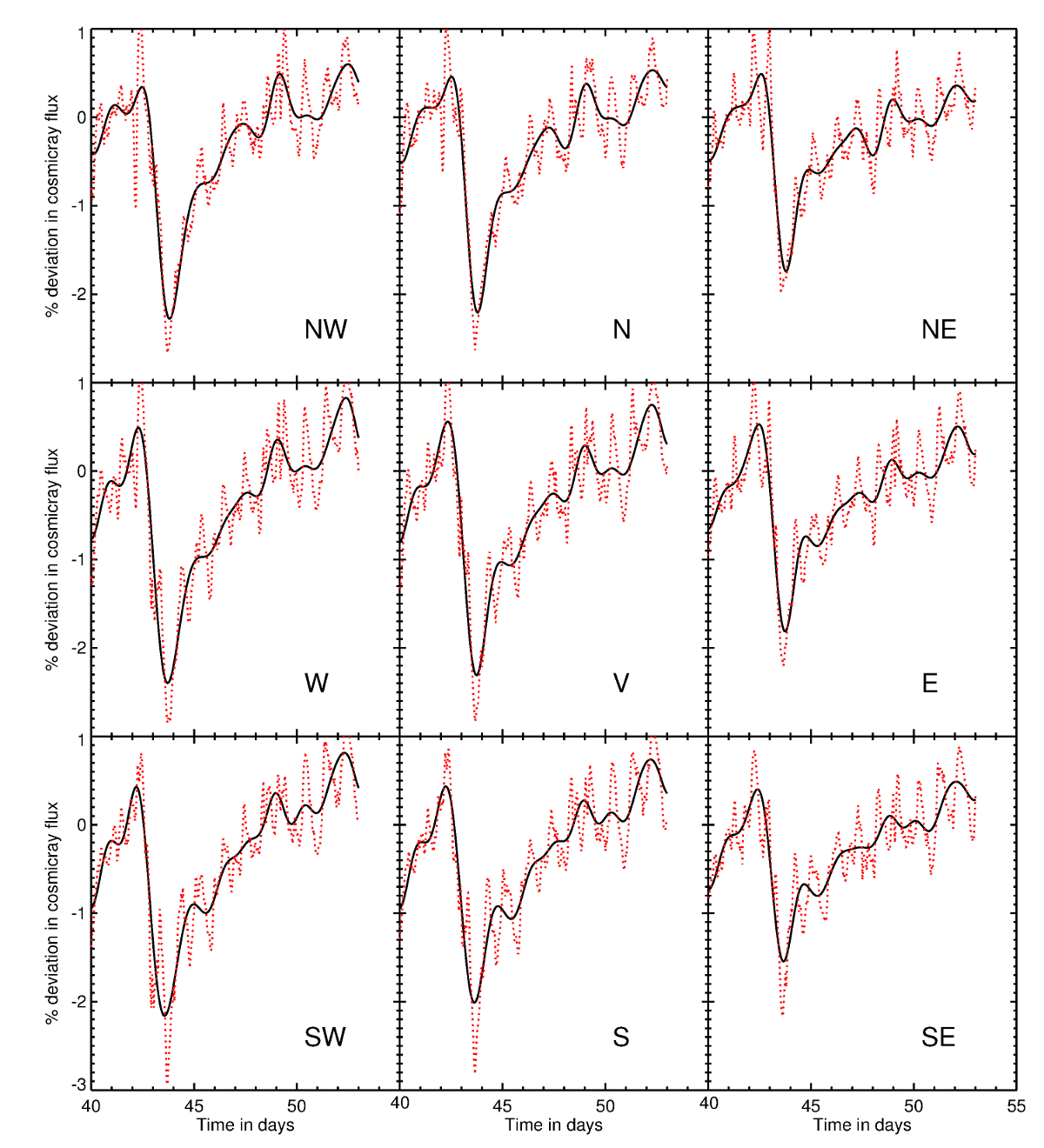}
\caption[FD event on 2001 April 11]{FD event observed by GRAPES 3 \index{GRAPES-3} on 2001 April 11. The nine panels shows the nine different bin of the muon telescope. The X axis shows the time in days starting from 2001 March 1, and Y axis shows the percentage deviation of the cosmic ray intensity. The black line shows the filtered data after using the low pass filter explained in the section \ref{DA} and the red dotted line shows the unfiltered data.} \label{apr11unf}

\end{figure}

It may be noted that the smoothing may tend to change the amplitude of the decrease and possibly shift the onset time for the FD by a few hours in some cases. 
 However, it is often difficult to determine whether these differences are artifacts of smoothing or the unfiltered data showed a different amplitude because a diurnal oscillation happened to have the right phase, so as to enhance or reduce the amplitude of the FD \index{Forbush decreases}.
 Some fluctuations in the muon flux could be due to FD and associated events. It is unlikely that these will be periodic in nature, and are not likely to be affected by the filter\index{GRAPES-3!filter}. The differences in amplitude caused by filter are not substantial, and are therefore unlikely to affect our results significantly. Thus in our work, we use the filtered data to study the characteristics of FD events.

An example of a FD observed by GRAPES-3 on 2001 April 11 is shown in the figure \ref{apr11unf}, the unfiltered data (dotted red line) are also shown in  figure along with the filtered data (solid black line ) for direct comparison. We repeated the calculations using both filtered and unfiltered data and find that the final results are not significantly different.  There can also be anisotropies intrinsic to the  CME \index{CME} /magnetic cloud \index{magnetic cloud} itself. These anisotropies \index{cosmic rays!anisotropy} can be arising from a $B\times \nabla N$ drift, where $B$ is the interplanetary magnetic field and $\nabla N$ denotes the cosmic-ray density gradient inside the  CME \index{CME}  \citep{Bieb98, mun03, mun05, kuw04}. Such anisotropies can potentially be ``mixed" with the diurnal anisotropy, and it is possible that there will still be some residual anisotropy even after the filter is applied. This is the case with one of the weaker events that we have studied in the chapter \ref{model}; namely 2003 November 20.

\section{Summary} 

 We have provided a short introduction to FDs in the section \S\ref{fode}. Characteristics such as the magnitude of FDs, their rigidity dependence, recovery characteristics, FD and CR anisotropies, Fd precursors, Solar associations and association with interplanetary magnetic fields are explained in this section. 

A brief introduction to the GRAPES-3 muon telescope is given in section \S\ref{grapes}.
The Data analysis methods used in order to identify the FDs in GRAPES-3 data are mentioned in section \S\ref{DA}.

\chapter{Forbush decrease models}
\label{model}

\noindent\makebox[\linewidth]{\rule{\textwidth}{3pt}} 
{\textit {The relative contributions of shocks and coronal mass ejections (CMEs) in causing FDs is a matter of debate. In this chapter we  identify the  major contributor in causing FDs.  We use multirigidity data from GRAPES-3 to check the validity of  two models:  i) the CME-only cumulative diffusion model, where we assume that the entire FD is caused only by the CME,  ii) the Shock-only model, where we assume the FD is caused only by the propagating diffusive barrier i.e, the shock. 
 }  }\\
\noindent\makebox[\linewidth]{\rule{\textwidth}{3pt}}

\section{Introduction}

Short-term decreases in the intensity of the galactic cosmic rays \index{cosmic rays} at the Earth are called Forbush decreases \index{Forbush decreases}. They are typically caused by the effects of interplanetary counterparts of coronal mass ejections (CMEs) from the Sun, and also by corotating interaction regions (CIRs) between the fast and slow solar wind \index{solar wind} streams from the Sun. In this chapter we  concentrate on CME driven Forbush decreases. The near-Earth manifestations of CMEs \index{CME} from the Sun typically have two major components: the interplanetary counterpart of the CME, commonly called an ICME, and the shock which propagates ahead of the CME. 

The relative contributions of shocks \index{shock} and ICMEs in causing Forbush decreases is a matter of debate. For instance, \citet{zha88, loc91, rea09}  argue against the contribution of magnetic clouds to Forbush decreases \index{Forbush decreases}. On the other hand, other studies (e.g., \citealp{bad86, sand90,kuw09}) concluded that magnetic clouds \index{magnetic cloud} can make an important contribution to FDs. {There have been recent conclusive associations of Forbush decreases with Earth-directed CMEs \index{CME} \citep{blanc13, oh12}. \citet{cane00} introduced the concept of a ``2-step'' FD, where the first step of the decrease is due to the shock and the second one is due to the ICME. Based on an extensive study of ICME-associated Forbush decreases at cosmic ray energies between 0.5 -- 450 MeV, \citet{rich11} conclude that shock and ICME effects are equally responsible for the Forbush decrease \index{Forbush decreases}. They also find that ICMEs that can be classified as magnetic clouds \index{magnetic cloud} are usually involved in the largest of the Forbush decreases they studied. From now on, we will use the term ``CME'' to denote the CME \index{CME} near the Sun, as well as its counterpart observed at the Earth.

In this chapter we have used Forbush decrease \index{Forbush decreases} data from  GRAPES-3   muon telescope \index{GRAPES-3!muon telescope} located at Ooty ($11.4^{\circ}$ N latitude, $76.7^{\circ}$ E longitude, and 2200 m altitude) in southern India. The  GRAPES-3 \index{GRAPES-3}  muon telescope records the flux of muons in nine independent directions (labelled NW, N, NE, W, V, E, SW, S and SE), and the geomagnetic cut-off rigidity \index{rigidity!cut-off rigidity} over this field of view varies from 12 to 42 GV. The details of this telescope are already discussed in chapter \ref{GRP3}. The  GRAPES-3 \index{GRAPES-3}  telescope observes the cosmic ray \index{cosmic rays} muon flux in nine different directions with varying cut-off rigidities simultaneously. 
The high muon counting rate measured by the  GRAPES-3 \index{GRAPES-3}  telescope results in extremely small statistical errors, allowing small changes in the intensity of the cosmic ray flux to be measured with high precision. Thus a small drop ($\sim$0.2\%) in the cosmic ray \index{cosmic rays} flux during a Forbush decrease \index{Forbush decreases} event can be reliably detected. This is possible even in the presence of the diurnal anisotropy of much larger magnitude ($\sim$1.0\%), through the filtering technique described in chapter \ref{GRP3}.

{A schematic of the CME, which is assumed to have a flux-rope \index{flux rope} geometry \citep{angelos13} together with the shock \index{shock}  it drives, is shown in figure~\ref{difn}. The shock drives turbulence \index{turbulence} ahead of it, and there is also turbulence in the CME \index{CME} sheath \index{sheath} region (e.g., \citealp{mano00,rich11}). }

\begin{figure}
\centering 
\includegraphics[width = 0.95\textwidth]{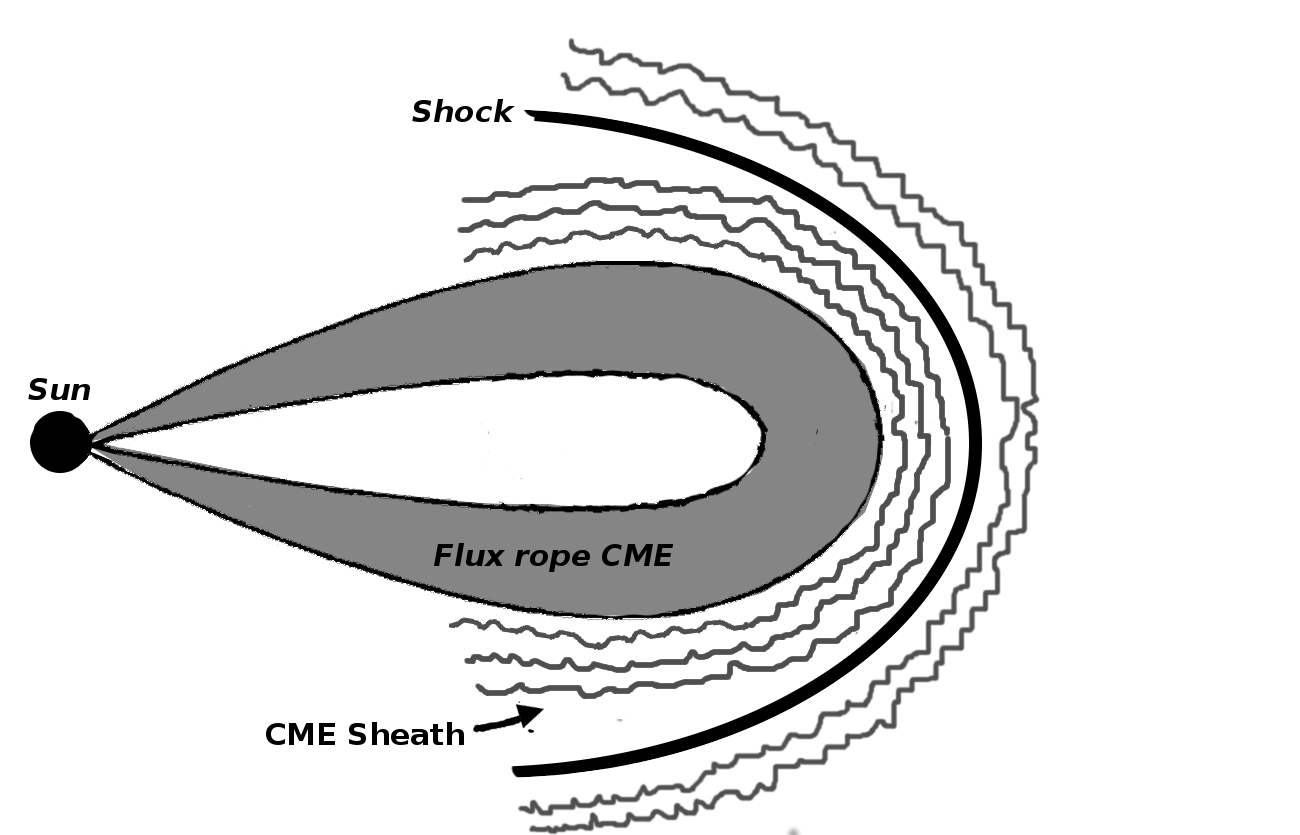}
\caption[A schematic of the CME-shock system]{A schematic of the CME-shock system. The CME \index{CME} is modelled as a flux rope \index{flux rope} structure. The undulating lines ahead of the shock denote MHD turbulence driven by the shock\index{shock}, while those in the CME \index{CME} sheath \index{sheath} region denote turbulence \index{turbulence} in that region.}
\label{difn}
\end{figure}

Instead of treating the entire system shown in figure~\ref{difn}, which would be rather involved, we consider two separate models. The first, that we call the ``CME-only'' model, is one where the Forbush decrease \index{Forbush decreases} is assumed to be exclusively due to the CME, which is progressively populated by high energy cosmic rays \index{cosmic rays} as it propagates from the Sun to the Earth (Figure \ref{crt}). The preliminary idea behind this model was first sketched by \citet{canr95} and developed in detail in \citet{sub09}. The work described here addresses multi-rigidity data, which is a major improvement over \citet{sub09}. We will describe several other salient improvements in the CME-only model \index{CME-only model} in subsequent sections. The second model, which we call the ``shock-only'' \index{shock-only model} model, is one  where the Forbush decrease is assumed to arise only due to a propagating diffusive barrier, which is the shock \index{shock}  driven by the CME \index{CME} (e.g., \citealp{wib98}). The diffusive barrier acts as a shield for the galactic cosmic ray \index{cosmic rays} flux, resulting in a lower cosmic ray density behind it. In treating these two models separately, we aim to identify which is the dominant contributor to the observed Forbush decrease\index{Forbush decreases} ; the CME, or the shock.

We identify Forbush decreases in the  GRAPES-3 \index{GRAPES-3}  data that can be associated with both near-Earth magnetic clouds \index{magnetic cloud} and shocks driven by them. We describe our event short-listing criteria in the next section \S \ref{ESC}. We then test the extent to which each of the models (the CME-only \index{CME-only model} and the shock-only model) satisfies the multi-rigidity Forbush decrease \index{Forbush decreases} data from the  GRAPES-3 \index{GRAPES-3}  muon telescope\index{GRAPES-3!muon telescope}. In the subsequent analysis we use the cut-off rather than  the median rigidity for the following reason. The cut-off rigidity \index{rigidity!cut-off rigidity} in a given direction represents the threshold rigidity of incoming cosmic rays \index{cosmic rays} and the magnitude \index{Forbush decreases!magnitude} of Forbush decrease is a sensitive function of it. On the other hand the median rigidity \index{rigidity!median rigidity}is comparatively insensitive to the magnitude of the Forbush decrease \index{Forbush decreases}.

\section{Event short-listing criteria} \label{ESC}
In our study we short list the Forbush decrease events based on characteristics of the Forbush decrease, magnetic clouds, and CMEs. These short-listing criteria  are described in the following subsections: \S \ref{ESC1}, \ref{ESC2} and \ref{VP}.

\subsection{First short-list: characteristics of the Forbush decrease} \label{ESC1}
We have examined all Forbush decrease \index{Forbush decreases} events observed by the  GRAPES-3 \index{GRAPES-3}  muon telescope \index{GRAPES-3!muon telescope} from 2001 to 2004. We then short-listed events that possess a relatively `clean' profile \index{Forbush decreases!profile} comprising a sudden decrease followed by a gradual exponential recovery and having magnitudes \index{Forbush decreases!magnitude} $> 0.25$~\%. While the figure of 0.25 \% might seem rather small by neutron monitor standards, we emphasize that the largest events observed with  GRAPES-3 \index{GRAPES-3}  have magnitudes of $\sim 1$~\%. This list, which contains 80 events, is called short-list 1. The events in short-list 1 are listed in  table \ref{SL1}

\begin{longtable}{|l|c|c|c|}
\caption [FD events short list 1]{\label{SL1} Forbush decrease \index{Forbush decreases} event short listed by FD profile \index{Forbush decreases!profile} and magnitude \index{Forbush decreases!magnitude} \index{Forbush decreases!minimum} \index{Forbush decreases!onset}}\\
\hline \hline
Event & Magnitude(Ver \%) & FD onset &  FD minimum  \\
\hline 
\endfirsthead
\caption[continuation of FD events short list 1]{Continuation of FD events short listed by FD profile and magnitude }\\
\hline \hline
Event & Magnitude(Ver \%) & FD onset &  FD minimum  \\ \hline
\endhead 
\hline \hline
\endfoot
13 Jan 2001 & 0.2946 &  13 Jan 2001, 01:00 & 13 Jan 2001, 22:00\\
3 Mar 2001 & 0.8967 & 3 Mar  2001, 04:00 & 5 Mar  2001, 23:00 \\
26 Mar 2001 & 1.1520 & 26 Mar  2001, 01:00 & 28 Mar  2001, 01:00\\
4 Apr 2001 & 1.1793 & 4 Apr  2001, 05:00 & 5 Apr  2001, 04:00 \\
7 Apr 2001 & 1.4009 & 7 Apr  2001, 08:00 & 8 Apr  2001, 22:00\\
11 Apr 2001 & 2.8630 & 11 Apr  2001, 09:00 & 12 Apr  2001, 18:00\\
27 Apr 2001 & 1.8250 & 27 Apr  2001, 21:00 & 29 Apr  2001, 02:00\\
27 May 2001 & 0.7825 & 27 May  2001, 06:00 & 28 May  2001, 03:00\\
1 Jun 2001 & 0.6630 & 1 Jun  2001, 19:00 & 2 Jun  2001, 20:00\\
6 Jun 2001 & 0.5928 & 6 Jun  2001, 09:00 & 10 Jun  2001, 12:00\\
1 Aug 2001 & 0.8595 & 31 Jul  2001, 01:00 & 3 Aug  2001, 07:00\\
13 Aug 2001 & 0.9220  & 13 Aug  2001, 03:00 & 14 Aug  2001, 04:00\\
17 Aug 2001 & 1.0296 & 16 Aug  2001, 19:00 &18 Aug  2001, 05:00\\
26 Aug 2001 & 2.0680 & 26 Aug  2001, 01:00 & 28 Aug  2001, 23:00\\
6 Sep 2001 & 0.2977 & 6 Sep  2001, 12:00 & 7 Sep  2001, 21:00\\
12 Sep 2001 & 0.7478 & 12 Sep  2001, 00:00 &14 Sep  2001, 08:00\\
25 Sep 2001 & 2.4309 & 23 Sep  2001, 06:00 & 26 Sep  2001, 08:00\\
29 Sep 2001 & 2.4385 & 25 Sep  2001, 05:00 & 31 Sep  2001, 01:00 \\
11 Oct 2001 & 1.3550 & 11 Oct  2001, 07:00 & 12 Oct  2001, 09:00 \\
27 Oct 2001  & 0.4828 & 27 Oct  2001, 11:00 & 28 Oct  2001, 13:00 \\
5 Nov 2001 & 2.7122 & 5 Nov  2001, 21:00 & 6 Nov  2001, 21:00\\
24 Nov 2001 & 1.5612 & 24 Nov  2001, 02:00 & 25 Nov  2001, 15:00\\
3 Dec2001 & 1.6951 & 2 Dec 2001, 21:00 & 4 Dec 2001, 17:00\\
14 Dec2001 & 1.7215 & 14 Dec 2001, 22:00 & 16 Dec 2001, 22:00\\
30 Dec2001 & 2.9233  & 8 Dec 2001, 01:00 & 1 Jan  2002, 00:00\\
10 Jan 2002 & 1.1771  & 10 Jan  2002, 05 & 11 Jan  2002, 10:00\\
27 Jan 2002 & 1.1685  & 24 Jan  2002, 14:00 & 30 Jan  2002, 00:00 \\
4 Mar 2002 & 0.7605 & 4 Mar  2002, 01:00 & 5 Mar  2002, 01:00 \\
23 May 2002 & 0.9313  & 23 May  2002, 02:00 & 23 May  2002, 23:00 \\
2 Jun 2002 & 0.5591 & 2 Jun  2002, 7:00 & 4 Jun  2002, 03:00 \\
17 Jun 2002 & 0.4833 & 17 Jun  2002, 00:00 & 19 Jun  2002, 21:00 \\
17 Jul 2002 & 1.9567 & 17 Jul  2002, 07:00  & 20 Jul  2002, 02:00 \\
22 Aug 2002 & 1.3254 & 21 Aug  2002, 23:00  & 28 Aug  2002, 14:00\\
7 Sep 2002 & 0.9708 & 7 Sep  2002, 14:00  & 8 Sep  2002, 13:00\\
23 Sep 2002 & 0.5684 & 20 Sep  2002, 01:00 & 23 Sep  2002, 22:00\\
30 Sep 2002 & 0.9725 & 30 Sep  2002, 10:00  & 31 Sep  2002, 08:00\\
16 Oct 2002 & 1.4910 &  14 Oct  2002, 15:00  & 20 Oct  2002, 22:00:00 \\
10 Nov 2002 & 1.8191 & 10 Nov  2002, 00:00  & 12 Nov  2002, 12:00\\
17 Nov 2002 & 1.8464 & 16 Nov  2002, 23:00  & 18 Nov  2002, 06:00\\
26 Nov 2002 & 0.5862 & 26 Nov  2002, 07:00  & 27 Nov  2002, 22:00\\
22 Dec2002 & 1.1280 & 21 Dec 2002, 23:00  & 23 Dec 2002, 05:00\\
9 Jan 2003 & 0.7571 & 9 Jan  2003, 00:00  & 10 Jan  2003, 17:00\\
12 Jan 2003 &  0.5653 & 12 Jan  2003, 20:00  & 14 Jan  2003, 20:00\\
23 Jan 2003 & 0.6375 & 23 Jan  2003, 03:00  & 24 Jan  2003, 8:00\\
30 Jan 2003 & 0.6397 & 30 Jan  2003, 01:00  & 30 Jan  2003, 22:00 \\
12 Feb 2003 & 0.5556 & 12 Feb  2003, 20:00 & 13 Feb  2003, 17:00\\
16 Feb 2003 & 0.7803 & 16 Feb  2003, 04:00  & 18 Feb  2003, 10:00\\
26 Mar 2003 & 1.2412 & 26 Mar  2003, 16:00  & 31 Mar  2003, 06:00\\
8 Apr 2003 & 0.8365 & 7 Apr  2003, 15:00  & 11 Apr  2003, 01:00\\
4 May 2003 & 0.4216 & 4 May  2003, 10:00  & 6 May  2003, 01:00\\
18 May 2003 & 0.6927 & 18 May  2003, 16:00  & 22 May  2003, 11:00\\
27 May 2003 & 0.8055 & 24 May  2003, 13:00  & 28 May  2003, 11:00\\
29 May 2003 & 2.4082 & 29 May  2003, 14:00  & 31 May  2003, 23:00\\
25 Jul 2003 & 0.5441 & 25 Jul  2003, 04:00  & 27 Jul  2003, 13:00\\
6 Aug 2003 & 0.5603 & 6 Aug  2003, 10:00 & 7 Aug  2003,  09:00 \\
16 Aug 2003 & 0.4796 & 15 Aug  2003, 01:00  & 18 Aug  2003, 02:00 \\
8 Sep 2003 & 0.7100 & 8 Sep  2003, 22:00  & 12 Sep  2003, 21:00 \\
21 Oct 2003 & 1.7469 & 21 Oct  2003, 12:00 & 22 Oct  2003, 20:00\\
29 Oct 2003 & 8.2729 & 28 Oct  2003, 22:00  & 29 Oct  2003, 23:00 \\
15 Nov 2003 & 1.1557 & 15 Nov  2003, 04:00  & 17 Nov  2003, 17:00  \\
20 Nov 2003 & 1.1642 & 20 Nov  2003, 07:00  & 24 Nov  2003, 04:00\\
7 Dec2003 & 0.6671 & 7 Dec 2003, 03:00  & 11 Dec 2003, 20:00\\
20 Dec2003 & 1.0800 & 19 Dec 2003, 22:00 & 22 Dec 2003, 18:00\\ 
27 Dec2003 & 0.6469 & 27 Dec 2003, 00:00 & 29 Dec 2003, 02:00\\
6 Jan 2004 & 1.3512 & 6 Jan  2004, 08:00  & 8 Jan  2004,  07:00\\
21 Jan 2004 & 2.3214 & 21 Jan  2004, 22:00  & 23 Jan  2004, 02:00 \\
11 Feb 2004 & 1.1520 & 11 Feb  2004, 17:00  & 14 Feb  2004, 02:00\\
26 Feb 2004 & 1.1324 & 25 Feb  2004, 01:00  &  1 Mar  2004, 02:00 \\
9 Mar 2004 & 0.7897 & 9 Mar  2004, 14:00  & 12 Mar  2004, 10:00\\
25 Mar 2004 & 0.8128 & 25 Mar  2004, 10:00  & 30 Mar  2004, 15:00 \\
29 May 2004 & 0.3545 & 29 May  2004, 06:00  & 31 May  2004, 07:00 \\
28 Jun 2004 & 0.5397 & 25 Jun  2004, 12:00  & 31 Jun  2004, 17:00  \\
26 Jul 2004 & 2.1298 & 26 Jul  2004, 14:00  & 27 Jul  2004, 11:00\\
28 Aug 2004 & 0.4903 & 20 Aug  2004, 05:00  & 22 Aug  2004, 04:00\\
30 Aug 2004 & 0.6924 &  30 Aug  2004, 22:00  & 2 Sep  2004, 02:00 \\
13 Sep 2004 & 1.1071 & 13 Sep  2004, 06:00  & 15 Sep  2004, 07:00 \\
21 Sep 2004 & 0.6585 & 21 Sep  2004, 02:00  &  22 Sep  2004, 01:00\\
6 Nov 2004 & 2.3208 & 5 Nov  2004, 01:00  & 10 Nov  2004, 10:00 \\
5 Dec2004 & 1.3709  & 5 Dec 2004, 02:00  & 6 Dec 2004, 02:00  \\
12 Dec2004 & 0.5874 & 12 Dec 2004, 19:00 & 14 Dec 2004, 01:00 \\ \hline
\end{longtable}

\subsection{Second short-list: Magnetic clouds} \label{ESC2}

We then correlate the events in short-list 1 with lists of magnetic clouds near the Earth observed by the WIND and ACE spacecraft. The list of magnetic clouds \index{magnetic cloud} are obtained from \citet{hut05, lyn03}  and Lara (private communication). From short-list 1  we select only those event which can be reasonably well connected with a near-Earth magnetic cloud and this set is labelled short-list 2 (Table~\ref{SL2}). The decrease minimum \index{Forbush decreases!minimum} for most of the FD events in short-list 2 lie between the start and the end of the near-Earth magnetic cloud.

\afterpage{
\begin{landscape}
\begin{table}
\caption [FD events short list 2]{Short-list 2 : Forbush decrease \index{Forbush decreases} events from short-list 1 associated with near-Earth magnetic clouds\index{magnetic cloud}.
{Magnitude(Ver) : Magnitude of Forbush decrease in vertical direction \index{Forbush decreases!magnitude} }
{FD onset : Time of FD onset in UT \index{Forbush decreases!onset}}
{FD minimum : Time of FD minimum in UT \index{Forbush decreases!minimum}}
{MC start : Magnetic cloud start time in UT  }
{MC stop : Magnetic cloud stop time in UT}} \label{SL2}
\centering
\begin{tabular}{lcccccc}
\hline \hline
Event & Magnitude    &  FD onset &  FD minimum & MC start  & MC stop \\
      & (Ver) ($\%$) &  (UT)     &  (UT)       & (UT)      &  (UT)\\
\hline \hline
04/04/01 & 1.1793 & 04/04/01, 08:04 & 05/04/01, 04:00 & 04/04/01, 20:50 & 05/04/01, 08:24\\
11/04/01 & 2.8630 & 11/04/01, 12:00 & 12/04/01, 18:00 & 11/04/01, 23:00 & 12/04/01, 18:00\\
17/08/01 & 1.0296 & 16/08/01, 22:34 & 18/08/01, 05:00 & 18/08/01, 00:00 & 18/08/01, 21:30\\
29/09/01 & 2.4385 & 29/09/01, 12:43 & 31/09/01, 01:00 & 29/09/01, 16:30 & 30/09/01, 11:30\\
24/11/01 & 1.5612 & 24/11/01, 03:21 & 25/11/01, 15:00 & 24/11/01, 17:00 & 25/11/01, 13:30\\
23/05/02 & 0.9313 & 23/05/02, 02:10 & 23/05/02, 23:00 & 23/05/02, 21:30 & 25/05/02, 18:00\\
07/09/02 & 0.9708 & 07/09/02, 14:52 & 08/09/02, 13:00 & 07/09/02, 17:00 & 08/09/02, 16:30\\
30/09/02 & 0.9725 & 30/09/02, 11:30 & 31/09/02, 08:00 & 30/09/02, 22:00 & 01/10/02, 16:30\\
20/11/03 & 1.1642 & 20/11/03, 10:48 & 24/11/03, 04:00 & 21/11/03, 06:10 & 22/11/03, 06:50\\
21/01/04 & 2.3214 & 21/01/04, 22:00 & 23/01/04, 02:00 & 22/01/04, 14:00 & 23/01/04, 14:00\\
26/07/04 & 2.1298 & 26/07/04, 15:36 & 27/07/04, 11:00 & 27/07/04, 02:00 & 28/07/04, 00:00\\ \hline
\end{tabular}

\end{table}
\end{landscape}}

\subsection{Third short-list: CME \index{CME} velocity profile} \label{VP} \index{CME!velocity}
Since we are looking for Forbush decrease \index{Forbush decreases} events that are associated with shocks \index{shock}  as well as CMEs, we examine the CME  catalogue (\href{http://cdaw.gsfc.nasa.gov/CME_list/}{ \textit{CDAW DATA CENTER} \footnote{$http://cdaw.gsfc.nasa.gov/CME \_ list /$}}) for a near-Sun CME \index{CME} that can reasonably correspond to the near-Earth magnetic cloud \index{magnetic cloud} that we associated with the Forbush decrease \index{Forbush decreases} in forming short-list 2. 
In \citet{sub09} it was assumed that the CME propagates with a constant speed from the Sun to the Earth. In this chapter, we adopt a more realistic model for the Sun-Earth propagation of CMEs.  We considered a two-step velocity profile that is described below. 

The data from the LASCO \index{coronagraph!LASCO} coronagraph aboard the SOHO spacecraft (\href{http://cdaw.gsfc.nasa.gov/CME_list/}{\textit{CDAW DATA CENTER}}) provide details about CME \index{CME!propagation} propagation up to a distance of $\approx$ 30 ${\rm R_{\odot}}$ from the Sun. 

A height time plot of the CME is obtained from the LASCO FOV. Assuming that the CME travels in the LASCO FOV with a constant acceleration we can fit the following velocity profile to the LASCO \index{coronagraph!LASCO} data points:
\begin{equation}
\rm V_{1} = v_{i} + a_{i} t \, , \,\,\,\,\,\,\,\,\,{\rm for}\,\,\,  R(t) \leq R_m  
\label{VP1}
\end{equation}
where ${\rm v_i}$ and ${\rm a_i}$ are the initial velocity and acceleration of CME \index{CME!acceleration} \index{CME!velocity} respectively, and $R_{m}$ is the heliocentric distance at which the CME \index{CME!observation}  is last observed in the LASCO field of view.

For distances $> {\rm R_{m}}$, we assume that the CME \index{CME!aerodynamic drag} dynamics are governed exclusively by the aerodynamic drag it experiences due to momentum coupling with the ambient solar wind\index{solar wind}. For heliocentric distances $> {\rm R_{m}}$, we therefore use the following widely used 1D differential equation to determine the CME \index{CME!velocity} velocity profile: (e.g., \citealp{bor09})
\begin{equation}
\rm m_{CME} V_{2} \frac {\partial V_{2}}{\partial R}  = \frac{1}{2} C_D \rho_{sw} A_{CME} (V_{2} - V_{sw})^2 \, , \,\,\,\,\,R(t) > R_{m}
\label{VP2}
\end{equation}

\begin{figure}
\centering 
\includegraphics[width = 0.8\textwidth]{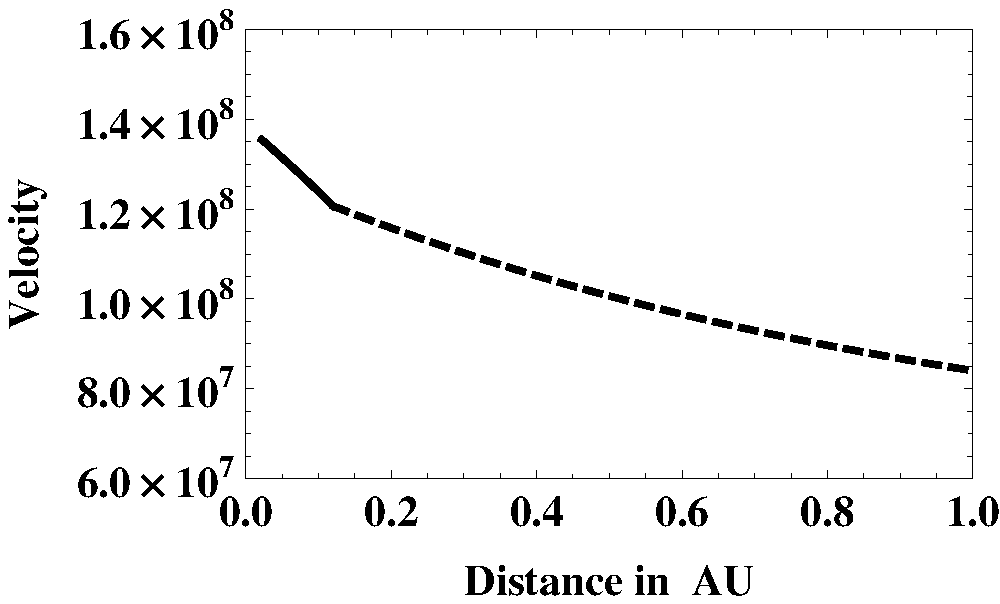}
\caption[Velocity profile of CME ]{A plot of the velocity profile for the CME \index{CME!velocity} corresponding to the 24 November 2001 FD event. The velocity of CME is given in units of $ cm \, s^{-1} $. The CME  was first observed in the LASCO FOV on 22 November 2001. The solid line shows the first stage governed by LASCO \index{coronagraph!LASCO} observations (Eq~\ref{VP1}), where the CME \index{CME!acceleration} is assumed to have a constant deceleration. The dashed line shows the second stage (Eq~\ref{VP2}), where the the CME \index{CME!aerodynamic drag} is assumed to experience an aerodynamic drag characterized by a constant ${\rm C_{D}}$.}
\label{velpro}
\end{figure}

where ${\rm m_{CME}}$ is the CME \index{CME!mass} mass, ${\rm C_D}$ is the dimensionless drag coefficient, ${\rm \rho_{sw}}$ is the solar wind density, ${\rm A_{CME} = \pi R_{CME}^2}$ is the cross-sectional area of the CME  and ${\rm V_{sw}}$ is the solar wind \index{solar wind}speed. The boundary condition used is ${\rm V_{2} = v_{m}}$ at ${\rm R(t) = R_{m}}$. The CME mass ${\rm m_{CME}}$ is assumed to be ${\rm 10^{15}}$ g and the solar wind speed ${\rm V_{sw}}$ is taken to be equal to 450 km/s. The solar wind density ${\rm \rho_{sw}} $ is given by the model of \citet{leb98}. 

The composite velocity profile for the CME \index{CME!velocity} is defined by

\begin{equation}
\rm V_{CME} = \begin{cases}
V_{1}\, ,  & \text{if $R(t) \leq R_{m}$} \\
V_{2}\, , & \text{ if $R(t) > R_{m}$}
\end{cases}
\label{VP3}
\end{equation}

The total travel time for the CME \index{CME} is ${\rm \int_{R_{i}}^{R_{f}} dR/V_{CME}}$, where ${\rm R_{i}}$ is the heliocentric radius at which the CME  is first detected and ${\rm R_{f}}$ is equal to 1 AU. We have used a constant drag coefficient ${\rm C_D}$ and adjusted its value so that the total travel time thus calculated matches the time elapsed between the first detection of the CME \index{CME!observation} in the LASCO \index{coronagraph!LASCO} FOV and its detection as a magnetic cloud near the Earth. We have retained only those events for which it is possible to find a constant ${\rm C_{D}}$ and this criterion is satisfied. We have adjusted the parameter ${\rm C_{D}}$ so that the final CME \index{CME} speed obtained from Eq~(\ref{VP2}) is close to the observed magnetic cloud \index{magnetic cloud} speed near the Earth. It is usually not possible to find a ${\rm C_{D}}$ that will yield an exact match for the velocities as well as the total travel times (e.g., \citealp{lara11} ).

If there are multiple CMEs  they may interact with each other as they propagate from the Sun to the Earth. We have eliminated magnetic clouds that could be associated with multiple CMEs in order to eliminate this possibility. This defines our final short-list, which we call short-list 3 (Table~\ref{SL3}). Figure~\ref{velpro} shows an example of the composite velocity profile (given by Eqs~\ref{VP1} and \ref{VP2}) for one such representative CME, which was first observed in the LASCO FOV on 22 November 2001, and resulted in a FD on 24 November 2001.

\afterpage{
\begin{landscape}
\begin{table}
\caption [FD events short list 3]{Short-list 3 : Forbush decrease \index{Forbush decreases} events from short-list 2 that are associated with near-Sun CMEs  and agree with the composite velocity profile (\S~\ref{VP}).
{Magnitude(Ver) : Magnitude of Forbush decrease in vertical direction \index{Forbush decreases!magnitude} }
{FD onset : Time of FD onset in UT \index{Forbush decreases!onset}}
{FD minimum : Time of FD minimum in UT \index{Forbush decreases!minimum}}
{MC start : Magnetic cloud \index{magnetic cloud} start time in UT  }
{MC stop : Magnetic cloud stop time in UT}
{CME near Sun : Time at which CME \index{CME!observation} was first observed in the LASCO FOV}}\label{SL3}
\centering
\begin{tabular}{lcccccc}
\hline \hline
Event & Magnitude     & FD onset &  FD min  & MC start & MC stop & CME \index{CME} near Sun \\
      & (Ver)  ($\%$) &  (UT)    &    (UT)  &  (UT)    &  (UT)   & (UT)  \\ \hline
11/04/01 & 2.8630 & 11/04/01, 12:00 & 12/04/01, 18:00 & 11/04/01, 23:00 & 12/04/01, 18:00 & 10/04/01, 05:30\\ 
17/08/01 & 1.0296 & 16/08/01, 22:34 & 18/08/01, 05:00 & 18/08/01, 00:00 & 18/08/01, 21:30 & 15/08/01, 23:54\\
24/11/01 & 1.5612 & 24/11/01, 03:21 & 25/11/01, 15:00 & 24/11/01, 17:00 & 25/11/01, 13:30 & 22/11/01, 22:48\\
 7/09/02 & 0.9708 &  7/09/02, 14:52 &  8/09/02, 13:00 &  7/09/02, 17:00 &  8/09/02, 16:30 & 05/09/01, 16:54\\
20/11/03 & 1.1642 & 20/11/03, 10:48 & 24/11/03, 04:00 & 21/11/03, 06:10 & 22/11/03, 06:50 & 18/11/03, 08:50\\
26/07/04 & 2.1298 & 26/07/04, 15:36 & 27/07/04, 11:00 & 27/07/04, 02:00 & 28/07/04, 00:00 & 25/07/04, 14:54 \\ \hline
\end{tabular}
\end{table}
\end{landscape}}

\section{Models for Forbush decreases} \label{Mod}
We apply two different models to the Forbush decrease events in Table~\ref{SL3}.
\begin{itemize}
\item The CME-only \index{CME-only model} cumulative diffusion model, which assumes that the Forbush decrease \index{Forbush decreases} owes its origin only to the CME, and the cosmic rays \index{cosmic rays} penetrate  into the CME (Figure \ref{crt}) by diffusing across the large-scale magnetic fields bounding it, aided by the turbulence \index{turbulence} in the CME sheath \index{sheath} region.
\item The shock only model, which assumes that the Forbush decrease is exclusively due to the shock. The shock \index{shock}  is approximated as a diffusive barrier and cosmic rays diffuse across the large-scale turbulent magnetic field \index{turbulence!turbulent magnetic field} compression at the shock.
\end{itemize}
Although both the shock and the CME \index{CME} are expected to contribute to the Forbush decrease \index{Forbush decreases}, our treatment seeks to determine which one of them is the dominant contributor at rigidities ranging from 14 to 24 GV. Before describing the models further we discuss the cross-field diffusion \index{diffusion!diffusion coefficient}coefficient.

\subsection{Cross-field diffusion coefficient} \label{crossDiff} \index{diffusion!diffusion coefficient}

 We use an isotropic  perpendicular diffusion coefficient ($D_{\perp}$) to characterize the penetration of cosmic rays \index{cosmic rays} across large scale, ordered  magnetic fields. We envisage a CME, with a flux rope \index{flux rope} structure, which propagates outwards from the Sun, driving a shock \index{shock}  ahead of it (see, e.g., \citealp{angelos13} ).  The flux rope CME-shock geometry is illustrated in Figure~\ref{difn}. The CME sheath \index{sheath} region between the CME \index{CME} and the shock is known to be turbulent (e.g., \citealp{mano00}). It is well accepted that the turbulent CME sheath region has a significant role to play in determining Forbush decreases \index{Forbush decreases} \citep{bad02, yu10}. The cross-field diffusion coefficient ${\rm D_{\perp}}$ characterizes the penetration of cosmic rays through the ordered, compressed large-scale magnetic field near the shock, as well as across the ordered magnetic field of the flux rope CME. In diffusing across the shock, the cosmic rays are affected by the turbulence ahead of the shock, and in diffusing across the magnetic fields bounding the flux rope \index{flux rope} CME, the cosmic rays \index{cosmic rays} are affected by the turbulence \index{turbulence} in the CME \index{CME} sheath region.

 The subject of charged particle diffusion \index{diffusion} across field lines in the presence of turbulence has a long history. Analytical treatments include the so-called ``classical'' scattering theory (e.g., \citealp{gj99} and references therein), and the non-linear guiding center theory \citep{matt03, shal10} for perpendicular diffusion. Numerical treatments include \citet{gj99}, \citet{Cass02}, \citet{cand04} and \citet{tau11}.  We find that the analytical fits to extensive numerical simulations provided by \citet{cand04} best suit our requirements. Their results not only reproduce the standard results of \citet{gj99} and \citet{Cass02} but also extend the regime of validity to include strong turbulence \index{turbulence}and high rigidities. Our approach is similar to that of \citet{eff12}, who adopt empirical expressions for the perpendicular diffusion coefficients.

In the formulation of \citet{cand04}, the perpendicular diffusion coefficient ${\rm D_{\perp}}$ \index{diffusion!diffusion coefficient} is a function of  the quantity ${\rm \rho}$ (which is closely related to the rigidity \index{rigidity} and indicates how tightly the proton is bound to the magnetic field) and the level of turbulence ${\rm \sigma^{2}}$\index{turbulence!turbulence level $\sigma$}. Our characterization of ${\rm D_{\perp} (\rho, \sigma^{2})}$ follows that of \citet{sub09}.  

The quantity $\rho$ is related to the rigidity $Rg$ by
\begin{equation}
\rm \rho = \frac{R_L}{L_{\rm max}} = \frac {Rg}{B_0 L_{\rm max}}
\label{E11}
\end{equation}

where $R_{L}$ is the Larmor radius, ${\rm B_0}$ is the strength of the relevant large-scale magnetic field. For the CME-only \index{CME-only model} model, $B_{0}$ refers to the large-scale magnetic field bounding the CME, and for the shock-only \index{shock-only model} model, it refers to the enhanced large-scale magnetic field at the shock.
In writing second step in Eq~(\ref{E11}), we have related the Larmor radius to the rigidity Rg by

\begin{equation}
\rm R_L(t) = \frac{Rg}{B_0}\, .
\label{rl}
\end{equation}

For the CME-only model, we adopt ${\rm L_{\rm max}=2\,R(T)}$, where ${\rm R(T)}$ is the radius of the near-Earth magnetic cloud\index{magnetic cloud}. This is in contrast with \citet{sub09}, where ${\rm L_{\rm max}}$ was taken to be ${\rm 10^6}$ km, which is the approximate value for the outer scale of the turbulent cascade in the solar wind\index{solar wind}. For the shock-only model, on the other hand, we assume that ${\rm L_{\rm max}}$ is equal to 1 AU.

{  The turbulence level \index{turbulence!turbulence level $\sigma$} ${\rm \sigma^2}$ is defined (as in \citealp{cand04} and \citealp{sub09}) to be 
\begin{equation}
 \rm \sigma^2 \equiv \frac{\langle {B_r}^2 \rangle}{{B_0}^2}
\label{E12}
\end{equation}

where ${\rm B_r}$ is the fluctuating part of the turbulent magnetic field \index{turbulence!turbulent magnetic field} and the angular braces denote an ensemble average.}

The parallel diffusion coefficient \index{diffusion!diffusion coefficient} due to scattering of particles along the mean magnetic field $D_{\parallel}$ is given by
\begin{equation}
D_{\parallel} = c\,L_{\rm max}\,\rho\,\frac{N_{\parallel}}{\sigma^{2}}\,\sqrt{\biggl (\frac{\rho}{\rho_{\parallel}} \biggr )^{2(1 - \gamma)} + 
\biggl (\frac{\rho}{\rho_{\parallel}} \biggr )^{2}} \, ,
\label{eq17}
\end{equation}
where $c$ is the speed of light and the quantities $N_{\parallel}$, $\gamma$ and $\rho_{\parallel}$ are constants specific for different kinds of turbulence\index{turbulence}. 

The cross-field diffusion coefficient
($D_{\perp}$) is related to the parallel one ($D_{\parallel}$) by

\begin{equation}
\frac{D_{\perp}}{D_{\parallel}} = \begin{cases}
N_{\perp}\,(\sigma^{2})^{a_{\perp}}\, , & \text{$\rho \leq 0.2$ }\\
N_{\perp}\,(\sigma^{2})^{a_{\perp}}\,\biggl (\frac{\rho}{0.2} \biggr )^{-2}\, , & \text{$\rho > 0.2$}
\end{cases}
\label{eq20}
\end{equation} 

The quantities $N_{\perp}$ and $a_{\perp}$ are constants specific to different kinds of turbulent spectra. In this work, we assume the Kolmogorov turbulence \index{turbulence}spectrum in our calculations. We use $N_{\parallel}$ = 1.7, $\gamma$ = 5/3, $\rho_{\parallel}$ =0.20, $N_{\perp}$ =0.025  and  $a_{\perp}$= 1.36    ( Table 1,  \citet{cand04} ). Eq~(\ref{eq17}) together with Eq~(\ref{eq20}) defines the cross-field diffusion coefficient \index{diffusion!diffusion coefficient} we use in this work.

\subsection{CME-only model} \label{cme-only}
The basic features of the CME-only model \index{CME-only model} are similar to that used in \citet{sub09}. Significant differences from  \citet{sub09} are highlighted later on in this section. There are practically no high-energy galactic cosmic rays \index{cosmic rays} inside the CME \index{CME} when it starts out near the Sun. Cosmic rays diffuse into it from the surroundings via cross-field diffusion across the closed magnetic field lines as it propagates through the heliosphere as shown schematically in Figure~\ref{crt}. Near the Earth, the difference between the (relatively lower) cosmic ray proton density inside the CME \index{CME} and that in the ambient medium appears as the Forbush decrease \index{Forbush decreases}. We next obtain an estimate of the cosmic ray \index{cosmic rays} proton density inside the CME  produced by the cumulative effect of diffusion\index{diffusion}. 

\begin{figure*}
\centering 
\includegraphics[width = 1.0\textwidth]{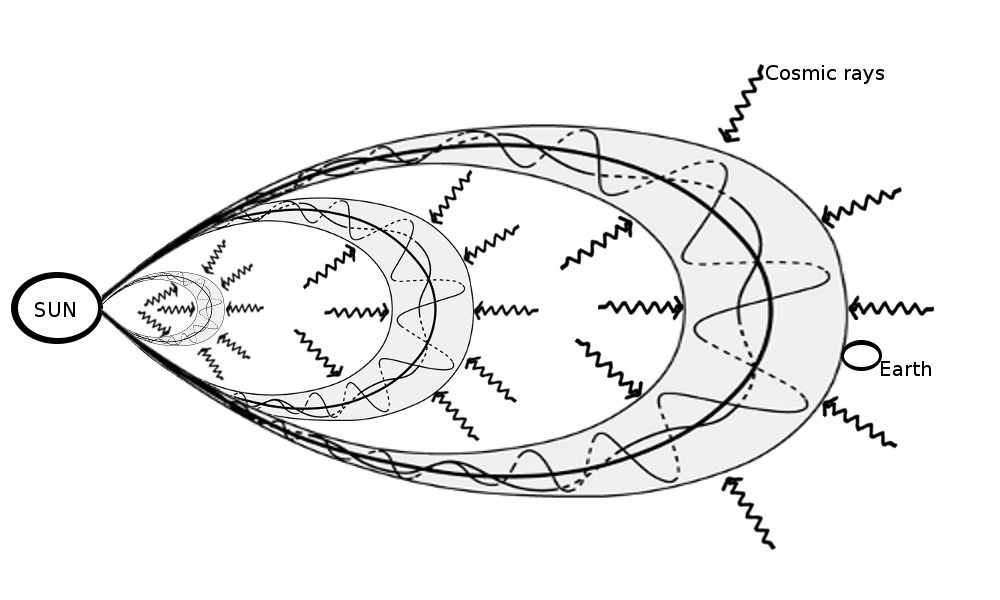}
\caption[A cartoon illustrating expanding flux rope \index{flux rope} and diffusing cosmic rays into it]{A cartoon illustrating a flux rope CME \index{CME} expanding and propagating away from the Sun. High energy galactic cosmic rays \index{cosmic rays} diffuse into the CME  across its bounding magnetic field.}
\label{crt}
\end{figure*}

The flux ${\rm F}$ \index{flux} of protons diffusing into the CME \index{CME} at a given time depends on the perpendicular diffusion coefficient ${\rm D_{\perp}}$ \index{diffusion!diffusion coefficient} and the density gradient \index{density gradient} ${\rm \partial N_a/\partial r}$, and can be written as  

\begin{equation}
 \rm {F \,\, {( cm^{-2} s^{-1})}} = \textrm{D}_{\perp} \frac {\partial N_a} {\partial r} \, .
\label{E1}
\end{equation}
{ As mentioned earlier, the perpendicular diffusion coefficient ${\rm D_{\perp}}$ characterizes diffusion across the (largely closed) magnetic fields bounding the CME \index{CME} and ${\rm N_a}$ is the ambient density of high energy protons.} The total number of cosmic ray \index{cosmic rays} protons that will have diffused into the CME \index{CME} after a time ${\rm T}$ is related to the diffusing flux by

\begin{equation}
\rm {U}_i = \int_{0}^{T} A(t) F(t) \,dt = \int_{0}^{T}  \textrm{D}_{\perp} A(t) \frac {\partial N_a} {\partial r}\,dt
\label{E2}
\end{equation}

where ${\rm A(t)}$ is the cross-sectional area of the CME \index{CME} at a given time ${\rm t}$. According to our convention, the CME \index{CME} is first observed in the LASCO \index{coronagraph!LASCO} FOV at ${\rm t=0}$ and it reaches the Earth at ${\rm t=T}$.
The ambient density gradient \index{density gradient} ${\rm \partial N_{a}/\partial r}$ is approximated by the following expression, the treatment of which is significantly different from that used in \citet{sub09}:


\begin{equation}
\rm \frac {\partial Na} {\partial r} \simeq \frac{Na}{L}\, ,
\label{E3}
\end{equation}
where $L$ is the gradient length-scale\index{density gradient!length-scale}. { Observations of the density gradient length-scale $L$ exist only  for a few rigidities.  \citet{heb08}  quote a value of $L^{-1}$ = 4.7 \% ${\rm AU}^{-1}$ for 1.2 GV protons. We take this as our reference value. In order to calculate $L$ for other rigidities (in the 14--24 GV range that we use here), we assume that $L \propto R_{L}^{1/3}$. This is broadly consistent with the observation \citep{desim11} that the density gradient length-scale is only weakly dependent on rigidity\index{rigidity}. Hence at 1 AU, the length scale becomes 

\begin{equation}
L = \kappa R_L^{1/3}
\end{equation}
where $\kappa$ is the proportionality constant. the value of $\kappa$ can be calculated from the observed value of length scale for 1.2 GV protons from \citet{heb08}.

 For a given rigidity\index{rigidity}, we also need to know $L$  from the Sun to the Earth. In order to do this, we recognize that $L$ near the CME/magnetic cloud \index{magnetic cloud} will not be the same as its value in the ambient solar wind\index{solar wind}. We use $ L \propto B_{\rm a}(t)/B_{\rm 0}(t)$, where $B_{\rm 0}(t)$ is the large-scale magnetic field bounding the CME \index{CME} at time $t$,  and $B_{\rm a}(t)$ is the (weaker) magnetic field in the ambient medium outside the CME at time $t$. Furthermore, while $B_{\rm 0}(t)$ varies according to Eq~(\ref{BFL}) below, the ambient field $B_{\rm a}(t)$ of the Parker spiral in the ecliptic plane varies inversely with heliocentric distance. } 

\begin{equation}
B_{\rm a}(t) = B_{\rm a}(T) \frac{H(T)}{H(t)} \label{ba}
\end{equation}

where $B_{\rm a}(T)$ is the ambient magnetic field measure in the space craft. $H(T)$ is the heliocentric distances at the time when CME reach space craft, which is 1 AU and $H(t)$ is the heliocentric distance at any time during the propagation of CME from Sun to Earth. 

Using equations \ref{ba}, \ref{BFL} and \ref{E5} we get

\begin{eqnarray}
\nonumber
 L(t) & = & \frac{B_{\rm a}(t)}{B_{\rm 0}(t)} \kappa R_L^{1/3} \\
      & = &   \kappa(t) R_L^{1/3} 
\end{eqnarray}

where the new time dependent proportionality constant $\kappa(t)$ is obtained by absorbing the time dependent part $\frac{B_{\rm a}(t)}{B_{\rm 0}(t)}$ in $\kappa$.

Now the density gradient \index{density gradient} becomes 

\begin{equation}
\rm \frac {\partial Na} {\partial r} \simeq \frac{Na}{\kappa(t) R_L^{1/3} }\, ,
\end{equation}

{We assume that the magnetic flux associated with the CME \index{CME} is ``frozen-in'' with it as it propagates. In other words, the product of the CME  magnetic field and the CME  cross-sectional area remains constant (e.g., \citealp{kmrst96, sv07}). One can therefore relate the CME \index{CME} magnetic field ${\rm B_{0}(t)}$ at a given time ${\rm t}$, to the value ${\rm B_{MC}}$ measured in the near-Earth magnetic cloud \index{magnetic cloud} using
\begin{equation} 
\rm B_0(t) = B_{MC} {\left[ \frac{R(T)}{R(t)} \right]}^2\, ,
\label{BFL}
\end{equation}
where ${\rm R(T)}$ is the radius of the magnetic cloud observed at the Earth and $R(t)$ is its radius at any other time $t$ during its passage from the Sun to the Earth. The CME \index{CME} radius ${\rm R(t)}$ and ${\rm R(T)}$ are related via Eq~(\ref{ER}) below. We emphasize that the magnetic field referred by equation~\ref{BFL} refers to the magnetic field bounding the CME, and not to the ambient magnetic field outside it.}

We model the CME  as an expanding cylindrical flux rope \index{flux rope} whose length increases with time as it propagates outwards. Its cross-sectional area at time $t$ is
\begin{equation}
\rm A(t) = 2 \pi L(t) R(t)
\label{E4}
\end{equation}
where ${\rm L(t)}$ is the length of the flux-rope \index{flux rope} cylinder at time ${\rm t}$, and is related to the height ${\rm H(t)}$ of the CME \index{CME} above the solar limb via
\begin{equation}
\rm L(t) = \pi H(t) \, .
\label{E5}
\end{equation}

We note that Eq~(\ref{E5}) differs from the definition used in \citet{sub09} by a factor of 2. We assume that the CMEs \index{CME} expand in a self-similar \index{self-similar} manner as they propagate outwards. 3D flux rope \index{flux rope} fittings to CMEs \index{CME} in the ${\rm \sim 2 - 20}$ ${\rm R_{\odot}}$ field of view using SECCHI/STEREO data validate this assumption ({ e.g., \citealp{poom10}}). We discuss self-similar  expansion further in Chapter \ref{fluxrope}. This assumption means that the radius of the ${\rm R(t)}$ of the flux rope is related to its heliocentric height ${\rm H(t)}$ by

\begin{equation}
\rm \frac{R(t)}{H(t)} = \frac{R(T)}{H(T)}
\label{ER}
\end{equation}
where ${\rm H(T)}$, the heliocentric height at time ${\rm T}$, is ${\rm  = 1\,AU}$ by definition, and ${\rm R(T)}$ is the measured radius of the magnetic cloud at the Earth.

As mentioned earlier, we consider a 2-stage velocity profile for CME \index{CME!propagation} propagation, expressed by Eqs~(\ref{VP1}) and (\ref{VP2}); this is substantially different from the constant speed profile adopted in \citet{sub09}.

Using Eqs (\ref{E3}), (\ref{E4}) and (\ref{E5})  in (\ref{E2}), we get the following expression for the total number of protons inside the CME \index{CME} when it arrives at the Earth: 

\begin{equation}
\rm \textit{U}_{i} = \int_{0}^{T} 2 \pi L(t) R(t) \textrm{D}_{\perp}  \frac{Na}{\kappa(t) {{R_L(t)} ^{1/3}}} \,dt \, .
\label{E7}
\end{equation}

The cosmic ray \index{cosmic rays} density inside the CME \index{CME} when it arrives at the Earth is 
\begin{equation}
\rm \textit{N}_i = \frac {\textit{U}_i}{\pi R(T)^2 L(T)} \, ,
\label{E8}
\end{equation} 
where ${\rm L(T)}$ \ and ${\rm R(T)}$ \ are the length and cross-sectional radius of the CME \index{CME} respectively at time T, when it reaches earth. When the CME \index{CME} arrives at the Earth, the relative difference between the cosmic ray \index{cosmic rays} density inside the CME and the ambient environment is manifested as the Forbush decrease \index{Forbush decreases}, whose magnitude \index{Forbush decreases!magnitude} ${\rm M}$ can be written as

\begin{eqnarray}
\rm M & = & \frac { N_a - N_i}{N_a}  =  \frac { \Delta N}{N_a} \\
 & = & 1 -\frac{  2 { \int_{0}^{T}} \frac{L(t) R(t) \textrm{D}_{\perp} }{ \kappa (t){{R_L} ^{1/3}}} \,dt} {R(T)^2 L(T)} 
\label{E9}
\end{eqnarray}

We compare the value of the Forbush decrease \index{Forbush decreases} magnitude \index{Forbush decreases!magnitude} ${\rm M}$ predicted by Eq~(\ref{E9}) with observations in \S~\ref{rslt}.

\begin{figure}
   \centering
      \includegraphics[width = 0.9\textwidth]{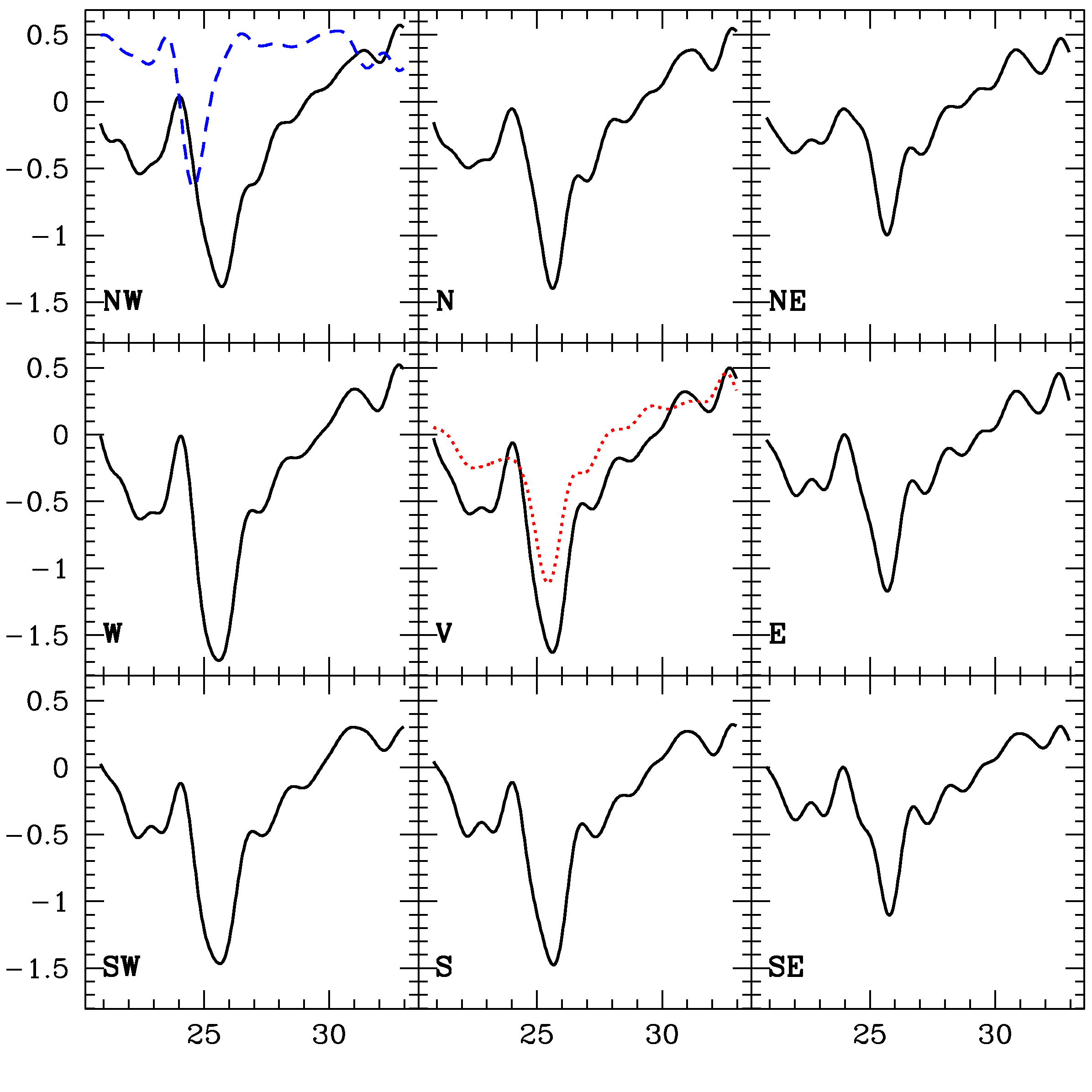}
   \caption[The muon flux along the nine directions, for the Forbush decrease on 2001 November 24]{The muon flux along the nine directions is shown for the Forbush decrease \index{Forbush decreases} on 2001 November 24. The fluxes are shown as percentage deviations from mean values. The solid black lines show the data after applying a low-pass filter (\citet{sub09}). The blue dashed line in the first panel shows the magnetic field observed in-situ by spacecraft. The magnetic field data are inverted (i.e., magnetic field peaks appear as troughs) and are scaled to fit in the panel. The red dotted line in middle panel shows the data from the Tibet neutron monitor scaled by a factor 3 to fit in the panel.}\label{FD24}%
    \end{figure}

An example of an FD event observed in all 9 bins of  GRAPES-3 \index{GRAPES-3}  is shown in Figure~\ref{FD24}. The x-axis is the time in days starting from 1 November 2001 and the y-axis gives the the percentage deviation of the muon flux from the pre-event mean.  The magnitude $M$ of the FD for a given rigidity  bin is the difference between the pre-event cosmic ray \index{cosmic rays} intensity and the intensity at the minimum \index{Forbush decreases!minimum} of the Forbush decrease \index{Forbush decreases}.

\subsection{Shock-only model} \label{Sonly} \index{shock-only model}

In this approach we assume that the FD is caused exclusively due to the shock\index{shock}, which is modelled as a propagating diffusive barrier.
The expression for the magnitude \index{Forbush decreases!magnitude} of the Forbush decrease \index{Forbush decreases} according to this model is  \citep{wib98}

\begin{equation}
\rm M \equiv \frac{U_a - U_{shock}}{U_a} = \frac{\Delta U}{U_a} = \frac{V_{sw} L_{shock}}{{D_{\perp}}^a} \left( \frac{{D_{\perp}}^a}{{D_{\perp}}^{shock}} -1 \right)
\label{E14}
\end{equation}

where ${\rm U_a}$  is the ambient cosmic ray \index{cosmic rays} density and ${\rm U_{shock}}$ is that inside the shock, ${\rm {D_{\perp}}^a}$ is the ambient perpendicular diffusion coefficient \index{diffusion!diffusion coefficient} and ${\rm {D_{\perp}}^{shock}}$ is that inside the shock, ${\rm V_{sw}}$ is the solar wind \index{solar wind}velocity and ${\rm L_{shock}}$ is the shock sheath \index{sheath} thickness. For each shock event, we examine the magnetic field data from the ACE and WIND spacecraft and estimate the shock sheath thickness ${\rm L_{shock}}$ to be the spatial extent of the magnetic field enhancement. An example is shown in Figure~\ref{Fshock}.

In computing ${\rm {D_{\perp}}^a}$ and ${\rm {D_{\perp}}^{shock}}$, we need to use different values for the proton rigidity ${\rm \rho}$ for the ambient medium and in the shock sheath; they are related to the proton rigidity $Rg$ \index{rigidity}by

\begin{eqnarray}
\rm {\rho}^a & = & \frac{Rg}{{B_0}^a L_{shock}} \\
\rm {\rho}^{shock} & = & \frac{Rg}{{B_0}^{shock} L_{shock}} \, ,
\label{E15}
\end{eqnarray}

where $B_{0}^{a}$ is the ambient magnetic field, $B_{0}^{shock}$ is the magnetic field inside the shock sheath\index{sheath}.

\begin{figure}
   \centering
      \includegraphics[width = 0.95\textwidth]{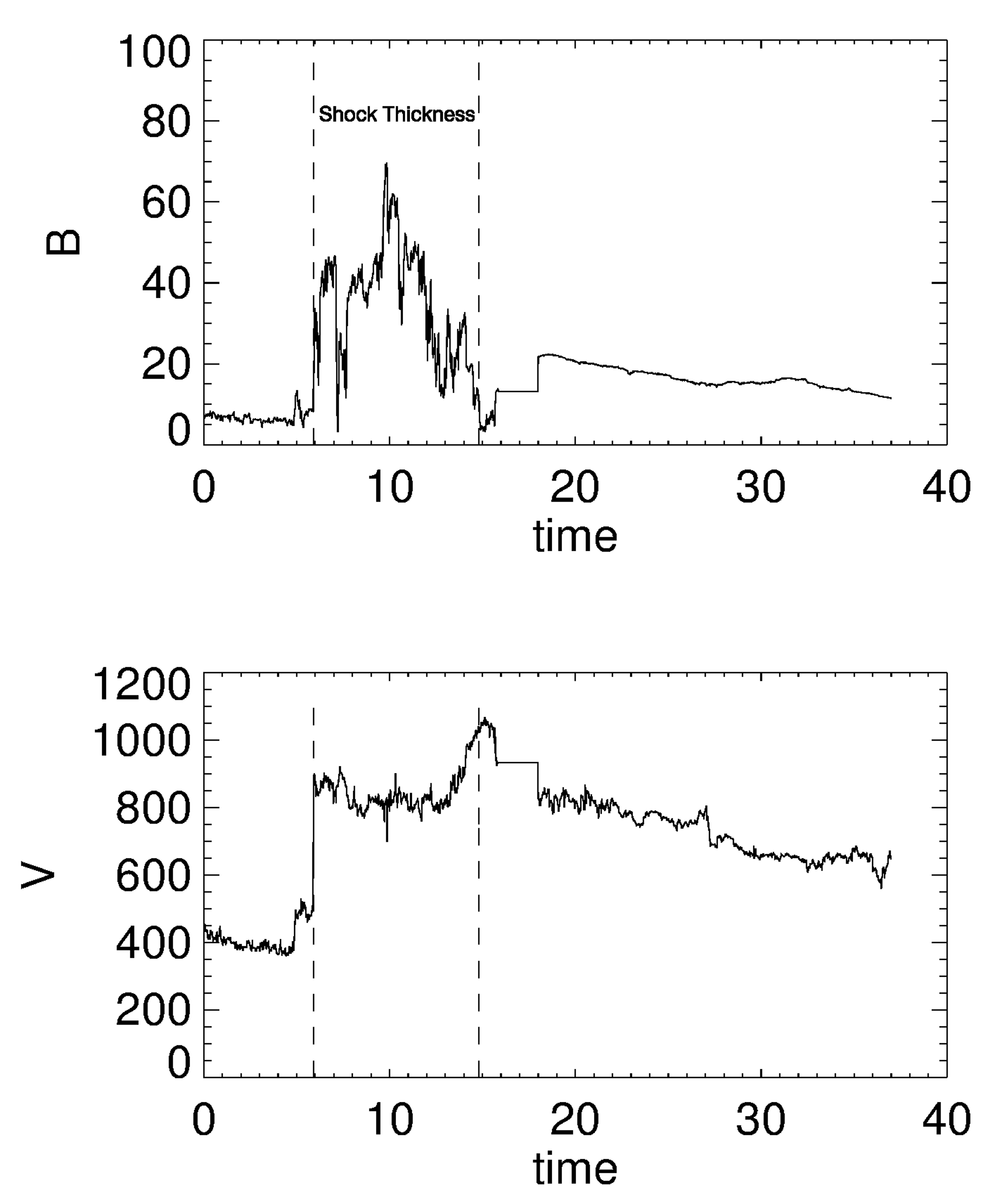}
   \caption[Interplanetary magnetic field and solar wind speed ]{Interplanetary magnetic field \index{Interplanetary magnetic field}and solar wind speed from the day 24 November 2001, The shock sheath \index{sheath} thickness is computed by multiplying the time interval inside the dotted lines by the solar wind speed}
              \label{Fshock}%
    \end{figure}

\section{Results} \label{rslt}
In this section we first describe various parameters needed for the CME-only \index{CME-only model} and the shock-only models \index{shock-only model} that are derived from observations for different events. Using these parameters, we then examine whether the notion of cosmic ray \index{cosmic rays} diffusion \index{diffusion} is valid for each model. Using the observationally determined parameters, we then obtain the (CME-only and shock-only) model that best reproduces the observed FD magnitude in each rigidity bin.  

\subsection{Details of events in Short-list 3 } 
In short-list 3 we finally chose 6 well observed events depending on the 3 short list criteria explained in section \S~\ref{ESC}. In this section we will explain each event  in detail. 
  
\subsubsection{11 April 2001} 
{ This main parameters for this event are listed in tables \ref{T11A} and \ref{TCME}. The cosmic ray intensity data from GRAPES-3 \index{GRAPES-3}  muon telescope \index{GRAPES-3!muon telescope} corresponding to this FD \index{Forbush decreases} event is shown in figure \ref{FD11A}. The  FD on-set was on 11 April 2001 and the minimum \index{Forbush decreases!minimum} of FD was on 12 April 2001. The magnitude of the FD in a given rigidity bin is the difference between the pre-event intensity of the cosmic rays \index{cosmic rays} and the intensity at the minimum of the Forbush decrease \index{Forbush decreases}. The FD magnitude\index{Forbush decreases!magnitude}, cut-off rigidity, FD on-set time and FD minimum time are given in  table \ref{T11A}. \\ }

This FD event is associated with a halo CME \index{CME!Halo}, which was first observed in LASCO \index{coronagraph!LASCO} FOV on 10 April 2001 ; 05:30 UT \href{http://cdaw.gsfc.nasa.gov/CME_list/}{ \textit{CDAW DATA CENTER} \footnote{$http://cdaw.gsfc.nasa.gov/CME \_ list /$}} at a distance 2.84 $R_{\odot}$. The CME \index{CME!observation} was last observed in the LASCO C2 FOV at $18.05 R_{\odot}$, where its speed was $V_{exp} =  2876 km s^{-1}$. The acceleration \index{CME!acceleration} in this FOV was $a_{i} = 211.60 m s^{-2}$. The velocity profile for this CME \index{CME!velocity} is fitted to the two-stage profile given by Eqs~(\ref{VP1}) and (\ref{VP2}) with a constant drag coefficient $C_D  =  0.325$.
The near-earth magnetic cloud \index{magnetic cloud} (MC) corresponding to this event was observed by the WIND spacecraft. It started at 23:00 UT on 11 April 2001 and end at 18:00 UT on 12 April 2001. The solar wind \index{solar wind}speed when the CME  reaches earth was $V_{sw}^{MC} = 725 km s^{-1}$. The radius of the magnetic cloud at earth is therefore $R(T) \, = \, 2.4795 \times 10^7 km$. The maximum magnetic field of the magnetic cloud when it reaches earth is $B_{MC} = 34.5 nT$. The time interval between the first observation of the halo CME \index{CME!observation} in the LASCO C2 coronagraph FOV and the detection of the magnetic cloud by the ACE spacecraft is 41.5 hours. The average speed of the magnetic cloud when it reaches the ACE spacecraft is $725 km s^{-1}$. Since the spacecraft is located around $ 1.5 \times 10^6 km$ from the Earth, we estimate that the magnetic cloud would have taken $\approx$ 0.6 hours to traverse the distance between ACE and the Earth. The total time is therefore $T = 41.5 + 0.6 = 42.1$ hours.

The interplanetary shock \index{shock}  reached the spacecraft at 14:06 UT on 11 April 2001, the solar wind \index{solar wind} speed at this time was $ V_{sw}^{Shock}  \, = 670 \, kms^{-1}$. The ambient solar wind magnetic field was $ B^a \, = 4.5 \, nT $ and the magnetic field in the shock-sheath \index{sheath} region was $B^{shock} \, = 32.5 \, nT $.

\begin{table}{\small
\caption[Observed parameters for FD on 11 April 2001.]{Observed parameters for FD \index{Forbush decreases} on 11 April 2001. \index{Forbush decreases!magnitude} \index{Forbush decreases!minimum}\index{Forbush decreases!onset}\index{rigidity!cut-off rigidity}}\label{T11A}
\begin{tabular}{l|cccc}
\hline \hline 
\multicolumn{5}{l}{  FD onset times are in UT, 11 April 2001}\\
\multicolumn{5}{l}{ FD minimum times are in UT, 12 April 2001}\\ \hline

Directions & FD Magnitude & Cut-off Rigidity & FD Onset & FD minimum \\
    & $(\%)$ & $(GV)$ & $(UT)$ & $(UT)$ \\ \hline
NW  & 1.02 & 15.5 & 12:57 & 19:00 \\
N   & 1.12 & 18.7 & 13:55 & 19:00 \\  
NE  & 0.91 & 24.0 & 16:04 & 19:00 \\  
W   & 1.38 & 14.3 & 11:16 & 17:00 \\
V   & 1.40 & 17.2 & 12:00 & 18:00 \\ 
E   & 1.05 & 22.4 & 12:57 & 18:00 \\ 
SW  & 1.36 & 14.4 & 08:24 & 14:00 \\
S   & 1.29 & 17.6 & 10:20 & 15:00 \\
SE  & 0.93 & 22.4 & 12:28 & 16:00 \\
\hline \hline
\end{tabular}}
\end{table}

\begin{figure}
\centering
\includegraphics[width= 0.95\textwidth]{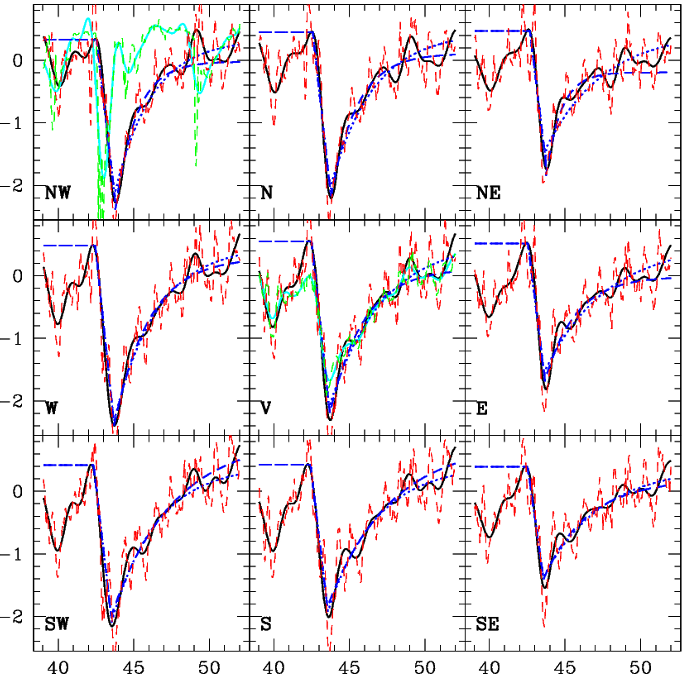}
\caption[ FD on 2001 April 11] {The muon flux in the nine directions is shown for the Forbush decrease \index{Forbush decreases} on 2001 April 11. The fluxes are shown as percentage deviation from mean values. The dashed red lines show the unfiltered data, while solid black lines show the data after applying a low-pass filter. The dotted blue lines show the fits to filtered data. The green lines in middle panel shows the data from the Tibet neutron monitor scaled by a factor 3 to fit in the panel. The green line in the top left panel shows the magnetic field variation. since magnetic field increases during Forbush decrease, we calculated average value and percentaged the deviation  of the quantity  $|100 - B|$ then scaled to a factor of 10 to fit in panel} \label{FD11A}
\end{figure}

\subsubsection{17 August 2001} 
{The salient parameters for this event are listed in tables \ref{T17A} and \ref{TCME}. The cosmic ray intensity data corresponding to this event is shown in the figure \ref{FD17A}. The FD on-set was observed on 17 August 2001 in the four bins (NW, N, NE, and SW), while for the other bins (W, V, E, S and SE) FD on-set was observed at the final hours of 16 August 2001. Minimum of this FD was on 18 August 2001. The FD magnitude\index{Forbush decreases!magnitude}, cut-off rigidity, FD on-set time and FD minimum \index{Forbush decreases!minimum} time are given in the table \ref{T17A}. \\ }


This event was associated with a halo CME \index{CME!Halo} which was first observed at 23:54 UT on 15 August 2001 \href{http://cdaw.gsfc.nasa.gov/CME_list/}{ \textit{CDAW DATA CENTER} \footnote{$http://cdaw.gsfc.nasa.gov/CME \_ list /$}}. The CME \index{CME!observation} is first observed by the coronagraph  \index{coronagraph} at the height $3.38 R_{\odot}$ and last observation of the CME  by the coronagraph was at height $25.91 R_{\odot}$, where the speed of the CME \index{CME!velocity} was $V_{exp} \ = \ 1413 \ km \ s^{-1}$, and the initial acceleration \index{CME!acceleration} was $a_i \ = \ -31.66 \ m \ s^{-2}$ in the coronagraph field of view.  There after we assume that  it follows the constant drag profile with  drag coefficient $C_D \ = \ 0.163 $. The near-Earth magnetic cloud  \index{magnetic cloud} was observed by the WIND space craft, which started at 00:00 UT of 18 August 2001 and end at 21:30 UT of the same day. The radius of the magnetic cloud as measured from WIND spacecraft is $R(T) \, =   \,  2.322 \times 10^7 \ km$,where we have assumed that magnetic cloud moves with the maximum in-situ speed of the ambient solar wind behind the shock \index{shock}  $V_{sw}^{MC} = 600 \ km \ s^{-1}$. The maximum magnetic field of the magnetic cloud when it reaches earth was $25.6 \ nT$. The total time of travel for the CME  is calculated using the time at which the CME  is first observed in SOHO and when it reaches the WIND spacecraft and the time taken for the CME  to travel from the space craft to earth using  the velocity  $V_{s} = 600 \ km \ s^{-1}$. We find $T \, = \, 44.79$ hours.

The interplanetary shock \index{shock} reached the spacecraft at 11:00 UT on 17 August 2001 and the solar wind \index{solar wind} speed at this time was $ V_{sw}^{Shock}  \, = 501 \, kms^{-1}$. The ambient solar wind magnetic field was $ B^a \, =5 \, nT $ and the magnetic field in the shock-sheath \index{sheath} region was $B^{shock} \, = 33 \, nT $. 

\begin{table}{\small
\caption[Observed parameters for FD on 17 August 2001.]{Observed parameters for FD \index{Forbush decreases} on 17 August 2001.\index{Forbush decreases!magnitude} \index{Forbush decreases!minimum}\index{Forbush decreases!onset} \index{rigidity!cut-off rigidity}}\label{T17A}
\begin{tabular}{l|cccc} 
\hline \hline 
\multicolumn{5}{l}{  FD onset times are  in UT, 17 August 2001}\\
\multicolumn{5}{l}{The $^1$ FD onset times are in UT, 16 August 2001}\\
\multicolumn{5}{l}{ FD minimum times are in UT, 18 August 2001}\\ \hline

Directions & FD Magnitude & Cut-off Rigidity & FD Onset & FD minimum \\
    & $(\%)$ & $(GV)$ & $(UT)$ & $(UT)$ \\ \hline
NW  & 1.166 & 15.5  & 04:19     & 4:00  \\
N   & 1.121 & 18.7  & 01:55     & 3:00  \\
NE  & 0.827 & 24.0  & 01:12     & 2:00  \\
W   & 0.974 & 14.3  & 23:17$^1$ & 6:00  \\
V   & 1.029 & 17.2  & 22:34$^1$ & 5:00  \\
E   & 0.788 & 22.4  & 22:05$^1$ & 5:00  \\
SW  & 1.120 & 14.4  & 00:00     & 23:00 \\ 
S   & 1.072 & 17.6  & 23:31$^1$ & 22:00 \\ 
SE  & 0.796 & 22.4  & 23:17$^1$ & 20:00 \\
\hline \hline
\end{tabular}}
\end{table}

\begin{figure}
\centering
\includegraphics[width= 0.95\textwidth]{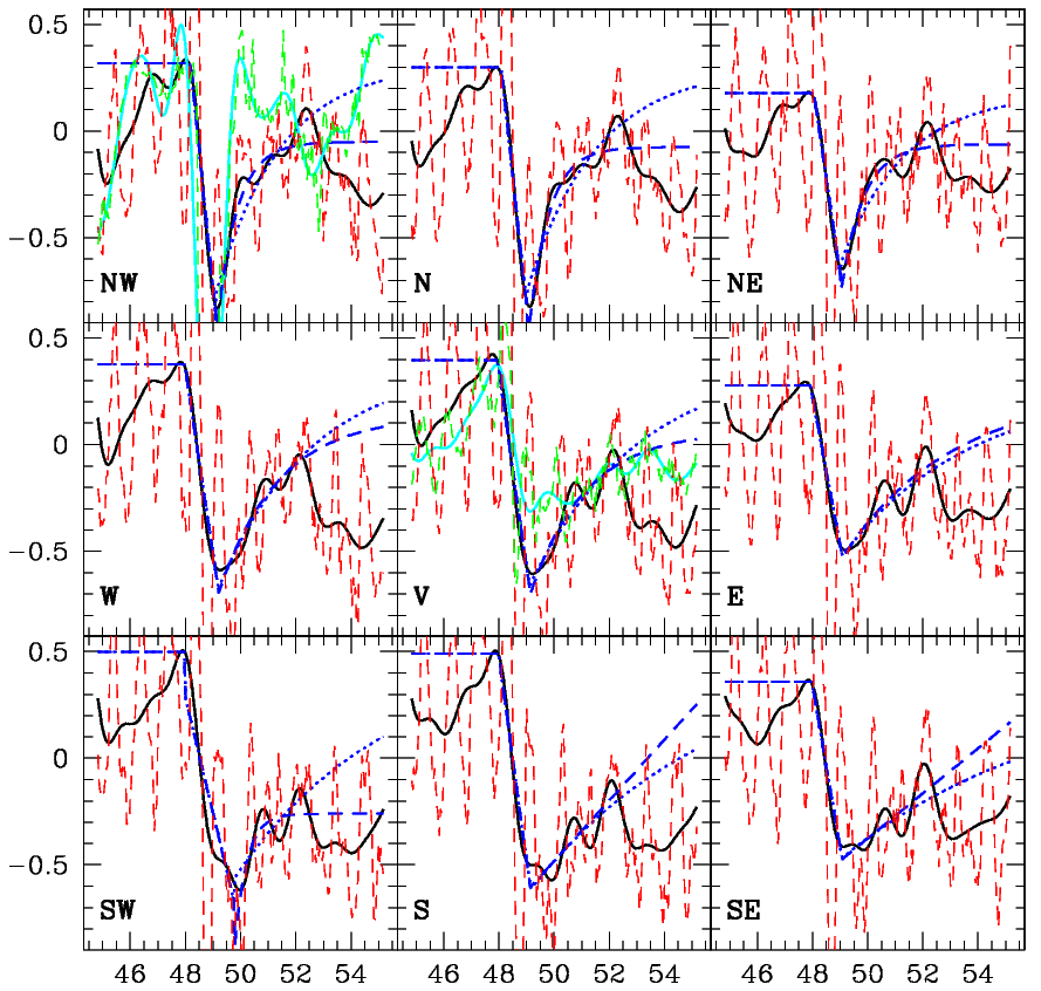}
\caption[FD on 2001 August 17]{The muon flux in the nine directions is shown for the Forbush decrease  \index{Forbush decreases} on 2001 August 17. The fluxes are shown as percentage deviation from mean values. The dashed red lines show the unfiltered data, while solid black lines show the data after applying a low-pass filter. The dotted blue lines show the fits to filtered data. The green lines in middle panel shows the data from the Tibet neutron monitor scaled by a factor 3 to fit in the panel. The green line in the top left panel shows the magnetic field variation. since magnetic field increases during Forbush decrease, we calculated average value and percentaged the deviation of the quantity  $|100 - B|$  then scaled to a factor of 10 to fit in panel} \label{FD17A}
\end{figure}

\subsubsection{24 November 2001} 

The main parameters for this event are listed in tables \ref{T24N} and \ref{TCME}. This FD was observed in  GRAPES-3 \index{GRAPES-3}  muon telescope \index{GRAPES-3!muon telescope} on 24 November 2001. The FD start was on 24 November  2001 and the minimum of the FD profile was reached on 25 November 2001 for all the bins. The cosmic ray intensity data corresponding to this event is shown in the figure \ref{FD24N}.    The FD magnitude\index{Forbush decreases!magnitude}, cut-off rigidity, FD on-set time and FD minimum \index{Forbush decreases!minimum} time are given in the table \ref{T24N}. \\

This event is caused by the halo CME \index{CME!Halo} that started at 22:48 UT on 2001 November 22 \href{http://cdaw.gsfc.nasa.gov/CME_list/}{ \textit{CDAW DATA CENTER} \footnote{$http://cdaw.gsfc.nasa.gov/CME \_ list /$}}. The CME \index{CME!observation} is first observed at a height $4.77 R_{\odot}$ and the last observations was at height $25.92 R_{\odot}$, where the speed of the CME \index{CME!velocity} was $V_{exp} \ = \ 1371 \ km \ s^{-1}$, and the initial acceleration was $a_i \ = \ -12.81 \ m \ s^{-2}$. The magnetic cloud \index{magnetic cloud} starts at  24 November 2001 at 17:00 UT and end at 25 November 2001 at 13:00 UT  \citep{hut05}. The velocity profile for this CME \index{CME!acceleration} is considered to follow the initial acceleration up to the last observational height in the  coronagraph \index{coronagraph} field of view, followed by the constant drag profile with   $C_D \ = \ 0.09 $.The solar wind \index{solar wind} speed when the CME  reaches earth was $V_{sw}^{MC} \ = \ 730 \ km \ s^{-1}$. The radius of magnetic cloud at earth is then, $R(T) \ = \ 10 \times 3600 \times 730  = 2.628 \times 10^7 \ km$.

 The time elapsed between the first observation of the halo CME  and the detection of the magnetic cloud by the ACE spacecraft is 40.5 hours. The average speed of the magnetic cloud by the time it reaches the ACE spacecraft is $730 \ km \ s^{-1}$, and the spacecraft is located around $ 1.5 \times 10^6 \ km$ from the Earth. We therefore estimate that the magnetic cloud would have taken $\sim 0.6$ hours to traverse the distance between ACE and the Earth. The total time $T = 40.5+0.6 = 41.1$ hours. The maximum magnetic field of the magnetic cloud when it reaches earth is $B_{MC} = 20 nT$.

The interplanetary shock \index{shock} reached the spacecraft at 06:00 UT on 24 November  2001, the solar wind speed at this time was $ V_{sw}^{Shock}  \, = 948 \, kms^{-1}$. The ambient solar wind magnetic field was $ B^a \, = 5 \, nT $ and the magnetic field in the shock-sheath \index{sheath} region was $B^{shock} \, = 41.5 \, nT $. 

\begin{table}{\small
\caption[Observed parameters for FD on 24 November 2001.]{Observed parameters for FD \index{Forbush decreases} on 24 November 2001. \index{Forbush decreases!magnitude} \index{Forbush decreases!minimum} \index{Forbush decreases!onset}\index{rigidity!cut-off rigidity}}\label{T24N}
\begin{tabular}{l|cccc}
\hline \hline 
\multicolumn{5}{l}{ FD onset times are in UT, 24 November 2001}\\
\multicolumn{5}{l}{  FD minimum  times are in UT, 25 November 2001}\\ \hline

Directions & FD Magnitude & Cut-off Rigidity & FD Onset & FD minimum \\
    & $(\%)$ & $(GV)$ & $(UT)$ & $(UT)$ \\ \hline
NW  & 1.42 & 15.5  & 03:07 & 17:00 \\ 
N   & 1.34 & 18.7  & 03:21 & 15:00 \\ 
NE  & 0.94 & 24.0  & 03:07 & 16:00 \\ 
W   & 1.67 & 14.3  & 04:05 & 14:00 \\ 
V   & 1.56 & 17.2  & 03:21 & 15:00 \\ 
E   & 1.16 & 22.4  & 02:52 & 17:00 \\ 
SW  & 1.34 & 14.4  & 04:05 & 15:00 \\ 
S   & 1.36 & 17.6  & 03:07 & 16:00 \\ 
SE  & 1.10 & 22.4  & 01:24 & 19:00 \\
\hline \hline
\end{tabular}}
\end{table}

\begin{figure}
\centering
\includegraphics[width= 0.95\textwidth]{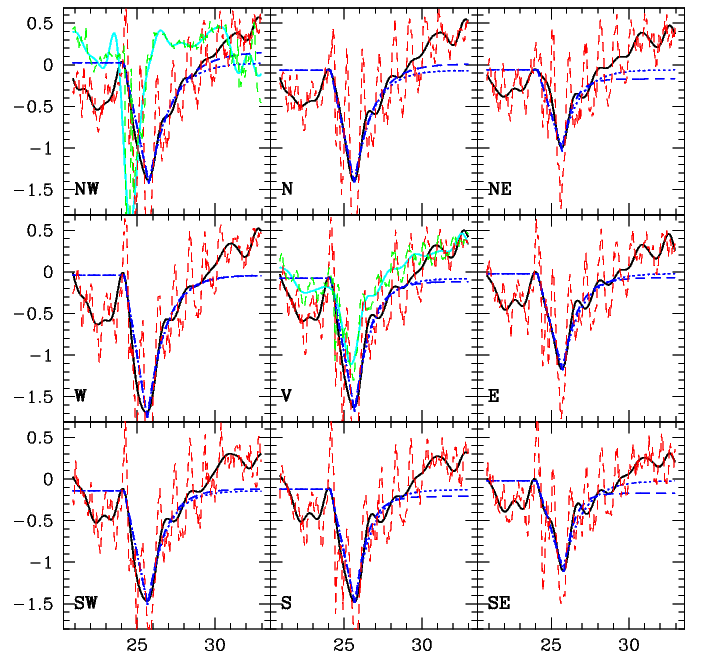}
\caption[FD on 2001 November 24]{The muon flux in the nine directions is shown for the Forbush decrease \index{Forbush decreases} on 2001 November 24. The fluxes are shown as percentage deviation from mean values. The dashed red lines show the unfiltered data, while solid black lines show the data after applying a low-pass filter. The dotted blue lines show the fits to filtered data. The green lines in middle panel shows the data from the Tibet neutron monitor scaled by a factor 3 to fit in the panel. The green line in the top left panel shows the magnetic field variation. since magnetic field increases during Forbush decrease, we calculated average value and percentaged the deviation of the quantity  $|100 - B|$  then scaled to a factor of 10 to fit in panel} \label{FD24N}
\end{figure}

\subsubsection{7 September 2002} 
The main parameters for this event are listed in tables \ref{T7S} and \ref{TCME}. This FD \index{Forbush decreases} was observed in  GRAPES-3 \index{GRAPES-3}  muon telescope \index{GRAPES-3!muon telescope} on 7 September 2002. FD on-set was on  7 September 2002 and the minimum \index{Forbush decreases!minimum} of FD profile was reached on 8 September 2002. The cosmic ray intensity data corresponding to this event is shown in the figure \ref{FD7S}.  The FD magnitude\index{Forbush decreases!magnitude}, cut-off rigidity, FD on-set time and FD minimum time are given in the table \ref{T7S}. \\

The Halo CME \index{CME!Halo} started at 16:54 on 5 September 2002. The CME \index{CME!observation} was first observed by the coronagraph \index{coronagraph} at a height of  $4.12 R_{\odot}$ and the last observation was at height $16.97 R_{\odot}$ with a speed of $V_{exp} \ = \ 1855 \ km \ s^{-1}$. The initial  acceleration \index{CME!acceleration} was $a_i \ = \ 43.01 \ m \ s^{-2}$.The near-earth magnetic cloud \index{magnetic cloud} was observed by ACE spacecraft. The magnetic cloud started at 17:00 UT on 7 September 2002 and end at 16:30 UT on 8 September 2002. The Velocity profile for this CME \index{CME!velocity} considered to follow the initial acceleration up to the last observational height by the coronagraph, then after it followed the constant drag profile with the drag coefficient $C_D \ = \ 0.312 $. The solar wind \index{solar wind} speed at the time of arrival of magnetic cloud was $V_{sw}^{MC} = 544 \ km \ s^{-1}$, then the radius of magnetic cloud is calculated as  $R(T) \, = \, 2.301 \times 10^7 \ km$.

The time elapsed between the first observation of the halo CME  and the detection of the magnetic cloud by the ACE spacecraft is 48.1 hours. The average speed of the magnetic cloud by the time it reaches the ACE spacecraft is $544 \ km \ s^{-1}$, and the spacecraft is located around $ 1.5 \times 10^6 \ km$ from the Earth. We therefore estimate that the magnetic cloud would have taken $\sim$ 0.76 hours to traverse the distance between ACE and the Earth. The total time $T = 48.86$ hours. The maximum magnetic field of the magnetic cloud when it reaches earth is $B_{MC} = 22.9 nT$.

The interplanetary shock \index{shock} reached the spacecraft at 14:20 UT on 7 September 2002, the solar wind \index{solar wind} speed at this time was $ V_{sw}^{Shock}  \, = 550 \, kms^{-1}$. The ambient solar wind magnetic field was $ B^a \, = 5.8 \, nT $ and the magnetic field in the shock-sheath \index{sheath} region was $B^{shock} \, = 23 \, nT $. 

\begin{table}{\small
\caption[Observed parameters for FD on 7 September 2002.]{Observed parameters for FD \index{Forbush decreases} on 7 September 2002. \index{Forbush decreases!magnitude} \index{Forbush decreases!minimum}\index{Forbush decreases!onset}\index{rigidity!cut-off rigidity}}\label{T7S}
\begin{tabular}{l|cccc}
\hline \hline 
\multicolumn{5}{l}{ FD onset times are in UT, 7 September 2002}\\
\multicolumn{5}{l}{ FD minimum times are in UT, 8 September 2002}\\ \hline

Directions & FD Magnitude & Cut-off Rigidity & FD Onset & FD minimum \\
    & $(\%)$ & $(GV)$ & $(UT)$ & $(UT)$ \\ \hline
NW  & 0.408  & 15.5   & 17:03  & 15:00  \\
N   & 0.593  & 18.7   & 17:17  & 14:00  \\
NE  & 0.633  & 24.0   & 16:05  & 14:00  \\
W   & 0.893  & 14.3   & 14:52  & 13:00  \\
V   & 0.971  & 17.2   & 14:52  & 13:00  \\
E   & 0.830  & 22.4   & 15:07  & 14:00  \\
SW  & 1.076  & 14.4   & 15:50  & 15:00  \\
S   & 1.065  & 17.6   & 16:19  & 16:00  \\
SE  & 0.840  & 22.4   & 17:03  & 16:00  \\
\hline \hline
\end{tabular}}
\end{table}

\begin{figure}
\centering
\includegraphics[width= 0.95\textwidth]{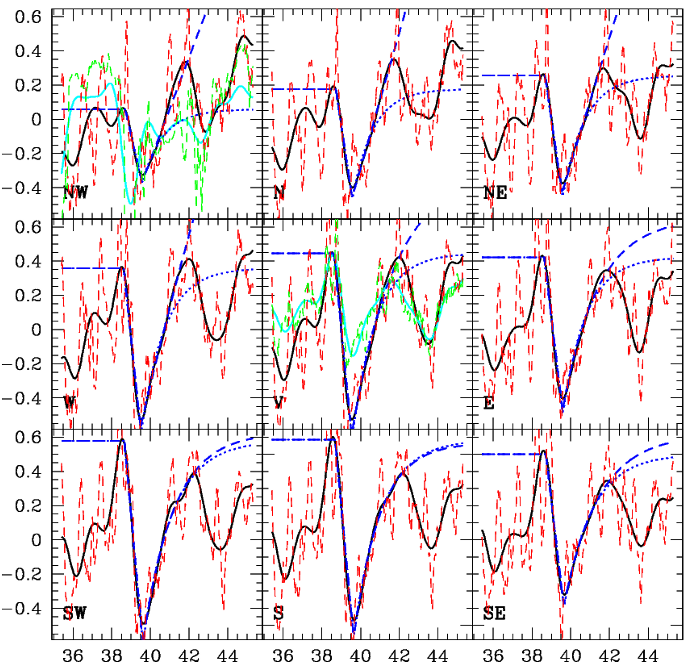}
\caption[FD on 2002 September 07]{The muon flux in the nine directions is shown for the Forbush decrease \index{Forbush decreases} on 2002 September 7. The fluxes are shown as percentage deviation from mean values. The dashed red lines show the unfiltered data, while solid black lines show the data after applying a low-pass filter. The dotted blue lines show the fits to filtered data. The green lines in middle panel shows the data from the Tibet neutron monitor scaled by a factor 3 to fit in the panel. The green line in the top left panel shows the magnetic field variation. since magnetic field increases during Forbush decrease, we calculated average value and percentaged the deviation of the quantity  $|100 - B|$  then scaled to a factor of 10 to fit in panel} \label{FD7S}
\end{figure}

\subsubsection{20 November 2003} 
The salient parameters for this event are listed in the tables \ref{T20N} and \ref{TCME}. This FD \index{Forbush decreases} was observed in  GRAPES-3 \index{GRAPES-3}  muon telescope \index{GRAPES-3!muon telescope} on 20 November 2003. The cosmic ray intensity data corresponding to this event is shown in the figure \ref{FD20N}.  The FD magnitude\index{Forbush decreases!magnitude}, cut-off rigidity, FD on-set time and FD minimum \index{Forbush decreases!minimum} time are given in the table \ref{T20N}. \\

The Halo CME \index{CME!Halo} started at 8:50 UT on 18 November 2003. The CME \index{CME!observation} was first observed by the coronagraph \index{coronagraph} at a height of  $6.3 R_{\odot}$ and the last observation was at height $27.5  R_{\odot}$ with a speed of $V_{exp} \ = \ 1645 \ km \ s^{-1}$. The initial acceleration \index{CME!acceleration} was $a_i \ = \ -3.29 \ m \ s^{-2}$.The near-earth magnetic cloud \index{magnetic cloud} was observed by ACE spacecraft. The magnetic cloud started at 10:06 UT on 20 November 2003 and ended at 00:24 UT on 21 November 2003. The Velocity profile for this CME \index{CME!velocity} considered to follow the initial acceleration up to the last observational height by the coronagraph, then after it followed the constant drag profile with the drag coefficient $C_D \ = \ 0.333 $. The solar wind \index{solar wind} speed at the time of arrival of magnetic cloud was $V_{sw}^{MC} = 750 \ km \ s^{-1}$, then the radius of magnetic cloud is calculated as  $R(T) \ = 1.89 \times 10^7 \ km$ \citep{wang06}.
 The total time travel of the CME was $42.1$ hours. The maximum magnetic field of the magnetic cloud when it reaches earth is $B_{MC} = 50 nT$. The interplanetary shock \index{shock}  reached spacecraft at 7:30 UT on 20 November 2003.

\begin{table}{\small
\caption[Observed parameters for FD on 20 November 2003.]{Observed parameters for FD \index{Forbush decreases} on 20 November 2003. \index{Forbush decreases!magnitude}\index{Forbush decreases!minimum}\index{Forbush decreases!onset}\index{rigidity!cut-off rigidity}}\label{T20N}
\begin{tabular}{l|cccc}
\hline \hline 
\multicolumn{5}{l}{  FD onset times are  in UT, 20 November 2003}\\
\multicolumn{5}{l}{ The $^1$ FD onset times are in UT, 19 November 2003}\\
\multicolumn{5}{l}{  The $^2$ FD onset times are  in UT, 21 November 2003}\\
\multicolumn{5}{l}{ FD minimum times are in UT, 24 November 2003}\\  \hline

Directions & FD Magnitude & Cut-off Rigidity & FD Onset & FD minimum \\
    & $(\%)$ & $(GV)$ & $(UT)$ & $(UT)$ \\ \hline
NW  & 0.95  & 15.5  & 21:22 $^1$ & 5:00  \\
N   & 0.93  & 18.7  & 08:10 $^2$ & 4:00  \\
NE  & 0.81  & 24.0  & 02:53 $^2$ & 2:00  \\
W   & 1.19  & 14.3  & 01:55      & 4:00  \\
V   & 1.16  & 17.2  & 10:48      & 4:00  \\
E   & 0.97  & 22.4  & 20:38      & 4:00  \\
SW  & 1.17  & 14.4  & 06:58      & 3:00  \\
S   & 1.20  & 17.6  & 10:19      & 3:00  \\
SE  & 0.93  & 22.4  & 15:07      & 2:00  \\
\hline \hline

\end{tabular}}
\end{table}

\begin{figure}
\centering
\includegraphics[width= 0.95\textwidth]{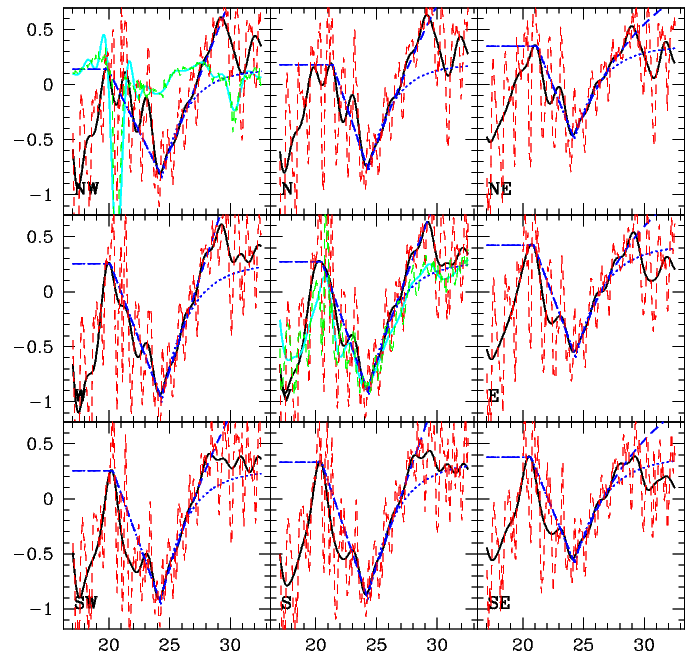}
\caption[FD on 2003 November 20]{The muon flux in the nine directions is shown for the Forbush decrease  \index{Forbush decreases} on 2003 November 20. The fluxes are shown as percentage deviation from mean values. The dashed red lines show the unfiltered data, while solid black lines show the data after applying a low-pass filter. The dotted blue lines show the fits to filtered data. The green lines in middle panel shows the data from the Tibet neutron monitor scaled by a factor 3 to fit in the panel. The green line in the top left panel shows the magnetic field variation. since magnetic field increases during Forbush decrease, we calculated average value and percentaged the deviation of the quantity  $|100 - B|$  then scaled to a factor of 10 to fit in panel} \label{FD20N}
\end{figure}

\subsubsection{26 July 2004} 
The main parameters for this event are listed in tables \ref{T26J} and \ref{TCME}. This FD was observed in  GRAPES-3 \index{GRAPES-3}  muon telescope \index{GRAPES-3!muon telescope} on 26 July 2004.  The FD on-set was  on 26 July 2004  and the minimum of the FD profile \index{Forbush decreases!profile} was reached on 27 July 2004. The cosmic ray intensity data corresponding to this event is shown in the figure \ref{FD26J}. The FD magnitude\index{Forbush decreases!magnitude}, cut-off rigidity \index{rigidity!cut-off rigidity}, FD on-set time and FD minimum \index{Forbush decreases!minimum} time are given in the table \ref{T26J}. \\

This event was associated with a halo CME \index{CME!Halo} which was first observed at 14:54 UT on 25 July 2004 \href{http://cdaw.gsfc.nasa.gov/CME_list/}{ \textit{CDAW DATA CENTER} \footnote{$http://cdaw.gsfc.nasa.gov/CME \_ list /$}}.The CME \index{CME!observation} is first seen at a height $4.22 R_{\odot}$ and last observed at a height of $21.86 R_{\odot}$ with a \index{CME!velocity}speed of  $V_{exp} \ = \ 1366 \ km \ s^{-1}$, and the initial acceleration \index{CME!acceleration} was $a_i \ = \ 7.0 \ m \ s^{-2}$ in the coronagraph \index{coronagraph} field of view. There after we assume that it follows the constant drag profile with drag coefficient $C_D \ = \ 0.016 $.
 The near-Earth magnetic cloud \index{magnetic cloud} was observed in the spacecraft at 02:00 UT on 27 July 2004 and end at 24:00 UT on the same day. The radius of the magnetic cloud as measured from ACE spacecraft is $R(T) \, =  \, 3.564 \times 10^7 \ km $,where we have assumed that magnetic cloud moves with the maximum in-situ speed of the ambient solar wind behind the shock $V_{sw}^{MC} = 900 \ km \ s^{-1}$. The maximum magnetic field of the magnetic cloud when it reaches earth was $25.3 \ nT$. The total time of travel for the CME  is calculated using the time at which the CME  is first observed in SOHO and the the when it reaches the ACE spacecraft and the time taken for the CME  to travel from space craft to earth using the velocity  $V_{sw}^{MC} = 900 \ km \ s^{-1}$. We find $T \, = \, 35.71$ hours.

The interplanetary shock \index{shock}  reached the spacecraft at 22:20 UT on 26 July 2004, the solar wind \index{solar wind} speed at this time was $ V_{sw}^{Shock}  \, = 893 \, kms^{-1}$. The ambient solar wind magnetic field was $ B^a \, = 5 \, nT $ and the magnetic field in the shock-sheath \index{sheath} region was $B^{shock} \, = 26.1 \, nT $. 

\begin{table} {\small
\caption[Observed parameters for FD on 26 July 2004. ]{Observed parameters for FD \index{Forbush decreases} on 26 July 2004. \index{Forbush decreases!magnitude}\index{Forbush decreases!minimum}\index{Forbush decreases!onset}\index{rigidity!cut-off rigidity}}\label{T26J}
\begin{tabular}{l|cccc}
\hline \hline 
\multicolumn{5}{l}{  FD onset times are  in UT, 26 July 2004}\\
\multicolumn{5}{l}{  FD minimum times are  in UT, 27 July 2004}\\ \hline

Directions & FD Magnitude & Cut-off Rigidity & FD Onset & FD minimum \\
    & $(\%)$ & $(GV)$ & $(UT)$ & $(UT)$ \\ \hline
NW  & 2.06  & 15.5  & 14:24 & 10:00 \\
N   & 2.04  & 18.7  & 15:22 & 11:00 \\
NE  & 1.67  & 24.0  & 17:17 & 13:00 \\
W   & 2.40  & 14.3  & 14:10 & 10:00 \\
V   & 2.12  & 17.2  & 15:36 & 11:00 \\
E   & 1.62  & 22.4  & 18:00 & 13:00 \\
SW  & 1.77  & 14.4  & 16:05 & 12:00 \\
S   & 1.67  & 17.6  & 18:29 & 14:00 \\
SE  & 1.40  & 22.4  & 20:38 & 15:00 \\
\hline \hline
\end{tabular} }
\end{table}

\begin{figure}
\centering
\includegraphics[width= 0.95\textwidth]{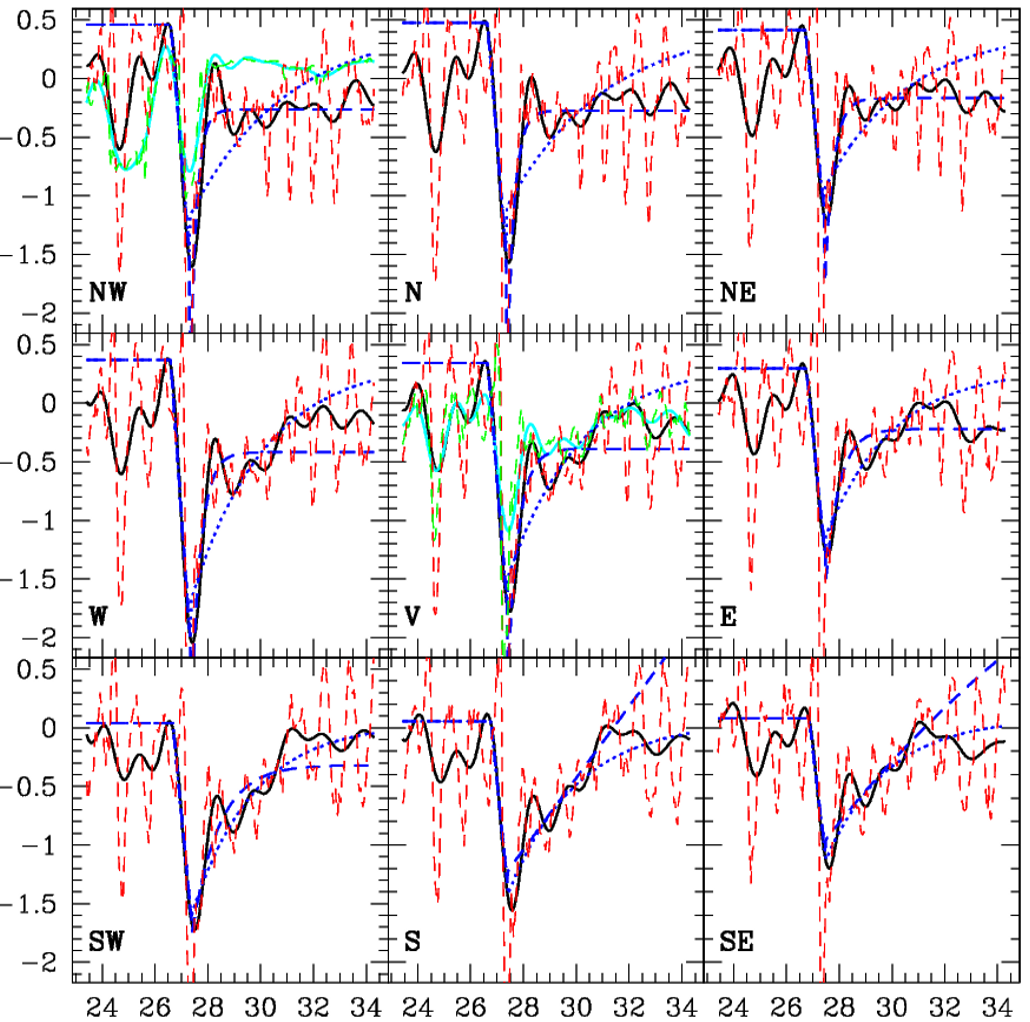}
\caption[FD on 2004 July 26]{The muon flux in the nine directions is shown for the Forbush decrease \index{Forbush decreases} on 2004 July 26. The fluxes are shown as percentage deviation from mean values. The dashed red lines show the unfiltered data, while solid black lines show the data after applying a low-pass filter. The dotted blue lines show the fits to filtered data. The green lines in middle panel shows the data from the Tibet neutron monitor scaled by a factor 3 to fit in the panel. The green line in the top left panel shows the magnetic field variation. since magnetic field increases during Forbush decrease, we calculated average value and percentaged the deviation of the quantity  $|100 - B|$  then scaled to a factor of 10 to fit in panel} \label{FD26J}
\end{figure}

\subsection{Summary of observationally derived parameters for events in short-list 3}

Table~\ref{TCME} contains the observationally determined parameters for each of the CMEs and their corresponding shocks in the final short-list (Table~\ref{SL3}).

The quantity ``First obs'' denotes the time (in UT) when the CME was first observed in the LASCO \index{coronagraph!LASCO} FOV, while ${\rm R_{first}}$ is the distance (in units of  ${\rm R_{\odot}}$) at which CME was first observed \index{CME!observation} in LASCO FOV and ${\rm R_{last}}$ is the  distance at which the CME was last observed in LASCO FOV. The quantity ${\rm V_{exp}}$ is the speed of CME \index{CME!velocity} at ${\rm R_{last}}$ (in units of ${\rm kms^{-1}}$) and  ${\rm a_i}$ is the  acceleration of CME \index{CME!acceleration} in the LASCO FOV  (in units of ${\rm ms^{-2}}$). The quantity ${\rm C_D}$ is the (constant) dimensionless drag coefficient used for the velocity profile (\S~{\ref{VP}}). The quantities MC start and MC end denote the start and end times of the magnetic cloud \index{magnetic cloud} in UT. The quantity ${\rm V_{sw}^{MC}}$ is solar wind speed  at the Earth (in units of ${\rm kms^{-1}}$) just ahead of the arrival of the magnetic cloud, and ${\rm R_{MC}}$ is the radius of the magnetic cloud (in units of ${\rm km }$). The quantity ${\rm B_{MC}}$ is the peak magnetic field inside the magnetic cloud (in units of ${\rm nT}$). The quantity ${\rm T_{total}}$ is the Sun-Earth travel time (in hours) taken by the CME  to travel from Sun to Earth. 
The quantity Shock \index{shock} arrival denotes the time (in UT) when the shock is detected near the Earth. The quantities ${\rm B^{a}}$ and ${\rm B^{shock}}$ represent the magnetic fields (in nT) in the ambient solar wind \index{solar wind} and inside shock sheath \index{sheath} region respectively (see Fig \ref{Fshock} for an example). The quantity ${\rm V_{sw}^{shock}}$ represents the near-Earth shock speed in ${\rm kms^{-1}}$. 

\afterpage{
\begin{landscape} 
\begin{table*} { \small
\caption[Observed  parameters of  CME $\&$ Shock for different events]{ Observed  parameters of  CME $\&$ Shock for different events}\label{TCME}
\centering
\begin{tabular}{l|c|c|c|c|c|c}
\hline \hline
 Event & 11 Apr 2001 & 17 Aug 2001 & 24 Nov 2001 & 7 Sep 2002 & 20 Nov 2003 & 26 Jul 2004 \\ \hline \hline 
\multicolumn{7}{l} { CME details }\\ \hline
 First obs.   ( UT) & $10/04/01$, 05:30 & $15/08/01$, 23:54 & $22/11/01$, 22:48 & $5/09/02$, 16:54 & $18/11/03$, 8:50 & $25/07/04$, 14:54 \\ \hline
${\rm R_{first}}$ $\, (R_{\odot})$ & 2.84  & 3.38 & 4.77  & 4.12  & 6.3  & 4.22 \\ \hline
${\rm R_{last}}$  $\, (R_{\odot})$ & 18.05  & 25.91 & 25.92  & 16.97 & 27.5 & 21.86\\ \hline
${\rm V_{exp}}$  $(km\, s^{-1})$& 2876 & 1413 & 1371 & 1855 & 1645 & 1366 \\ \hline
${\rm a_i}$ $(m\, s^{-2})$ & 211.60 & -31.66 & -12.81 &  43.01 & -3.29 & 7.0 \\  \hline
${\rm C_D}$ & 0.325 & 0.163 & 0.09 & 0.312 & 0.333 & 0.016 \\ \hline
MC start   (UT) & $11/04/01$, 23:00 & $18/08/01$, 00:00 & $24/11/01$, 17:00 & $7/09/02$, 17:00 & $20/11/03$, 10:06 & $27/07/04$, 2:00 \\ \hline
MC end    (UT)& $12/04/01$, 18:00 &  $18/08/01$, 21:30 & $25/11/01$, 13:00 & $8/09/02$, 16:30 & $21/11/03$, 00:24 & $27/07/04$, 24:00  \\ \hline
${\rm V_{sw}^{MC}}$ $(km\, s^{-1})$ & 725 & 600  & 730 & 544 & 750 & 900 \\ \hline
${\rm R_{MC}}$  $(km\,) $& $2.48 \times 10^7 $  & $2.32 \times 10^7 $ & $2.63 \times 10^7 $ & $2.30 \times 10^7 $ & $1.89 \times 10^7 $ & $3.56 \times 10^7 $\\ \hline
${\rm B_{MC}}$   $(nT)$ & 34.5  & 25.6  & 20  & 22.9 & 50  & 25.3  \\ \hline
${\rm T_{total}}$ $(hours)$ & 42.1  & 44.9 & 41.1 & 48.86  & 42.1 & 35.71 \\ \hline \hline
\multicolumn{7}{l} {Shock }\\ \hline
Shock arrival (UT)& $11/04/01$, 14:06 & $17/08/01$, 11:00 & $24/11/01$, 6:00 & $7/09/02$, 14:20 & $20/11/03$, 7:30 & $26/7/04$, 22:20 \\ \hline
${\rm B^{a}}$ $(nT)$ & 4.5 & 5 & 5 & 5.8 & 7 & 5 \\ \hline
${\rm B^{shock}}$ $(nT)$& 32.5 & 33 & 41.5 & 23 & & 26.1\\ \hline
${\rm V_{sw}^{shock}}$  $(km\, s^{-1})$& 670 & 501 & 948 & 550 & & 893\\ \hline \hline
\hline 

\end{tabular}}
\end{table*} 
\end{landscape}}

\subsection{Is the notion of cosmic ray diffusion valid?} \label{NoD1}
In order for the diffusion \index{diffusion} approximation to be valid, the Larmor radius of a typical cosmic ray proton ($r_L$) needs to be substantially smaller than the CME size for the CME-only model\index{CME-only model}. For the shock-only model\index{shock-only model}, the proton Larmor radius needs to be substantially smaller than the CME-shock stand-off distance.

\subsubsection{${\rm r_{\rm L}/R_{\rm CME}}$ for CME-only model} \index{CME-only model}
{\citet{kub10} have simulated the process of cosmic ray diffusion into an ideal flux rope CME \index{CME} in the presence of MHD turbulence. They find that, if the quantity ${\rm f_0(t) \equiv R_{L}(t)/R(t)}$ is small, cosmic ray \index{cosmic rays} penetration into the flux rope is dominated by diffusion via turbulent irregularities. Other effects such as gradient drift due to the curvature of the magnetic field are unimportant under these conditions. Figure~\ref{Fig1} shows the quantity ${\rm f_0(t) \equiv R_{L}(t)/R(t)}$ for 12 and 24 GV protons, for each of the CMEs \index{CME} in our final short-list (Table~\ref{SL3}). The Larmor radius ${\rm R_{L}(t)}$ is defined by Eqs~(\ref{rl}) and (\ref{BFL}) and the CME  radius ${\rm R(t)}$ is defined in Eq~(\ref{ER}). Clearly, ${\rm f_{0} \ll 1}$ all through the Sun-Earth passage of the CMEs, and this means that the role of MHD turbulence \index{turbulence} in aiding penetration of cosmic rays into the flux rope structure is expected to be important. }

\begin{figure*}[h]
   \centering
      \includegraphics[width = 1.0\textwidth]{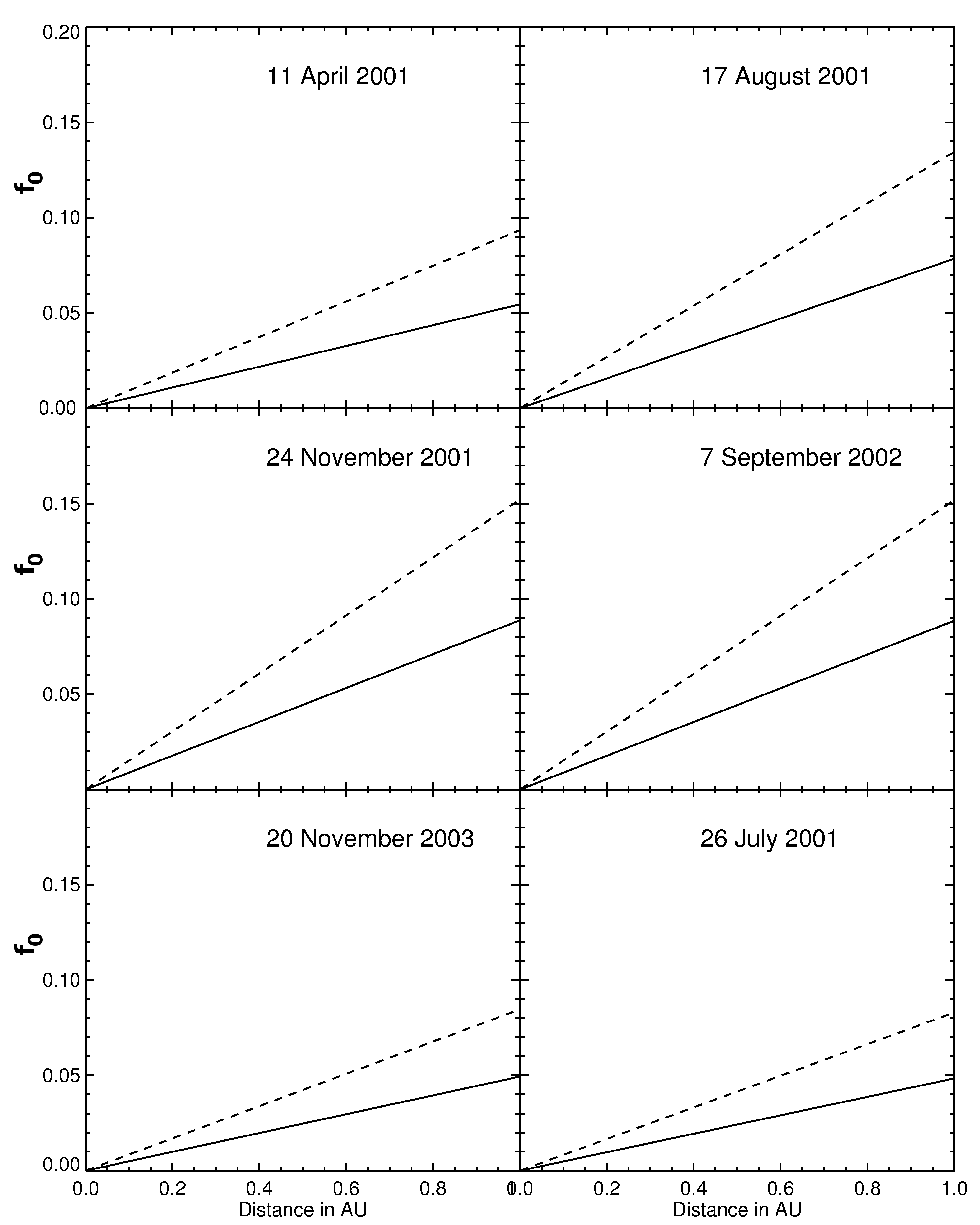}
   \caption[The quantity ${f_{0} \equiv \rm R_{L}/R}$ for the propagating CME]{The quantity ${f_{0} \equiv \rm R_{L}/R}$ versus distance as a CME \index{CME} propagates from the Sun to the Earth. The dashed and continuous lines represent 24 GV and 12 GV protons, respectively.}
              \label{Fig1}%
    \end{figure*}

\subsubsection{Ratio of ${\rm r_{\rm L}}$ to shock CME stand-off distance for shock-only model}

Table~\ref{TNRL} shows the ratio of the Larmor radius of the 14 GV and 24 GV cosmic rays \index{cosmic rays} to the shock-CME stand-off distance for each of the events in our final short-list. Assuming that the shock-CME stand-off distance is representative of the shock sheath \index{sheath} thickness, the numbers in Table~\ref{TNRL} lead us to conclude that the idea of cosmic ray diffusion would be valid for the shock-only \index{shock-only model} model as well.
                                  
\begin{table}
\caption{Comparison of shock CME stand-off distance and Larmor radius of high energy particles}\label{TNRL}
\centering
\begin{tabular}{l|cc}
\hline
Event & ${\rm R_l/L}$ & $ {\rm R_l/L}$ \\
     & (14 GV) & (24 GV) \\ \hline
 11 Apr 2001 & 0.0661460  & 0.113393 \\ 
 17 Aug 2001 & 0.0603128  & 0.103393 \\
 24 Nov 2001 & 0.0299540  & 0.0513498 \\
 07 Sep 2002 & 0.385241   & 0.660413  \\
 20 Nov 2003 & 0.321034   & 0.550344  \\
 26 Jul 2004 & 0.151961   & 0.260504  \\ \hline
\noalign{\smallskip}
\end{tabular}
\end{table}

\subsection{Fitting the CME-only and shock-only models to multi-rigidity FD data}


Using the observational parameters listed in Table~\ref{TCME}, we have computed the magnitude \index{Forbush decreases!magnitude} of the Forbush decrease \index{Forbush decreases} using the CME-only \index{CME-only model} (\S~\ref{cme-only}) and shock-only models \index{shock-only model} (\S~\ref{Sonly}). The only free parameter in our model is the ratio of the energy density in the random magnetic fields to that in the large scale magnetic field ${\rm \sigma^2 \equiv \langle {B_{\rm turb}}^2/{B_{0}}^2 \rangle}$. Figure \ref{Call} shows the best fits of the CME-only model to the multi-rigidity data. The only free parameter in the model is ${\rm \sigma^2 \equiv \langle {B_{\rm turb}}^2/{B_{0}}^2 \rangle}$, and the best fit is chosen by minimizing the $\chi^2$ with respect to $\sigma^2$\index{turbulence!turbulence level $\sigma$}. For each FD event, the $\ast$ symbols denote the observed FD \index{Forbush decreases} magnitude for a given rigidity bin. The dashed line denotes the FD magnitude predicted by the CME-only model.  We define the chi-square statistic as  
\begin{equation}
\rm \chi^2_ = \sum _i \frac {(E_i - D_i)^2} {var_i}
\label{ch2}
\end{equation}
where ${\rm E_i}$ is the value predicted by the theoretical model  ${\rm D_i}$ is the corresponding  GRAPES-3 \index{GRAPES-3}  data point and ${\rm var_i}$ is the variance for the corresponding data points. The ${\rm \chi ^2}$ values  obtained after minimizing with respect to $\sigma^2$ are listed in Table~\ref{Chi2}. 

\begin{figure*}
   \centering
      \includegraphics[width = 1.0\textwidth]{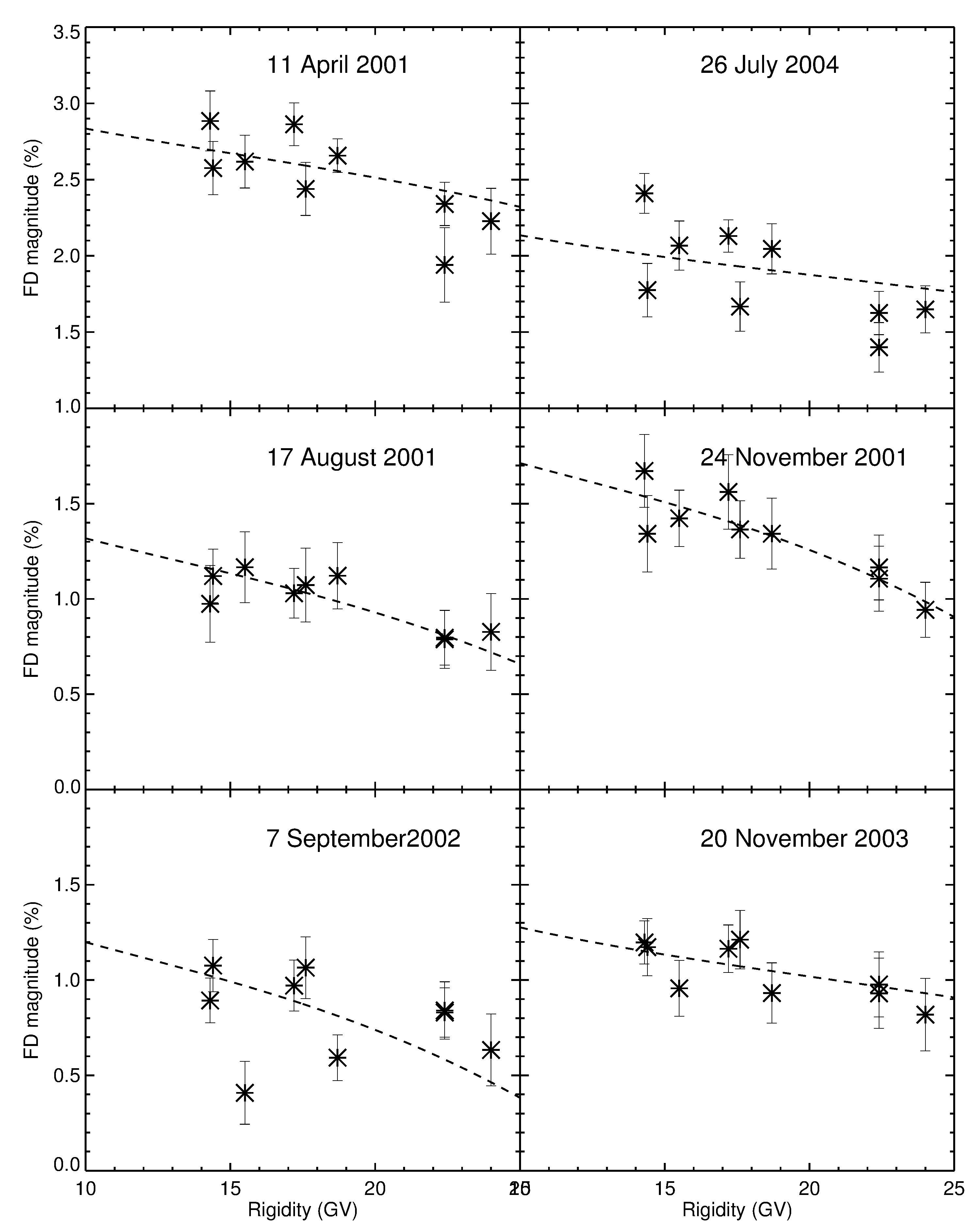}
   \caption[FD magnitude predicted by CME-only model]{The $\ast$ symbols show the Forbush decrease \index{Forbush decreases!magnitude} magnitude observed with  GRAPES-3\index{GRAPES-3}. The dashed line is obtained using the CME-only model}
              \label{Call}%
\end{figure*}

\begin{table}
\caption[Minimum $\chi^2$ values for the CME-only model]{{ Minimum} ${\rm \chi^2} ${ values for the CME-only \index{CME-only model} model fits to  GRAPES-3 \index{GRAPES-3}  data}\label{Chi2}}
\centering
\begin{tabular}{l c}
\hline \hline
Event & ${\rm \chi^2}$  \\
\hline
11 April 2001 & 11.135 \\
17 August 2001 & 1.963\\
24 November 2001 & 2.457\\
7 September 2002 & 25.537 \\
20 November 2003 & 3.661 \\
26 July 2004 & 27.480 \\
\hline \hline
\end{tabular}
\end{table}

The entries in the row $\rm {\sigma_{CME}}$ in Table~\ref{Sig} denote the square roots of the turbulence parameter $\sigma^{2}$ that we have used for the model fits for each event. These values represent the level \index{turbulence!turbulence level $\sigma$} of turbulence \index{turbulence} in the sheath \index{sheath} region immediately ahead of the CME, through which the cosmic rays \index{cosmic rays} must traverse in order to diffuse into the CME. By comparison, the value of $\sigma$ \index{turbulence!turbulence level $\sigma$} for the quiescent solar wind \index{solar wind} ranges from 6--15\% \citep{spa02}. The CME-only model \index{CME-only model} thus implies that the sheath region ahead of the CME \index{CME} is only a little more turbulent than the quiescent solar wind, except for the 26 July 2004 event, where the speed of CME \index{CME} at the Earth was much higher than that for the other events.

We have carried out a similar exercise for the shock-only model ({\S~\ref{Sonly}}). For each event, we have used the observationally obtained parameters pertaining to the shock \index{shock} listed in Table~\ref{TCME}. Since this model needs the turbulence levels in both the ambient medium as well as the shock sheath region to be specified, we have assumed that the turbulence level \index{turbulence!turbulence level $\sigma$} inside the shock sheath region is twice that in the ambient medium. We find that it is not possible to fit the shock-only \index{shock-only model} model to the multi-rigidity data using values for the turbulence parameter that are reasonably close to that in the quiescent solar wind. For each event, the column called $\rm {\sigma_{Shock}}$  in Table~\ref{Sig} denotes the turbulence level in the shock sheath region that are required to obtain a reasonable fit to the data. Clearly, these values are an order of magnitude higher than those observed in the quiet solar wind. 

\begin{table}
\caption[Turbulence levels in the sheath region]{Turbulence levels  in the sheath region required by the models}\label{Sig}
\centering
\begin{tabular}{l|cc}
\hline \hline
Event & CME-only model & shock-only model \\
 & ${\rm \sigma _{mc}}$& ${\rm \sigma _{Shock}}$ \\
\hline
11 April 2001 & 9.375 \% & 100 \% \\
17 August 2001 & 13.492 \%  & 180 \%   \\
24 November 2001 & 28.389 \%  &   400 \% \\
7 September 2002 & 13.379\% &100\% \\
20 November 2003 &  6.6571 \% &  400 \% \\
26 July 2004 & 46.197 \%  &  200 \% \\
\hline \hline
\end{tabular}
\end{table}

\section{Summary and Conclusion}
Our main aim in this chapter is to determine whether Forbush decreases \index{Forbush decreases} due to cosmic rays \index{cosmic rays} of rigidities ranging from  14 to 24 GV are caused primarily by the CME, or by the shock associated with it. We examine this question in the context of multi-rigidity Forbush decrease data from the  GRAPES-3 \index{GRAPES-3}  instrument.
We use a carefully selected sample of FD \index{Forbush decreases} events from  GRAPES-3 \index{GRAPES-3}  that are associated with both CMEs \index{CME} and shocks\index{shock}.

We consider two models, the CME-only cumulative diffusion model \index{CME-only model} (\S~\ref{cme-only}) and the shock-only model \index{shock-only model} (\S~\ref{Sonly}) to understand the Forbush decreases. 

\begin{itemize}
\item  In the CME-only cumulative diffusion model, we envisage the CME \index{CME} as an expanding bubble bounded by large-scale turbulent magnetic fields. The CME \index{CME} starts out from near the Sun with practically no high energy cosmic rays inside it. As it travels toward the Earth, high energy cosmic rays diffuse into the CME \index{CME} across the large-scale turbulent magnetic fields \index{turbulence!turbulent magnetic field} bounding it. The diffusion coefficient \index{diffusion!diffusion coefficient} is a function of the rigidity of the cosmic ray particles as well as the level of MHD turbulence in the vicinity of the CME \index{CME} (the sheath \index{sheath} region). Despite the progressive diffusion \index{diffusion}of cosmic rays into it, the cosmic ray density inside the CME \index{CME} is still lower than the ambient density when it reaches the Earth. When the CME \index{CME} engulfs the Earth, this density difference causes the Forbush decrease \index{Forbush decreases} observed by cosmic ray detectors.
\item In the shock-only model\index{shock-only model}, we consider the shock as a propagating diffusive barrier. It acts as an umbrella against cosmic rays, and the cosmic ray \index{cosmic rays} density behind the ``umbrella'' is lower than that ahead of it. This difference in intensity is manifested as Forbush decrease.
\end{itemize}

We have obtained a list of Forbush decrease \index{Forbush decreases} events observed by the  GRAPES-3 \index{GRAPES-3}  instrument using the short-listing criteria described in \S~ \ref{ESC}. For each of these short-listed events, we have used observationally derived parameters listed in Table~\ref{TCME} for both the models. The only free parameter was the level of MHD turbulence (defined as the square root of the energy density in the turbulent magnetic fluctuations to that in the large-scale magnetic field) in the sheath region. 

Figure~\ref{Call} shows the results of the CME-only cumulative diffusion model fits to multi-rigidity data for each of the short-listed events. We use the turbulence level in the shock sheath region as the free parameter in our models to fit the observed FD magnitudes\index{Forbush decreases!magnitude}. Table~\ref{Sig} summarizes the values of these turbulence levels that we have used for each of the FD events in the final short-list. These values may be compared with the estimate of 6--15 \% for the turbulence level in the quiescent solar wind \citep{spa02}. We thus find that a good model fit using the CME-only  cumulative diffusion model requires a turbulence level in the sheath region that is typically only a little higher than that in the quiet solar wind. On the other hand, a good fit using the shock-only model demands a turbulence level in the shock sheath region that is often an order of magnitude higher than that in the quiet solar wind\index{solar wind}, which is unrealistic. The results summarized in Table~\ref{Sig} imply that, for FDs \index{Forbush decreases} involving protons of rigidities ranging from 14 to 24 GV, the CME-only cumulative diffusion  model is a viable one, while the shock-only model is not. Given the remarkably good fits to multi-rigidity data (Figure~\ref{Call}, the reasonable turbulence levels in the sheath region demanded by the CME-only  cumulative diffusion model (Table~\ref{Sig}) and because the FD minima usually occur well within the magnetic cloud (Tables~\ref{T11A},\ref{T17A},\ref{T24N},\ref{T7S},\ref{T20N} \$ \ref{T26J}), we conclude that CMEs \index{CME} are the dominant cause of the FDs observed by  GRAPES-3\index{GRAPES-3}.

\chapter[Relation of FD with IP magnetic field]{Relation of Forbush decreases with Interplanetary magnetic field compressions}
\label{corrIP}

\noindent\makebox[\linewidth]{\rule{\textwidth}{3pt}} 
{\textit {The CME-only model of the last chapter envisaged the FD as arising out of the cumulative diffusion of high energy protons into the CME through the turbulent magnetic field of the sheath region. However, the diffusion was assumed to occur across an idealized thin boundary region. In this chapter we go beyond this assumption and investigate the magnetic structure of the CME sheath. We examine the relation between the magnetic field compression profile and the FD profile, and interpret it in terms of cross-field diffusion of protons. }  }\\
\noindent\makebox[\linewidth]{\rule{\textwidth}{3pt}}

\section{Introduction}
Solar transients like CMEs cause enhancements in the interplanetary (IP) magnetic field\index{Interplanetary magnetic field}. The near-Earth Manifestation of a CME \index{CME} from the Sun typically has two major components: i) the interplanetary counterpart of CME (commonly called an ICME), and ii) the shock, which is driven ahead of it. Both the shock \index{shock} and the ICME will cause significant enhancement in the IP magnetic fields. Some ICMEs, which possess some well defined criteria such as reduction in plasma temperature and smooth rotation of magnetic field  are called magnetic clouds \index{magnetic cloud} (e.g., \citealp{bur81, botsch98}). The relative contribution of shock and magnetic cloud for the Forbush decrease is a matter of debate. \citet{zha88, loc91, rea09}  argue against the contribution of magnetic clouds to Forbush decreases \index{Forbush decreases} whereas \citet{bad86, sand90,kuw09}  concluded that magnetic clouds can make an important contribution to FDs\index{Forbush decreases}. 



Correlations \index{correlation} between parameters characterising FDs and solar wind \index{solar wind} parameters has been a subject of considerable study. \citet{bel01}  and \citet{kan10} maintain that there is a reasonable correlation between the FD magnitude \index{Forbush decreases!magnitude} and the product of maximum magnetic field and maximum solar wind speed. \citet{dum12} also found reasonable correlations between the FD magnitude $|FD|$, and duration with the solar wind parameters such as the amplitude of magnetic field enhancement $B$, amplitude of the magnetic field fluctuations $\delta B$, maximum solar wind speed associated with the disturbance $v$, duration of the disturbance $t_B$. 
We note that the Forbush decrease magnitude also depends strongly on other solar wind \index{solar wind} parameters like velocity of the CME, turbulence level \index{turbulence!turbulence level $\sigma$} in the magnetic field, size of the CME etc. The contributions of these parameters are explained in the CME-only \index{CME-only model} cumulative diffusion model in the section \ref{cme-only}

In chapter \ref{model} we described the CME-only cumulative diffusion model \index{CME-only model} for Forbush decreases, \index{Forbush decreases} where the cumulative effects of diffusion \index{diffusion} of cosmic ray \index{cosmic rays} protons through the turbulent sheath \index{sheath} region as the CME propagated from the Sun to the Earth was invoked to explain the FD magnitude\index{Forbush decreases!magnitude}. However, the diffusion was envisaged to occur across an idealized thin boundary. In this chapter we examine the detailed relationship between the FD profile \index{Forbush decreases!profile} and the IP \index{Interplanetary magnetic field!compression}magnetic field compression. 
  
\section{Data Analysis}\label{data}

In our studies we identified the Forbush decreases (FDs) \index{Forbush decreases} using the data from GRAPES-3 muon telescope. Details of this muon telescope \index{GRAPES-3!muon telescope} and the data analysis used to identify the FD are explained in chapter \ref{GRP3}. We examined all the Forbush decrease events observed by the GRAPES-3 muon telescope during the years 2001 - 2004.  As an example, the percentage variation of the muon flux data for the 24 November 2001 FD event is shown in figure \ref{unf}. The dotted lines are the unfiltered data and the solid lines are the filtered data after using the low pass filter to remove the frequencies more than 1/day.

\begin{figure}[h]

\includegraphics[width = 0.95\columnwidth]{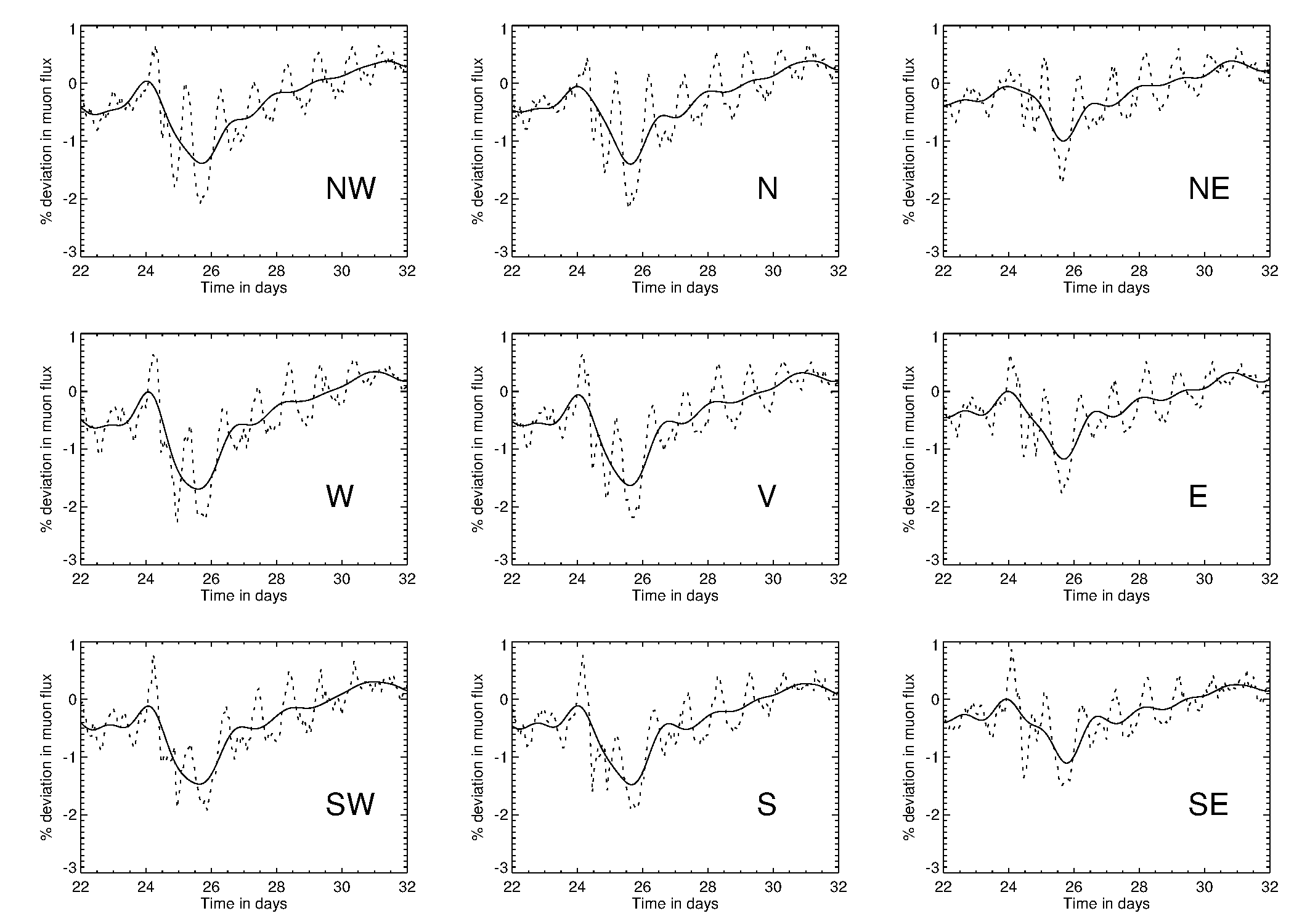}
\caption[Filtered and unfiltered data on 24 November 2001]{Forbush decrease \index{Forbush decreases} event on 24 November 2004. The figure shows the percentage deviation of the muon flux for different bins in different panels, the solid line shows the percentage deviation for the filtered data and the dotted line shows the same for the unfiltered data. }\label{unf}
\end{figure}

The FDs we study  are associated with near-Earth CME \index{CME} counterparts, which contribute to significant increases  in the interplanetary magnetic fields. We intend to investigate the relation between these IP magnetic field \index{Interplanetary magnetic field!enhancement} enhancements and  FDs. We used the IP magnetic field data observed by the ACE and WIND spacecraft available from the   \href{http://omniweb.gsfc.nasa.gov/}{ \textit{OMNI} \footnote{$http://omniweb.gsfc.nasa.gov/$}}   database. We used hourly resolution data of $B_{total}$, $B_x$, $B_y$, $B_z$ magnetic fields in the  geocentric solar ecliptic (GSE) coordinate system.  $B_{total}$ is the scalar magnetic field, $B_x$ is the magnetic field along the Sun-Earth line in the ecliptic plane pointing towards Sun, $B_z$ is the magnetic field parallel to the ecliptic north pole and $B_y$ is the magnetic field in the ecliptic plane pointing towards dusk. For consistency with  the muon flux data  we have applied the same low-pass filter to the magnetic field data as we did to the muon flux, which removes any oscillations having frequency $>$ 1/day. 
Since FD \index{Forbush decreases} events are associated with enhancements in the IP \index{Interplanetary magnetic field} magnetic field, we use the quantity $100-|B|$ and calculate the average value and percentage deviation of this quantity over the same data interval as the FD. This effectively ``flips'' the magnetic field increase and makes it look like a decrease, enabling easy comparison with the FD profile\index{Forbush decreases!profile}.  Figure \ref{FDi} shows the Forbush decrease event on 24 November 2004 together with the IP magnetic field \index{Interplanetary magnetic field} data processed in this manner. The black solid line is the percentage deviation of cosmic ray \index{cosmic rays} intensity in each of the bins. The red, blue, green and orange lines are the percentage deviation of interplanetary magnetic fields $B_{total}$, $B_z$,  $B_y $ and  $B_x$ respectively.

\begin{figure}[h]

\includegraphics[width = 0.95\columnwidth]{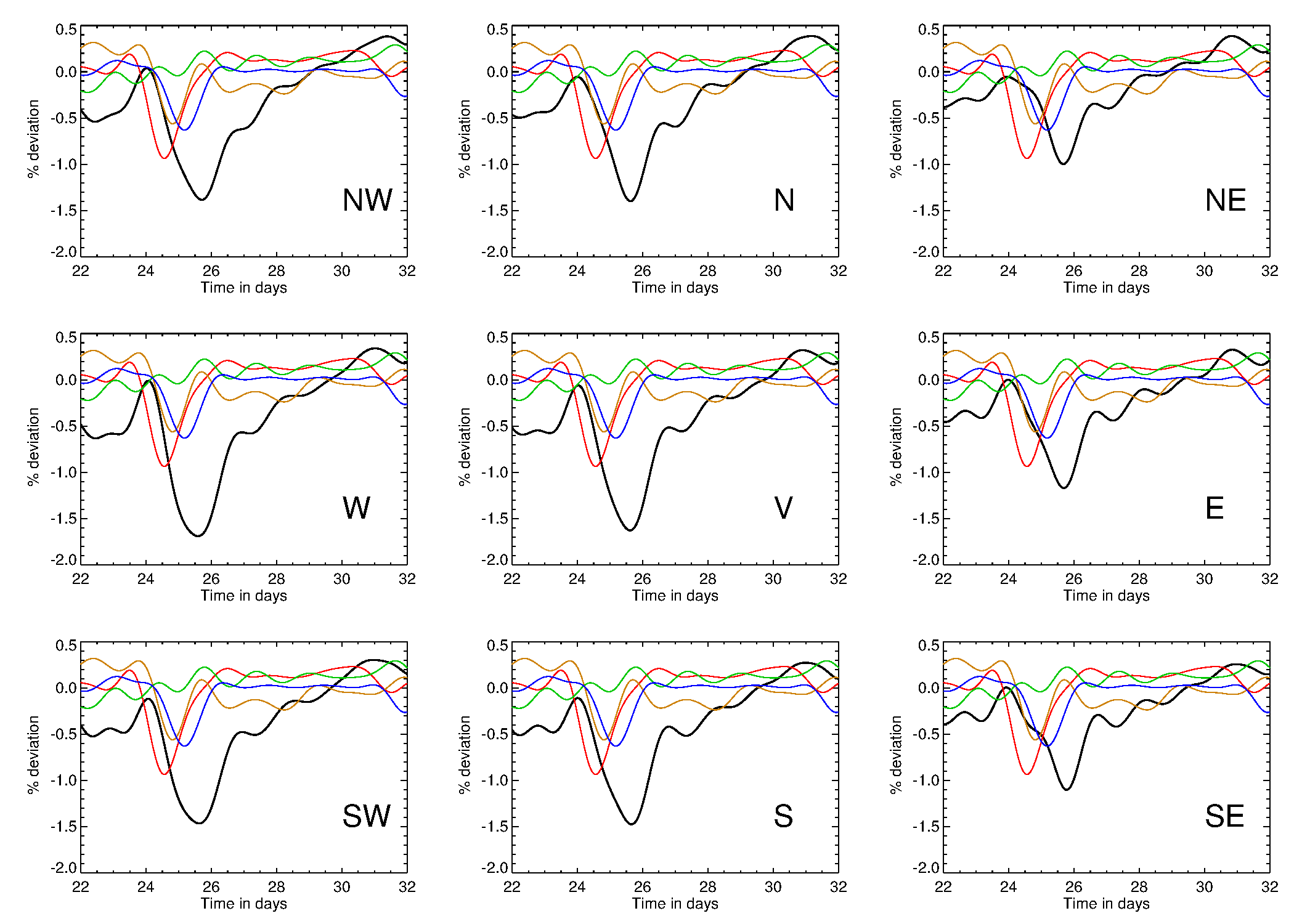}
\caption[FD on 24 November 2001]{Forbush decrease event \index{Forbush decreases} on 24 November 2004. Figure shows the Forbush decrease event and the magnetic fields for all the 9 bins in GRAPES-3 muon telescope \index{GRAPES-3!muon telescope}. The black solid line is the percentage deviation of cosmic ray \index{cosmic rays} intensity in each of the bins. The red, blue, green and orange lines are the percentage deviation of interplanetary magnetic fields $B_{total}$, $B_z$,  $B_y $ and  $B_x$ respectively, which are scaled down by a factor of 10 to fit in the frame. }\label{FDi}
\end{figure}

\section{Correlation of FD magnitude with peak IP magnetic field} \index{correlation}

Before studying the detailed relationship between the IP magnetic field \index{Interplanetary magnetic field} and FD profiles\index{Forbush decreases!profile}, we examine the relation of the peak IP  magnetic field to the FD magnitudes\index{Forbush decreases!magnitude}. We restrict our attention to FD events in short-list 1 (table \ref{SL1}) in chapter \ref{model}.
The FD magnitude for a given bin is calculated as the difference between the pre-event intensity of the cosmic rays and the intensity at the minimum \index{Forbush decreases!minimum} of the decrease. We examine the corresponding interplanetary magnetic field  during these events. We call the $B_y$ and $B_z$ ``perpendicular'' fields, because they are tangential to a flux rope \index{flux rope} CME approaching the Earth. They are perpendicular to a typical cosmic ray \index{cosmic rays} proton that seeks to enter the CME radially; it will therefore have to cross these perpendicular fields. We study  the relation between the FD magnitude and the peak of the total magnetic field $B_{total} = \,(B_x^2 + B_y^2 + B_z^2)^{1/2}$ and the peak of the net perpendicular magnetic field $B_p \, = \,(B_y^2 + B_z^2)^{1/2}$.   

The correlation coefficients of  the peak $B_{total}$  with FD magnitude for different bins are listed in table \ref{corrTt} and shown in  figure \ref{corBT}.  The correlation coefficients \index{correlation!coefficient} of the peak $B_p$  with FD magnitude are listed in table \ref{corrTp} and shown in  figure \ref{corBP}. We find that the correlation coefficient between peak $B_p$ and peak $B_{total}$ with the FD magnitude ranges from 62\% to 72\%. From the CME-only cumulative diffusion \index{CME-only model} model described in section \ref{cme-only} we know that the FD magnitude depends on various parameters associated with CME, like velocity of CME\index{CME!velocity}, turbulence level \index{turbulence!turbulence level $\sigma$} in the magnetic field and the size of the CME.  Its thus not surprising that the FD magnitude \index{Forbush decreases!magnitude} correlates only moderately with the peak value of the IP magnetic field.

\begin{figure}[h]
\centering

\includegraphics[width =0.95\columnwidth]{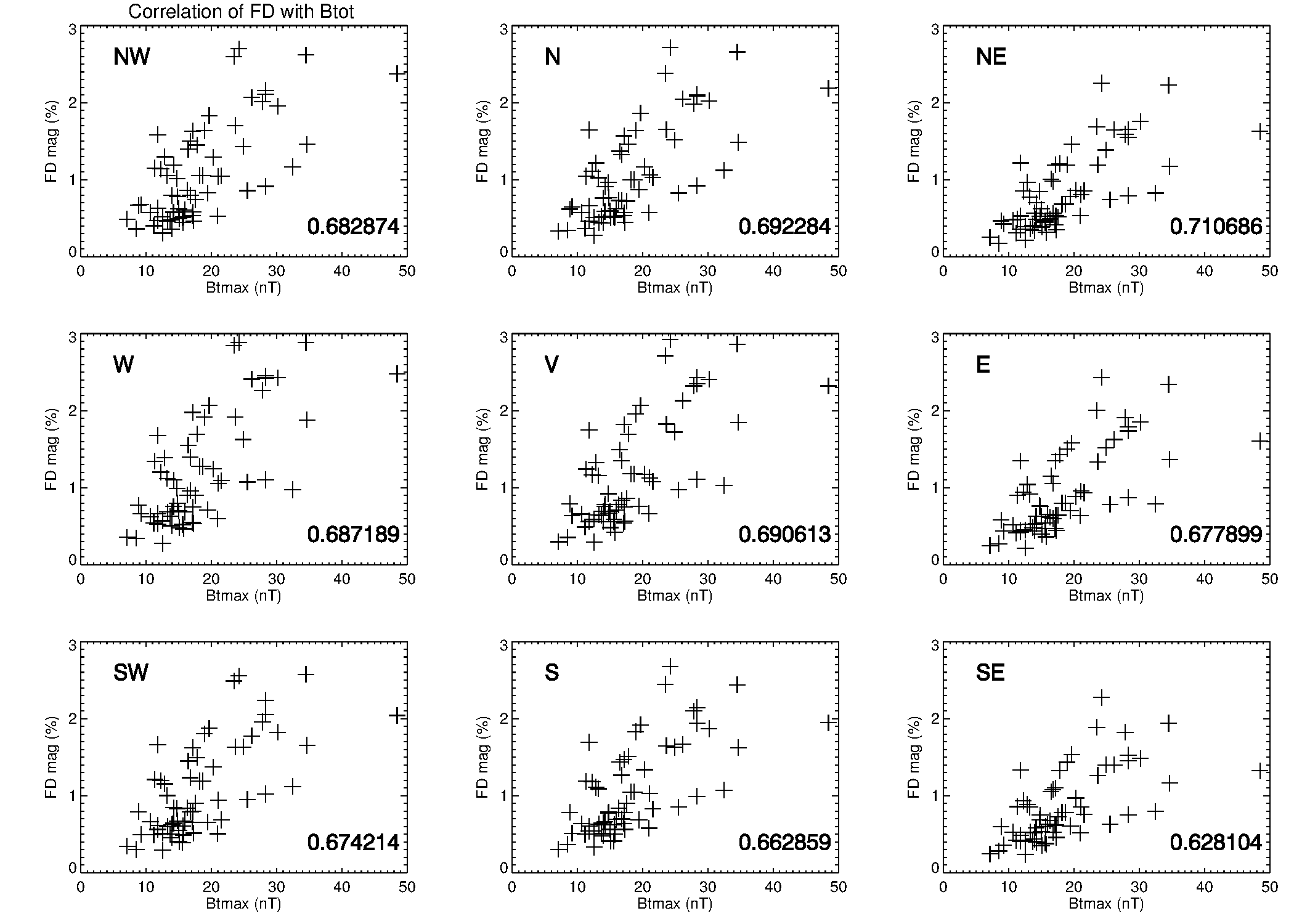}
\caption[Correlation of the Forbush decrease magnitude with the maximum total IP magnetic field]{Correlation \index{correlation} of maximum total magnetic field in the magnetic field enhancement to Forbush decrease magnitude\index{Forbush decreases!magnitude}.  }\label{corBT}
\end{figure}

\begin{table}[h]
\centering
\caption[Correlation of the Forbush decrease magnitude with the maximum total IP magnetic field]{Correlation \index{correlation} of the Forbush  decrease magnitude \index{Forbush decreases!magnitude} with the maximum total IP magnetic field\index{Interplanetary magnetic field}. For each bin of GRAPES 3 the correlation is calculated. }\label{corrTt}
\begin{tabular}{l c c} \hline \hline
Direction &Cut-off Rigidity (GV) &  Correlation \\ \hline 
NW & 15.5 & 0.682874 \\
N  & 18.7 & 0.692284 \\ 
NE & 24.0 & 0.710686 \\
W  & 14.3 & 0.687189 \\
V  & 17.2 & 0.690613 \\
E  & 22.4 & 0.677899 \\ 
SW & 14.4 & 0.674214 \\
S  & 17.6 & 0.662859 \\ 
SE & 22.4 & 0.628104 \\ \hline \hline
\end{tabular}
\end{table}

\begin{figure}[h]
\centering

\includegraphics[width =0.95\columnwidth]{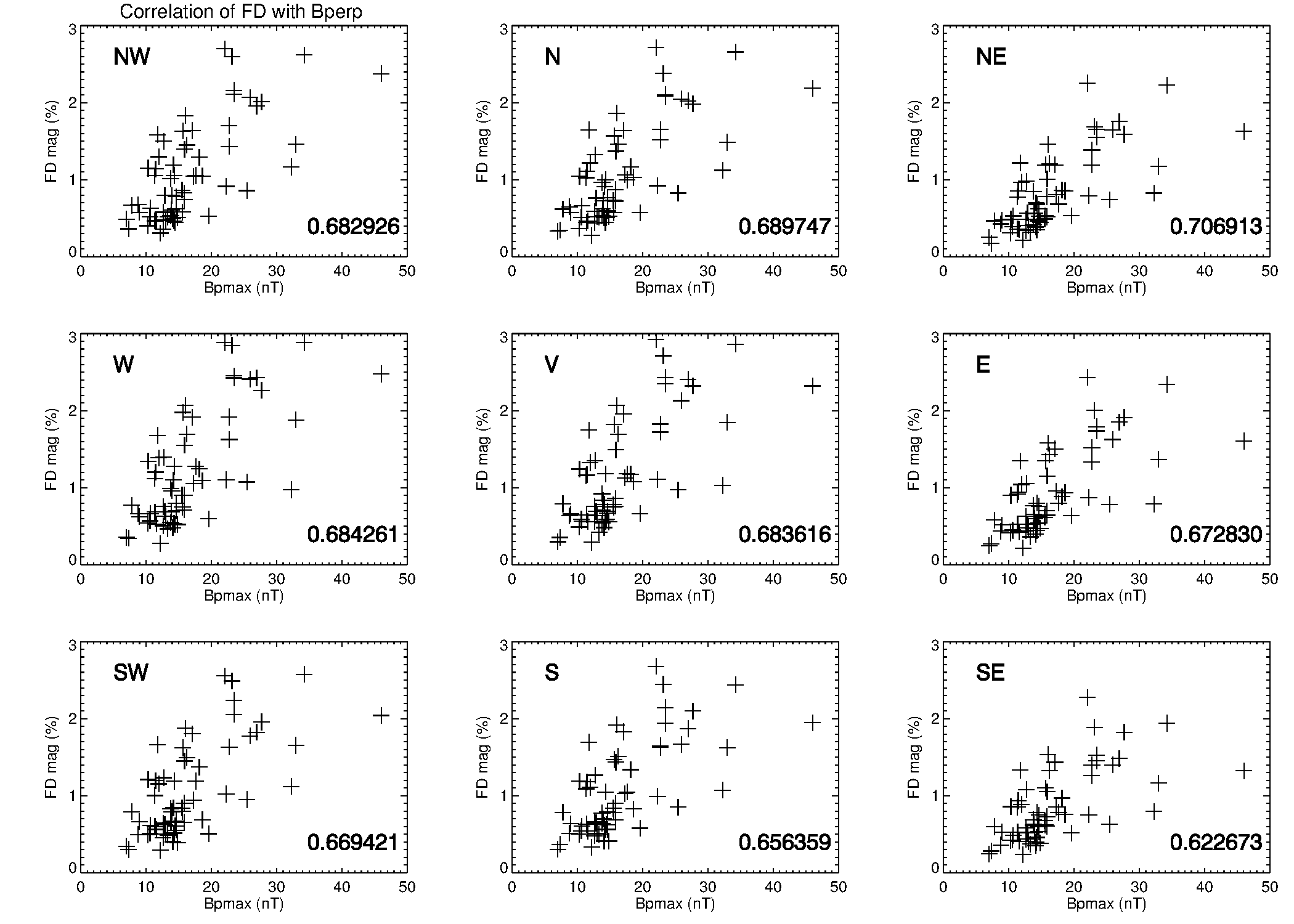}
\caption[Correlation of the Forbush decrease magnitude with the maximum perpendicular IP magnetic field]{Correlation \index{correlation}of maximum perpendicular magnetic field in the magnetic field enhancement to Forbush decrease  magnitude\index{Forbush decreases!magnitude}.  }\label{corBP}
\end{figure}

\begin{table}[h]
\centering
\caption[Correlation of the Forbush decrease magnitude with the maximum perpendicular IP magnetic field]{Correlation \index{correlation} of the Forbush decrease magnitude \index{Forbush decreases!magnitude} with the maximum perpendicular IP magnetic field\index{Interplanetary magnetic field}. For each bin of GRAPES 3 the correlation is calculated. }\label{corrTp}
\begin{tabular}{l c c} \hline \hline
Direction & Cut-off Rigidity (GV) & Correlation \\ \hline 
NW & 15.5 & 0.682926 \\
N  & 18.7 & 0.689747 \\
NE & 24.0 & 0.706913 \\
W  & 14.3 & 0.684261 \\
V  & 17.2 & 0.683616 \\
E  & 22.4 & 0.672830 \\ 
SW & 14.4 & 0.669421 \\
S  & 17.6 & 0.656359 \\
SE & 22.4 & 0.622673 \\ \hline \hline
\end{tabular}
\end{table}

\section{IP magnetic field compression} \label{ipmc}
\index{Interplanetary magnetic field!compression}
As mentioned earlier, we  consider FD events associated with the magnetic field enhancements that are due to the shock \index{shock} propagating ahead of the ICME as well as the ICME itself. ICMEs which posses certain well defined criteria such as reduction in plasma temperature and smooth rotation of magnetic field are called magnetic clouds \index{magnetic cloud} \citep{bur81, botsch98}. In this section we examine the magnetic field compression associated with Forbush decrease events. An example of this is shown in figure \ref{timi}, where the nine different panels shows the cosmic ray \index{cosmic rays} flux (FD profile) \index{Forbush decreases!profile} of nine different bins of GRAPES-3 muon telescope \index{GRAPES-3!muon telescope}  which shows the  FD observed on 24 November 2001. The black line denotes the percentage deviation of the cosmic ray intensity, while the red dotted line is the percentage deviation of the total magnetic field $|100 - B|$ as explained in section \ref{data}. The vertical brown, magenta, blue, green and black lines are the timings of the FD onset\index{Forbush decreases!onset}, shock arrival, magnetic cloud start, magnetic cloud end and FD minimum \index{Forbush decreases!minimum} respectively. The red vertical line corresponds to the maximum of the magnetic field compression. It is clear that the magnetic field compression responsible for the FD is in the sheath \index{sheath} region; the region between the shock \index{shock} and magnetic cloud. This is in agreement with \citet{rich11}. 

\begin{figure}[h]
\centering
\includegraphics[width =1.0\columnwidth]{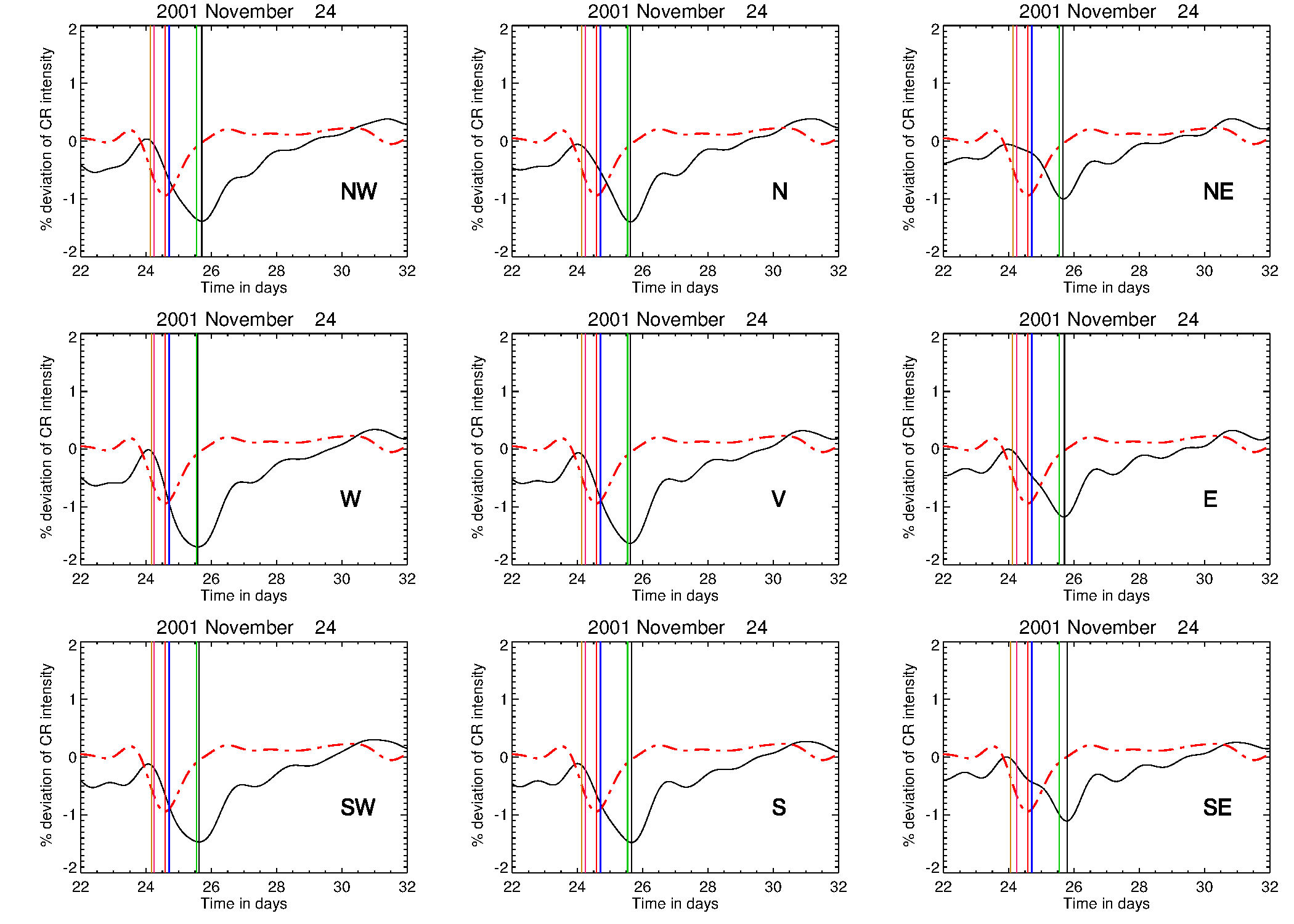}
\caption[Forbush decrease on 24 November 2001 with timings of shock and magnetic cloud.]{Forbush decrease \index{Forbush decreases} event on 24 November 2001. The black line denotes the percentage deviation of the cosmic ray \index{cosmic rays} intensity, the red dotted line the percentage deviation of the total magnetic field $|100 - B|$ as explained in section \ref{data} which is scaled down to fit in the frame. The vertical brown, magenta, blue, green and black lines denote the timings corresponding to the FD onset\index{Forbush decreases!onset}, shock \index{shock} arrival, magnetic cloud \index{magnetic cloud} start, magnetic cloud end and FD minimum \index{Forbush decreases!minimum} respectively. The red vertical line corresponds to the maximum of the magnetic field compression. }\label{timi}
\end{figure}

The CME-only model described in the section \ref{cme-only}   deals with the diffusion \index{diffusion} of cosmic rays \index{cosmic rays} though the turbulent magnetic field \index{turbulence!turbulent magnetic field} in the sheath \index{sheath} region. The cross-field diffusion coefficient \index{diffusion!diffusion coefficient}   depends on the rigidity \index{rigidity} of the proton and the turbulence level \index{turbulence!turbulence level $\sigma$} in the magnetic field; e.g., \citep{cand04}. The turbulence level in the magnetic field is an important parameter in this context. We have calculated the turbulence level using one minute averaged data from the ACE/WIND spacecraft available from the \href{http://omniweb.gsfc.nasa.gov/}{ \textit{OMNI} \footnote{$http://omniweb.gsfc.nasa.gov/$}} data base. In order to calculate the turbulence level $\sigma$ we use a 1 hour running average of the magnetic field ($B_0$) and the fluctuation of the IP magnetic field \index{Interplanetary magnetic field} around this average ($B_{tur} = B -B_0$). We define the quantity $\sigma$ as
\begin{equation}
\sigma \, = \,\left( \frac {\langle B_{tur}^2 \rangle}{B_0^2}\right) ^{0.5} \label{dysig}
\end{equation}

where $\langle B_{tur}^2 \rangle$ denotes the average of $B_{tur}^2$ over the 1 hour window.
Figure \ref{tur} shows a representative event. The first panel shows the one minute average magnetic field for 21-30 November 2001. The second panel in this figure shows the turbulence level $\sigma$ calculated for this event. The vertical red, blue and green lines correspond to the timings of shock arrival, start and end of the magnetic cloud\index{magnetic cloud}. We observed that the magnetic field compression responsible for the FD \index{Forbush decreases} occurs in the shock sheath \index{sheath} region, region between the shock and the magnetic cloud. The turbulence level \index{turbulence!turbulence level $\sigma$} enhancement also occurs in this region.

We studied  FD events listed in short-list 2 ({table \ref{SL2}}) in chapter \ref{model}, which have a good FD profile \index{Forbush decreases!profile} and FD magnitude \index{Forbush decreases!magnitude} $> 0.25$~\%, and are also associated with a near-Earth shock \index{shock} and magnetic cloud. We eliminated the event on 29 September 2001 from this list, since it was associated with many IP magnetic field enhancements\index{Interplanetary magnetic field!enhancement}, which could be due to multiple Halo and partial halo CMEs\index{CME!Halo}.  
We examine the the magnetic field enhancement and the turbulence level \index{turbulence!turbulence level $\sigma$} ($\sigma$) for these selected events. The timings of the shock,  maximum of the magnetic field compression, magnetic cloud \index{magnetic cloud} start and end timings along with the FD onset \index{Forbush decreases!onset}times for different bins are given in  table \ref{tabtim}. The peak of the magnetic field enhancement in the smoothed data generally occurs before the start of the magnetic cloud or at the start of the magnetic cloud, whereas in the unsmoothed data the enhancement lies in the sheath \index{sheath} region. We note that the smoothing procedure using the low pass filter shifts the maximum by a small amount (-5 to 10 hours).   

\afterpage{
\begin{landscape}
\begin{table*}{ \small
\caption[Timings of FD onset, shock and magnetic cloud]{FD onset \index{Forbush decreases!onset} timings for different bins together with the shock \index{shock} arrival time, time of maximum magnetic field enhancement,  magnetic cloud \index{magnetic cloud} start and end timings}\label{tabtim}
\begin{tabular}{l|ccccccccc|cccc} \hline \hline
Event & \multicolumn{9}{|c|} {FD onset} & Shock &Maximum of  & MC & MC \\   \cline{2-10}
      & NW & N & NE & W & V & E & SW & S & SE & arrival &Mag. compre. & Start & end \\ \hline 
2001 Apr 04  & 04.30  & 04.33 & 04.29 & 04.3 &04.34 &04.32 &04.37 & 04.34 &04.25 & 04.61 & 04.79 & 04.87 & 05.35 \\
2001 Apr 11  & 11.54 & 11.58 & 11.67 & 11.47 &11.50 &11.54 &11.35 & 11.43 &11.52 & 11.58 & 12.00 &  11.958 & 12.75 \\
2001 Aug 17  & 17.18 & 17.08 & 17.05 & 16.97 &16.94 &16.92 &17.00 & 16.98 &16.97 & 17.45 & 17.87 & 18.00 & 18.896 \\
2001 Nov 24  & 24.13 & 24.14 & 24.13 & 24.17 &24.14 &24.12 &24.17 & 24.13 &24.06 & 24.25 & 24.58 &  24.708 & 25.541 \\
2002 May 23  & 23.13 & 23.08 & 23.00 & 23.17 &23.09 &23.04 &23.21 & 23.13 &23.09 & 23.44 & 23.58 &  23.896 & 25.75 \\
2002 Sep 07  & 07.71 & 07.72 & 07.67 & 07.62 &07.62 &07.63 &07.66 & 07.68 &07.71 & 07.6  & 08.00 & 07.708 & 08.6875 \\
2002 Sep 30  & 30.56 & 30.45 & 30.43 & 30.56 &30.48 &30.42 &30.52 & 30.45 &30.36 & 30.31 & 31.16 & 30.917 & 31.6875 \\
2003 Nov 20  & 19.89 & 21.34 & 21.12 & 20.08 &20.45 &20.86 &20.29 & 20.43 &20.63 & 20.31 & 20.67 & 21.26 & 22.29 \\
2004 Jan 21  & 22.09 & 22.03 & 21.96 & 21.98 &22.00 &21.97 &21.83 & 21.91 &21.90 & 22.09 & 22.50 &  22.58 & 23.58 \\
2004 Jul 26  & 26.60 & 26.64 & 26.75 & 26.59 &26.65 &26.75 &26.67 & 26.77 &26.86 & 26.93 & 27.29 & 27.08 & 28.00 \\  \hline \hline
\end{tabular} }
\end{table*}
\end{landscape}
}

\begin{figure*}[h]
\includegraphics[width =0.95\columnwidth]{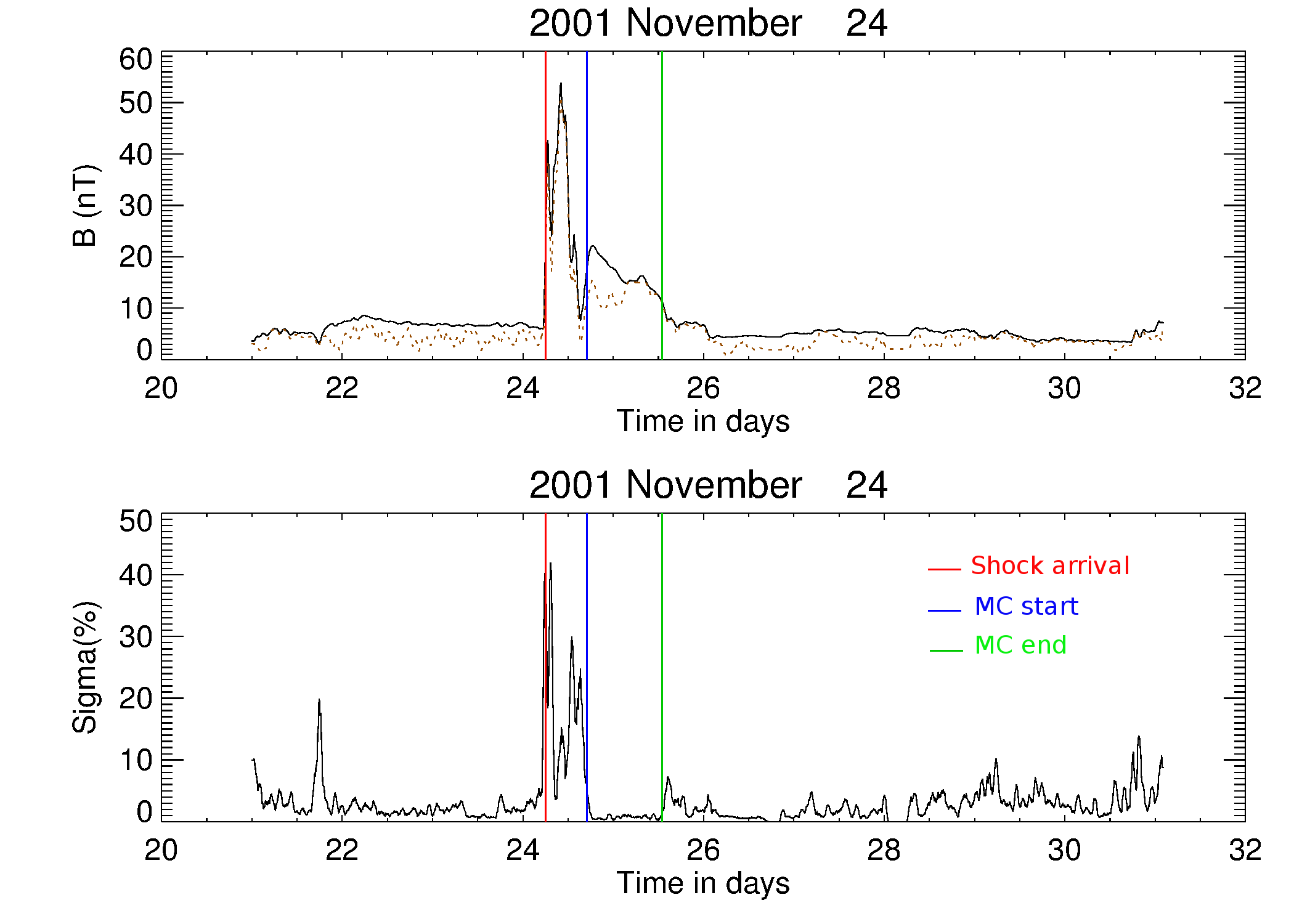}
\caption[Magnetic field compressions on 24 November 2001]{ Magnetic field compression associated with the Forbush decrease \index{Forbush decreases} event on 24 November 2001. In the first panel the black line denotes $B_{total}$ and the brown dotted line denotes $B_p$. The black line in the second panel shows the turbulence level  for $B_{total}$. In both panels the vertical lines red corresponds to the shock \index{shock} arrival time, blue correspond to magnetic cloud start time and green correspond to magnetic cloud \index{magnetic cloud} end time. \label{tur} }
\end{figure*}

It is clear from the figures \ref{timi} , \ref {tur} and table \ref{tabtim} that the peak of the magnetic field compression responsible for the FD lies in the sheath \index{sheath} region, and the turbulence level \index{turbulence!turbulence level $\sigma$} is also  enhanced in this region. This is in broad agreement with \citet{rich11}


\section{How similar are the FD and the IP magnetic field profiles?}

One of the near-Earth effects of a CME \index{CME} is the compression of (and consequent increase in) the interplanetary magnetic field. IP magnetic fields \index{Interplanetary magnetic field} measured by spacecraft such as  WIND and ACE can detect these magnetic field compressions. 
We investigate the relation of these magnetic field compressions to the FD  \index{Forbush decreases} profile. We work with the hourly resolution interplanetary magnetic field data from the ACE and WIND spacecraft obtained from the  \href{http://omniweb.gsfc.nasa.gov/}{ \textit{OMNI} \footnote{$http://omniweb.gsfc.nasa.gov/$}} database. Applying the low-pass filter, described in section \S~\ref{DA}, to this data yields a combined magnetic field compression comprising the  shock \index{shock} and ICME/magnetic cloud\index{magnetic cloud}. A visual comparison of the FD profile \index{Forbush decreases!profile} with the magnetic field compression often reveals remarkable similarities.
In order to quantify the similarity between the profiles, we studied the cross correlation \index{correlation} of the cosmic ray intensity profile with the IP magnetic field profile. In order to do so, we shift the magnetic field profile (with respect to the FD profile) by amounts ranging from -36 hours to 12 hours. We identify the peak correlation value and the shift corresponding to this value is considered to be the time lag between the IP magnetic field \index{Interplanetary magnetic field} and the cosmic ray FD profile. Most of the FD events  exhibit correlations $\geq 60 \%$ with at least one of the four IP magnetic field components ($B_{total}$, $B_x$, $B_y$, $B_z$). An example of the crosscorrelation of FD profile (cosmic ray flux) with $B_{total}$, for 23 May 2002 is shown in the figure \ref{23may}, where the correlation lag is -13 hours. In other words, the IP magnetic field profile precedes the FD profile by 13 hours. The first panel in this figure is the percentage deviation of the cosmic ray flux and the  $B_{total}$. The percentage deviation of $B_{total}$ is scaled to fit in the frame. The second frame shows the same percentage deviations but the magnetic field is shifted by the peak correlation lag and the last panel shows the correlation coefficients \index{correlation!coefficient} corresponding to different lags. In  further discussion we consider only those events showing high crosscorrelation ( $\geq 70 \%$) for a  lag between  -36 to 12 hours.  The short-listed events are listed in table \ref{T1}.

\begin{figure}[h]
\centering
  
\includegraphics[width =0.95\columnwidth]{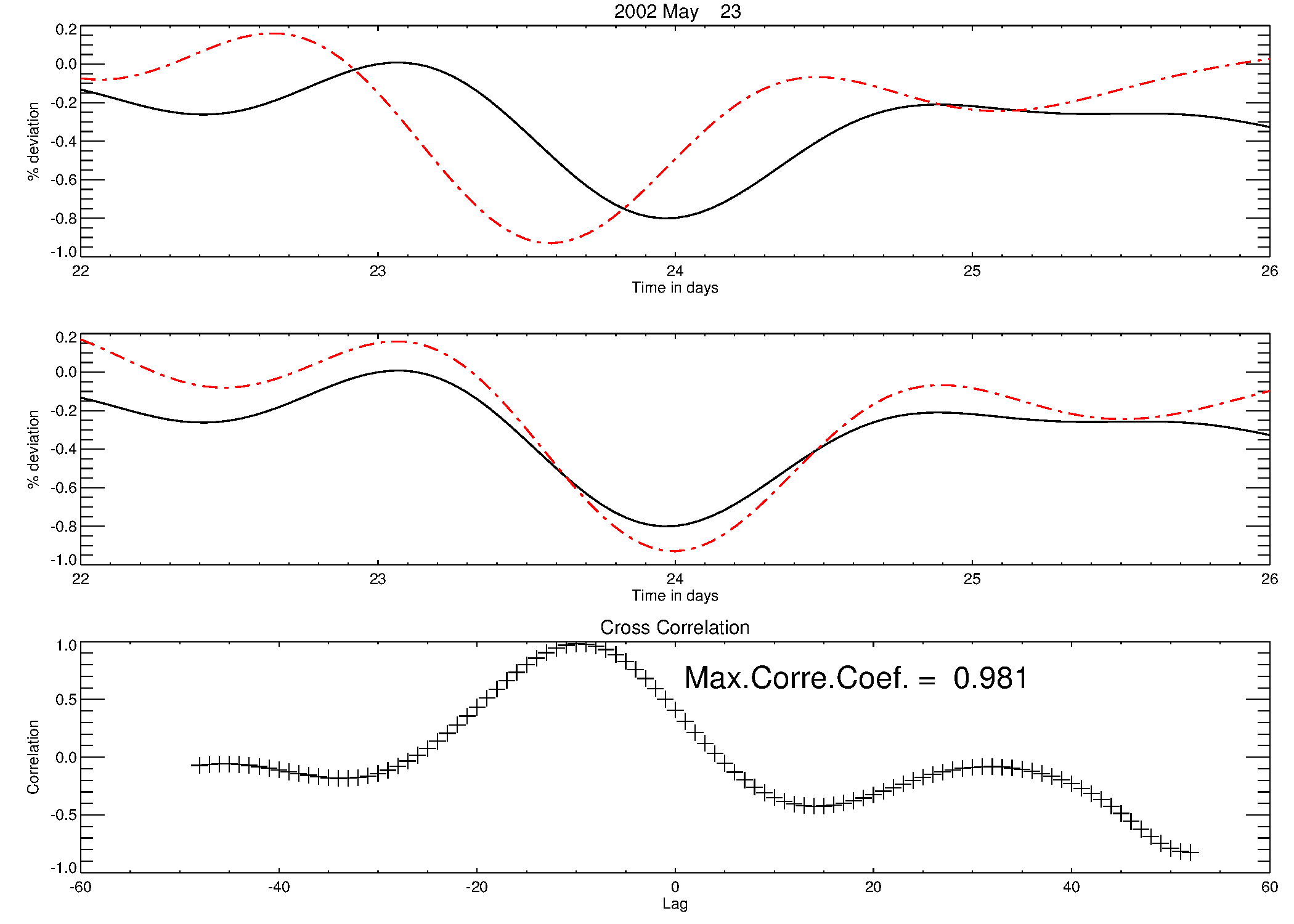}
\caption[Crosscorrelation of cosmic ray flux with magnetic field.]{Crosscorrelation \index{correlation} of the cosmic ray flux with the $B_{total}$. First panel showing the percentage deviation of cosmic ray \index{cosmic rays} flux and the magnetic field (scaled to fit in the frame). The second panel shows the same with the magnetic field shifted to right correspond to the lag and the third panel shows the correlation coefficient \index{correlation!coefficient} for different lags. }\label{23may} 
\end{figure}

\begin{table}
\centering
\caption[List of Forbush decrease \index{Forbush decreases} events with correlation \index{correlation} $\geq 70 \%$]{List of Forbush decrease \index{Forbush decreases} events with correlation $\geq 70 \%$, The `-' entries denoted events that have low correlation values for lags between -36 and 12 hours} \label{T1}
\begin{tabular}{|l|c|c|c|c|} \hline \hline
Event & \multicolumn{4}{|c|} {Correlation (\%)}\\ \cline{2-5}
      & $ B_{total} $ & $ B_{x} $ & $ B_{y} $ & $ B_{z} $ \\ \hline \hline
13 Jan 2001 & 97.2 & -  & 95.8 & 96.6 \\
26 Mar 2001 & 70.3 & - & - & 39.1 \\
4 Apr 2001 & 92.8 & 97.3 & 77.2 & 63.4 \\
7 Apr 2001 & 94.3 & 92.7 & 71.9 & 54.2 \\
 11 Apr 2001 & 77.7 &  - & -  & 79.0 \\
27 May 2001 & 66.5 & 27.8 & 65.0 & 75.7 \\
 1 Jun 2001 & 77.1 & 70.5 & -  & 54.5 \\
13 Aug 2001 & 51.8 & - & 97.5 & 70.0 \\
17 Aug 2001 & 84.6 &  - & 31.2 & 58.0 \\
6 Sep 2001 & 68.8 & 87.0 & 64.7 & 45.1\\
 12 Sep 2001  & 79.2 & - & - & 86.2 \\
29 Sep 2001 & 70.3 & - & 58.1 & - \\
5 Nov 2001 & 88.3 & 64.6 & 34.8 & -  \\
24 Nov 2001 & 85.3 & 32.4 & 41.0  & 77.1 \\
14 Dec 2001 & 74.7 & 42.7 & 69.7 & 72.8 \\
23 May 2002 & 98.1 & 79.4 & 75.9 & 60.0   \\
 7 Sep 2002 & 77.1 & - & 49.6 & 87.4 \\
23 Sep 2002 & 60.4 & 87.9 &  41.4 & 93.1 \\
30 Sep 2002 & 81.1 & 58.7 & 72.1 &  75.7 \\
22 Dec 2002 & 73.4 & -  & 43.4 & 84.7 \\
9 Jan 2003 & 90.1 & 68.1 & - & 56.2 \\
 23 Jan 2003 & 70.9 & - & - & 75.4 \\
30 Jan 2003 & 94.8 & 84.4 & 42.7 & 95.7 \\
16 Feb 2003 & - & 31.3 & - & 74.3\\
26 Mar 2003 & 77.1 & 64.8 & - & - \\
4 May 2003 & 83.4 & -  & 84.7 & 80.7   \\
18 May 2003 &  86.5 & -  & - & -  \\
25 Jul 2003 & 95.3 & 53.3 & 73.3 &  41.6  \\
16 Aug 2003 & 71.6& 49.6 & 45.5 & 57.7 \\
21 Oct 2003 & 83.8 & 92.0 & 70.1 & 93.5    \\
 29 Oct 2003  & 78.8 & - & 43.1 & -   \\
27 Dec 2003 & 86.1 & - & 21.7 & 87.2 \\
 21 Jan 2004 & 77.9 & 78.2 & - & -  \\
29 May 2004 & 53.1  & 90.2 & - & 86.9 \\
26 Jul 2004 & 86.5 & 73.3 & 85.5 & 94.8 \\
30 Aug 2004 & -  & - & -  & 92.4 \\
5 Dec 2004 & 85.3 & - & 89.4 & 58.4 \\
12 Dec 2004 & 81.1 & 61.6  & 73.3 & 78.9 \\ \hline \hline
\end{tabular}
\end{table}

\subsection{$\perp$ diffusion of cosmic rays through the sheath magnetic field into the ICME}  

The lag we observed in the correlation \index{correlation} of the cosmic ray \index{cosmic rays} flux and the IP magnetic field \index{Interplanetary magnetic field} is due to the fact that the high energy protons do not respond to the magnetic field compressions immediately; they are subjected to the classical magnetic mirror effect arising from the gradient in the longitudinal magnetic field and to turbulent cross-field (also referred to as $\perp$) diffusion \index{diffusion}  (e.g; \citealp{kub10}). We concentrate here only on the cross-field diffusion of the high energy protons through the turbulent sheath \index{sheath} region between the shock \index{shock} and the CME\index{CME}. As discussed earlier, we have identified the IP magnetic field compression \index{Interplanetary magnetic field!compression} to comprise mainly of this sheath region; we therefore use the observed values of the mean field and turbulent fluctuations in the sheath region to calculate representative diffusion timescale for cosmic rays. The time delay between the IP magnetic field compression and the FD profile \index{Forbush decreases!profile} (i.e., the correlation lag) can be interpreted as the time taken by the particles to diffuse into the magnetic compression. Our approach may be contrasted with that of \citealp{kub10}, who use a computational approach to investigate cosmic ray dynamics (thus incorporating both the mirror effect and cross field diffusion) in a magnetic field configuration that comprises an idealized flux rope \index{flux rope} CME. They do not consider the sheath region, and neither do they use observations to guide their choice of magnetic field turbulence levels\index{turbulence!turbulence level $\sigma$}.

In order to calculate the cross-field diffusion timescale, we proceed as follows: considering the flux rope \index{flux rope} geometry of a near-Earth CME, the magnetic field along the Sun-Earth ($B_X$) represents the longitudinal magnetic field. The magnetic field $B_y$ and $B_z$ represent the perpendicular magnetic fields encountered by the diffusing protons. In our discussion we consider only perpendicular diffusion; we therefore choose  events which exhibit good correlation \index{correlation} with the $B_y$ and $B_z$ magnetic field compressions and poor correlation with compressions in $B_x$. The events short listed using these criteria are listed in table \ref{T2}.

\afterpage{
\begin{landscape}
\begin{table*}
\centering
\caption[Forbush decrease \index{Forbush decreases} events having good correlation \index{correlation} with the perpendicular IP magnetic fields.]{Forbush decrease \index{Forbush decreases} events having good correlation with the perpendicular IP magnetic fields. The CMEs associated with these events are also listed in this table. The correlation lags are given in hours.}\label{T2}
\begin{tabular}{|l|l|c|c|c|c|c|c|c|c|c|}\hline \hline
 Event  & CME & Time  & type & $V_m $&\multicolumn{2}{|c|} {$B_{total}$} & \multicolumn{2}{|c|} {$B_{y}$} & \multicolumn{2}{|c|} {$B_{z}$} \\ \cline{6-11}
& (near Sun) & (UT) & & $km \, s^{-1}$& Corr.(\%) & Lag (hrs) &  Corr.(\%) & Lag (hrs) &  Corr.(\%) & Lag (hrs)\\ \hline 
2001 Jan 13  & Jan 10 & 00:54 & Halo & 832 & 97.2 & -13 &  95.8 & -14 & 96.6 & -23 \\
2001 Apr 11  & Apr 10 & 05:30 & Halo & 2411 & 77.7 & -18  & - & -  & 79.0 & -5 \\   
2001 May 27  & May 25 & 04:06 & 354 & 569 & 66.5 & 0 &  65.0 & -3 & 75.7 & -21 \\
2001 Aug 13  & Aug 11 & 04:30 & 313 & 548 & 51.8 & -7 & 97.5 & -7 & 70.0 & -5\\
2001 Sep 12  & Sep 11 & 14:54 & Halo & 791 & 79.2 & -25 & - & -  & 86.2 & -1  \\
2001 Nov 24  & Nov 22 & 23:30 & Halo & 1437 & 85.3 & -21 & 41.0  & -31  & 77.1 & -14  \\ 
2001 Dec 14  & Dec 13 & 14:54 & Halo & 864 & 74.7 & -35 & 69.7 & -2 & 72.8 & -17\\ 
2002 Sep 07  & Sep 05 & 16:54 & Halo & 1748  & 77.1 & -18 & 49.6 & -24 & 87.4 & 3  \\
2002 Sep 30  & Sep 29 & 15:08 & 261 & 958 & 81.1 & -5 & 72.1 & 8 & 75.7 & -12\\
2002 Dec 22  & Dec 19 & 22:06 & Halo & 1092 &  73.4 & -15 & 43.4 & 0 & 84.7  & -12\\ 
2003 Jan 23  & Jan 22 & 05:06 & 338 & 875 & 70.9 & -21 & - & - & 75.4 & -28  \\
2003 Feb 16  & Feb 14 & 20:06 & 256 & 796 & - & - & - & - & 74.3 & -11  \\
2003 May 04  & May 02 & 12:26 & 222 & 595 & 83.4 & -8 & 84.7 & -10 & 80.7 & 0 \\
2003 Jul 25  & Jul 23 & 05:30 & 302 & 543  & 95.3 & -19 & 73.3 & -2 & 41.6  & 2  \\
2003 Dec 27  & Dec 25 & 09:06 & 257 & 178 & 86.1 & -35 & 21.7 & 5 & 87.2 & -3\\
2004 Aug 30  & Aug 29 & 02:30 & 274 & 1195 & - & - & -  & - & 92.4& 1\\
2004 Dec 05  & Dec 03 & 00:26 & Halo & 1216 & 85.3 & -12 & 89.4 & 8 & 58.4 & -13 \\
2004 Dec 12  & Dec 08 & 20:26 & Halo & 611  & 81.1 & -17 & 73.3 & -25& 78.9 & -13\\ \hline
\end{tabular} 
\end{table*}
\end{landscape}}

\section{Perpendicular diffusion coefficient ($D_{\perp}$)} \label{dperp} \index{diffusion!diffusion coefficient} 

The cross-field diffusion coefficient $D_{\perp}$  governs the diffusion of the ambient high-energy protons into the CME \index{CME} across the magnetic fields that enclose it.  The topic of cross-field diffusion of charged particles across magnetic field lines in presence of turbulence is subject of considerable research. Analytical treatments include classical scattering theory (e.g. \citealp{gj99}, and references therein) and non-linear guiding center theory \citep{matt03, shal10} for perpendicular diffusion. Numerical treatments of perpendicular diffusion of charged particle in turbulent magnetic field include \citet{gj99}, \citet{Cass02}, \citet{cand04}, \citet{tau11} and \citet{pot14}. We seek a concrete  prescription for $D_{\perp}$ that can incorporate observationally determined quantities. 
Accordingly, we consider two different models of $D_{\perp}$ in our work.     

\subsection{$D_{\perp}$, from Candia \& Roulet (2004)} \index{diffusion!diffusion coefficient} 

One $D_{\perp}$ prescription we use is given by \citet{cand04} and obtained from extensive Monte Carlo simulations of cosmic rays \index{cosmic rays} propagating through tangled magnetic fields. Their results reproduce the standard results of\citet{gj99} and \citet{Cass02}, and it also extends the regime of validity to include strong turbulence \index{turbulence} and high rigidities. The extent of cross-field diffusion of protons depends on several parameters, 1)  the proton rigidity\index{rigidity}, which indicates how tightly the proton is bound to the magnetic field.  2) the level \index{turbulence!turbulence level $\sigma$} of magnetic field turbulence, which can contribute to field line transport. The detailed description of this cross-field diffusion coefficient is given in \S~\ref{crossDiff} of chapter \ref{model}

\subsection{$D_{\perp}$, from  Potgieter et al. (2014)}

Another cross-field diffusion coefficient \index{diffusion!diffusion coefficient}  prescription we use is due to \citet{pot14}. According to this prescription, the diffusion coefficient parallel to the averaged background heliospheric magnetic field is given by
\begin{equation}
D_{\parallel} \, = \, {\left( D_{\parallel} \right)}_0 \beta \left( \frac{B_0}{B} \right) \left(\frac{P}{P_0} \right)^a \left( \frac{\left(\frac{P}{P_0} \right)^c + \left(\frac{P_k}{P_0} \right)^c}{1+\left(\frac{P_k}{P_0} \right)^c} \right)^{\frac{(b-a)}{c}}  
\end{equation}

where $ \beta = \frac{v}{c} $ , the ratio of the particle's speed to the speed of light. Here, ${\left( D_{\parallel} \right)}_0 = 10^{22}$ $ cm^2\, s^{-1}$. $ P$ and $B$ are the rigidity \index{rigidity} of the particle and magnetic field respectively, the constant $P_0 \, = \, 1GV $ and $B_0 \, = \, 1nT $ makes the corresponding quantities in the parentheses dimensionless. The parameter `a' is a power law index which changes with time as described in table 1 of \citet{pot14}, which together with `b = 1.95' determines the slope of the rigidity dependence respectively above and below a rigidity with the value $P_k$, whereas `c = 3.0' determines the smoothness of the transition. Thus the rigidity dependence of the diffusion coefficient is expressed as a combination of power laws. The value of $P_k$ determines where the break in power law occurs and the value of `a' determines the slope after $P_k$. the value of $P_k$ is also given in table 1 of \citet{pot14}.

The perpendicular diffusion coefficient is then given by

\begin{equation}
D_{\perp} \, = \,  0.02 \times D_{\parallel}  
\end{equation}

\section{The B field-FD lag: how many diffusion lengths?} \label{NdL} \index{diffusion!diffusion length} 
We have established that the FD profile \index{Forbush decreases!profile} is often very similar to that of the IP magnetic field compression\index{Interplanetary magnetic field!compression}, and lags it by a few hours.
We interpret the observed time lag between the IP magnetic field \index{Interplanetary magnetic field} and the FD profiles as the time taken by the protons to diffuse through the magnetic filed compression via cross-field diffusion.

The diffusion \index{diffusion}  time  for a single diffusion of high energy proton into the magnetic structure of CME is given by 

\begin{equation}
t_{diff} = \frac{D_{\perp}}{c V_{sw}} 
\end{equation}

where c is speed of light (which is the typical propagation speed for the highly relativistic galactic cosmic rays we are concerned with) and $V_{sw}$ is the solar wind speed \index{solar wind} ahead of the CME\index{CME}. 

We calculate $t_{diff}$ using the two different $D_{\perp}$ prescriptions described in the section \ref{dperp}. Furthermore, when using the $D_{\perp}$ from Candia \& Roulet (2004), we use two different ways for computing the turbulence level \index{turbulence!turbulence level $\sigma$} $\sigma$: in the first one, we assume a constant value of 15 \% for $\sigma$. In the second, we calculate $\sigma$ as a function of time using the one-minute averaged IP magnetic field data, as described in Eq~(\ref{dysig}). When using the Potegieter et al prescription for $D_{\perp}$, we use value of $\beta \, = \, v/c \, = \, 0.99$ (since we are considering high energy cosmic rays), $P_k \, = \, 4.0 \, GV $ and `a' = 0.56.

We used all the three methods described above to calculate $t_{diff}$. Using these values of $t_{diff}$ we calculated the number of diffusion \index{diffusion!diffusion length} lengths required to account for the observed time lag between the FD profile  and the IP magnetic field profile as,

\begin{equation}
{\rm No. \,  of \, Diffusions \, = \frac{Lag}{t_{diff}} }\label{noDs}
\end{equation}

The results for the number of diffusion \index{diffusion}  times needed to account for the observed lag between the IP magnetic field enhancement \index{Interplanetary magnetic field!enhancement} and the FD profile \index{Forbush decreases!profile} are shown in table \ref{T3}. These numbers are calculated using the peak value of the IP magnetic field profile. It is evident that the observed lags can be accounted for by a few tens to a few hundred diffusion times.

 
\section{Summary}
We studied all the Forbush decrease events observed by GRAPES-3 during the years 2001-2004. The magnetic field compression responsible for the FD as well as the turbulence level gets enhanced in the shock sheath \index{sheath} region.  The details regarding shock \index{shock} timing,  magnetic cloud start and end timings along with the FD onset time for different bins are given in the table \ref{tabtim}. 


\afterpage{
\begin{landscape}
\centering
\begin{longtable}{|l|c|c|c|c|c|c|c|c|c|c|c|c|c|}
\caption[Number of diffusions required for the observed lag]{ \label{T3} Numb er of diffusions required for the observed lag in Forbush decrease \index{Forbush decreases} events using the three different methods described in the section \ref{NdL}. $1^{st}$  stands for the method using the dynamic $\sigma$, $2^{nd}$  stands for the method using the constant $\sigma$ and $3^{rd}$  using the Potgeiter 2014,  } \\
\hline \hline
       & Rigidity &  \multicolumn{4}{|c|} {$B_{total}$}    &  \multicolumn{4}{|c|} {$B_{y}$} &  \multicolumn{4}{|c|} {$B_{z}$} \\ \cline{3-14}
 Event & of Proton& Lag & \multicolumn{3}{|c|} {No: of diffusions} & Lag & \multicolumn{3}{|c|} {No: of diffusions} & Lag & \multicolumn{3}{|c|} {No: of diffusions}\\ \cline{4-6} \cline{8-10} \cline{12-14}  
      & (GV) & & $1^{st}$ &$2^{nd}$ &$3^{rd}$ & & $1^{st}$ &$2^{nd}$ &$3^{rd}$ & & $1^{st}$ &$2^{nd}$ &$3^{rd}$   \\ \hline  \hline
\endfirsthead
\caption{Continuation of number of diffusions }\\
\hline  \hline
 & Rigidity &  \multicolumn{4}{|c|} {$B_{total}$}    &  \multicolumn{4}{|c|} {$B_{y}$} &  \multicolumn{4}{|c|} {$B_{z}$} \\ \cline{3-14}
 Event & of Proton& Lag & \multicolumn{3}{|c|} {No: of diffusions} & Lag & \multicolumn{3}{|c|} {No: of diffusions} & Lag & \multicolumn{3}{|c|} {No: of diffusions}\\ \cline{4-6} \cline{8-10} \cline{12-14} 
      & (GV) & & $1^{st}$ &$2^{nd}$ &$3^{rd}$ & & $1^{st}$ &$2^{nd}$ &$3^{rd}$ & & $1^{st}$ &$2^{nd}$ &$3^{rd}$   \\ \hline  \hline
\endhead 
\hline \hline
\endfoot
 \hline  \hline 
2001  Jan 13  &14.3 &  -13 & 384 & 173 & 168 &  -14 & 196 & 158 & 137 &  -23 &  97 & 112 & 107\\ 
              &24.0 &  -13 & 245 & 121 &  62 &  -14 & 117 &  95 &  50 &  -23 &  38 &  44 &  39\\ \hline
2001  Apr 11  &14.3 &  -18 & 656 & 340 & 583 &  - &   - &   - &   - &   -5 &  41 &  67 &  45 \\ 
              &24.0 &  -18 & 533 & 283 & 214 &  - &   - &   - &   - &   -5 &  25 &  47 &  16 \\  \hline 
2001  May 27  &14.3 &   0 &   0 &   0 &   0 &   -3 &  44 &  38 &  35 &  -21 & 111 & 177 & 117 \\ 
              &24.0 &   0 &   0 &   0 &   0 &   -3 &  29 &  25 &  13 &  -21 &  53 &  85 &  43 \\ \hline 
2001  Aug 13  &14.3 &   -7 & 216 & 106 & 121 &   -7 &  76 &  86 &  78 &   -5 &  26 &  58 &  37 \\
              &24.0 &   -7 & 165 &  82 &  45 &   -7 &  49 &  55 &  28 &   -5 &  15 &  35 &  14 \\ \hline 
2001  Sep 12  &14.3 &  -25 & 838 & 354 & 373 &  - &   - &   - &   - &   -1 &  17 &  13 &  12 \\
              &24.0 &  -25 & 583 & 263 & 137 &  - &   - &   - &   - &   -1 &  11 &   9 &   5 \\ \hline 
2001  Nov 24  &14.3 &  -21 & 822 & 405 & 719 &  -31 & 173 & 429 & 183 &  -14 & 260 & 226 & 230 \\
              &24.0 &  -21 & 666 & 338 & 264 &  -31 & 112 & 311 &  67 &  -14 & 199 & 181 &  84 \\  \hline 
2001  Dec 14  &14.3 &  -35 &1346 & 604 & 868 &   -2 &  56 &  28 &  29 &  -17 & 255 & 240 & 237 \\
              &24.0 &  -35 & 872 & 494 & 319 &   -2 &  28 &  21 &  11 &  -17 & 119 & 177 &  87 \\  \hline 
2002  Sep 07  &14.3 &  -18 & 776 & 298 & 398 &  -24 & 312 & 305 & 273 &   3 &  24 &  37 &  24 \\
              &24.0 &  -18 & 555 & 241 & 146 &  -24 & 172 & 205 & 100 &   3 &  14 &  25 &   9 \\ \hline 
2002  Sep 30  &14.3 &   -5 & 207 &  93 & 154 &   8 & 130 & 128 & 151 &  -12 & 198 & 185 & 167 \\
              &24.0 &   -5 & 171 &  77 &  57 &   8 &  93 & 102 &  56 &  -12 & 138 & 145 &  61 \\ \hline 
2002  Dec 22  &14.3 &  -15 & 481 & 244 & 316 &   0 &   0 &   0 &   0 &  -12 &  66 & 126 &  77 \\
	      &24.0 &  -15 & 363 & 196 & 116 &   0 &   0 &   0 &   0 &  -12 &  37 &  70 &  28 \\ \hline 
2003  Jan 23  &14.3 &  -21 & 509 & 270 & 256 &   - &   - &   - &   - &  -28 &  49 & 170 &  54 \\
              &24.0 &  -21 & 341 & 183 &  94 &   - &   - &   - &   - &  -28 &  21 &  71 &  20 \\ \hline 
2003  Feb 16  &14.3 &   - &   - &   - &   - &   - &   - &   - &   - &  -11 &  55 & 100 &  58 \\
              &24.0 &   - &   - &   - &   - &   - &   - &   - &   - &  -11 &  22 &  50 &  21 \\ \hline 
2003  May 04  &14.3 &   -8 & 252 & 127 & 157 &  -10 &  97 & 125 & 113 &   0 &   0 &   0 &   0 \\
              &24.0 &   -8 & 200 & 101 &  58 &  -10 &  64 &  82 &  41 &   0 &   0 &   0 &   0 \\ \hline 
2003  Jul 25  &14.3 &  -19 & 565 & 347 & 556 &   -2 &  32 &  31 &  31 &   2 &   9 &  22 &   8 \\
              &24.0 &  -19 & 434 & 288 & 204 &   -2 &  22 &  24 &  11 &   2 &   5 &  13 &   3 \\ \hline 
2003  Dec 27  &14.3 &  -35 &1210 & 506 & 545 &   5 &  41 &  52 &  43 &   -3 &  10 &  27 &  10 \\
              &24.0 &  -35 & 900 & 380 & 200 &   5 &  23 &  29 &  16 &   -3 &   5 &  13 &   4 \\ \hline 
2004  Aug 30  &14.3 &   - &   - &   - &   - &  - &   - &   - &   - &   1 &  10 &  12 &  11 \\
              &24.0 &   - &   - &   - &   - &  - &   - &   - &   - &   1 &   7 &   8 &   4 \\  \hline 
2004  Dec 05  &14.3 &  -12 & 326 & 209 & 307 &   8 & 107 & 112 &  93 &  -13 & 117 & 193 & 204 \\
              &24.0 &  -12 & 260 & 172 & 113 &   8 &  71 &  83 &  34 &  -13 &  89 & 148 &  75 \\ \hline 
2004  Dec 12  &14.3 &  -17 & 595 & 266 & 322 &  -25 & 243 & 322 & 288 &  -13 & 138 & 163 & 138 \\
              &24.0 &  -17 & 469 & 210 & 118 &  -25 & 163 & 218 & 106 &  -13 &  91 & 108 &  51 \\ \hline \hline 
\end{longtable}
\end{landscape}}

 We find that the FD profile looks remarkably similar to that of the IP magnetic compression \index{Interplanetary magnetic field!compression} and lags it by few hours. Considering cross-field diffusion \index{diffusion}  as the dominant mechanism causing the Forbush decrease, \index{Forbush decreases} we chose the events which have good correlations \index{correlation} with the enhancements in the perpendicular magnetic fields ( $B_y$ , $B_z$)  and not with the radial magnetic field ($B_x$). We have calculated the number of diffusions using eqn \ref{noDs} for 14.3 GV and 24.0 GV protons, using the peak value of the IP magnetic field compression. The number of diffusions corresponding the observed lag for the chosen events are listed in table \ref{T3}.

 For most of the events the observed lag corresponds to few tens to few hundred diffusions. It is to be noted that there are two exceptional events, 2001  December 14 and 2003  December  27, where the number of diffusions are $\sim$1000 using the time-varying $\sigma$ prescription. 
There are three events in this list which have no correlation \index{correlation} lag between the IP magnetic field \index{Interplanetary magnetic field} profile and FD profile\index{Forbush decreases!profile}. The FD on 2001 May 27 correlates well with $B_{total}$, the FD on 2002  December 22 correlates well with  $B_y$ and  the FD on 2003  May 04  correlates with the $B_z$ with no correlation  lag.

\section{Conclusion}
Our aim in this work is to understand the relation between the Forbush decrease \index{Forbush decreases} and the interplanetary magnetic fields. We found a reasonable correlation \index{correlation!coefficient} between the FD magnitude \index{Forbush decreases!magnitude} and the peak magnetic field of the magnetic field compression (table \ref{corrTt}, \ref{corrTp}). A detailed examination of the FD and IP magnetic field profiles quantitatively established the following: 
1) the importance of the turbulent sheath \index{sheath} region between the shock \index{shock} and ICME - the magnetic field enhancement responsible for the Forbush decrease is in the shock sheath region and the magnetic turbulence levels \index{turbulence!turbulence level $\sigma$} also get enhanced in this region, 
and 2) the viability of cross-field diffusion as the primary reason for Forbush decreases. We found that the FD and the IP magnetic field profiles are very similar, and the FD lags the magnetic field enhancement by a few hours (tables \ref{T1} and \ref{T2}). We find that the observed lag between the cosmic ray \index{cosmic rays} flux and the IP magnetic field  corresponds to few tens to few hundreds of diffusions (table \ref{T3}).

 \chapter[Self-similar expansion of CMEs]{Self-similar expansion of solar coronal mass ejections}
\label{fluxrope}

\noindent\makebox[\linewidth]{\rule{\textwidth}{3pt}} 
{\textit { In the previous chapters we discussed the effects of CMEs near the Earth. In this chapter we examine CME kinematics near the Sun. Specifically, we examine the evolution of the 3D structure of CMEs to draw conclusions regarding the manner in which they are driven.   }  }\\
\noindent\makebox[\linewidth]{\rule{\textwidth}{3pt}}

\section{Introduction}

The study of Earth-directed coronal mass ejections (CMEs) from the Sun are crucial for  spaceweather since these are typically responsible for most major geomagnetic storms. It is important to get a thorough understanding of the forces governing their initiation and propagation through the interplanetary medium.  CME kinematics exhibit a variety of characteristics \citep{yash04, webb12}. Some CMEs experience most of their acceleration within $\approx$ 1--2 $R_{\odot}$ above the solar limb, while others show evidence of being continuously driven throughout typical coronagraph \index{coronagraph} fields of view that extend upto $\approx$ 30 $R_{\odot}$ \citep{sv07}. CMEs \index{CME} whose mechanical energies are increasing through the coronagraph fields of view, are thought to be driven by Lorentz self-forces \index{Lorentz self-forces} (e.g., \citealp{sng13, olmetal, chkr, chn96, kmrst96}). 
 These ${\mathbf J} \times {\mathbf B}$ forces are usually thought to arise from  misaligned currents and magnetic fields contained within the CME. The other important force to be considered in  CME kinematics in the interplanetary region is the ``drag'' force, which contributes towards CME \index{CME!acceleration} deceleration. These drag forces are thought to be due to momentum coupling between the CMEs and the ambient solar wind (e.g., \citealp{gpl00, lew02, cgl04, vrs, slb12}). Recent studies (\citealp{slb12})  have attempted  a preliminary understanding of the physics underlying these drag forces\index{CME!aerodynamic drag}.  These authors used the viscous drag in a collisionless plasma to address the drag forces present in CME dynamics. { Save for some broad ideas (e.g. \citealp{kunch, chkun}) we do not  have a very good understanding of the details of the driving force}. We also don't have a clear idea of the typical heliocentric distance at which the driving force ceases to be important in comparison to the drag force.

In this chapter we  focus on the driving force acting on CMEs \index{CME} as they propagate through the coronagraph field of view. It is the magnetic energy contained by the CMEs \index{CME!energy}, which is generally thought to be responsible for propelling them. This concept of CME driving at the expense of magnetic energy was  quantitatively demonstrated by \citet{vr00}.  \citet{sv07} showed that, on the average, the magnetic energy contained in CMEs  can provide  for at least 74 \% of energy  what is required for their propagation from the Sun to the Earth. We identify a set of well observed CMEs observed by the SECCHI \index{coronagraph!SECCHI} coronagraphs \index{coronagraph}  \citep{how08} aboard the STEREO satellites \citep{kaiser08}. A large majority of CMEs \index{CME!observation} observed with coronagraphs are now confirmed to have a flux rope morphology (e.g., \citealp{angelos13, jie13}). Using this as a reference in our study we first  fit the graduated cylindrical shell (GCS) model for flux rope \index{flux rope} CMEs (\citealp{thr09}) to these well observed CMEs \index{CME} in order to obtain their 3D structure. One  result of this fitting procedure turns out to be the fact that the flux rope CMEs \index{CME} evolve self-similarly; i.e. in a manner such that the ratio of their minor to major radii remains constant. There has been some prior observational evidence for the fact that flux rope CMEs  expand in a self-similar \index{self-similar} manner in the coronagraph field of view  (e.g., \citealt{poom10, kilpua2012, robin2013}). However this work is the first systematic demonstration of self-similar expansion. Self-similarity has been invoked in a number of theories relating to CMEs \index{CME!propagation} propagation (e.g., \citealp{rust96, Dem, wang09}). In this chapter, we assume that the entire evolution \index{CME!evolution} of the flux rope CMEs \index{flux rope} in the coronagraph \index{coronagraph} field of view (i.e., expansion and translation) is due to Lorentz self-forces\index{Lorentz self-forces}. Using this assumption and the observed self-similar \index{self-similar} expansion, we draw conclusions regarding the extent to which the flux rope structures are non-force-free\index{non-force-free}. In section \ref{DseA} we describe the observational results, which we use to draw conclusions regarding the current and magnetic field configurations in section \ref{LSF}. Conclusions are drawn in section \ref{sum}

\section{Data analysis} \label{DseA}

\subsection{Graduated Cylindrical Shell Model} \label{GCSM}

The graduated cylindrical shell model (GCS) was developed by \citet{thr06} and \citet{thr09}. This model provides a method to analyse the 3D morphology, position and kinematics of CMEs \index{CME} using the  white-light remote sensing observations. The GCS model uses forward-modeling techniques. It allows the user to fit a geometric repesentation of a flux rope \index{flux rope} to CME \index{CME!observation} observations. The GCS is meant to reproduce the large-scale structure of flux-rope-like CMEs, which consists of a tubular section forming the main body of the structure attached to two cones that correspond to the ``legs'' of the CME. The resulting shape obtained by this model resembles a hollow croissant because electrons are placed only at the surface and the prominence material is not modelled. The geometry of the empirical flux rope model is shown in figure \ref{GCS}. Figures \ref{GCS}(a) and (b) show, respectively, a face-on and an edge-on representation of the model. The dash-dotted line represents the axis of the model and the solid line the outline of the shell, where the density is placed. Here $h$ is the height of the legs and $\alpha$ is the half-angle between the legs. The cross section of the model is a circular annulus of varying radius  $a$.  The distance from the center of the Sun to a point at the edge of the shell is $R$. The aspect ratio of the loop is given as $\kappa \, = \, \frac{a}{R} $ \citep{thr09}.

The model is positioned using the longitude, latitude and the rotation parameters. The origin of the flux rope is fixed at the center of the Sun. The size of the flux rope \index{flux rope} model is controlled using three parameters which define the apex height, foot point separation and the radius of the shell. 

\begin{figure}
\includegraphics[width = \columnwidth]{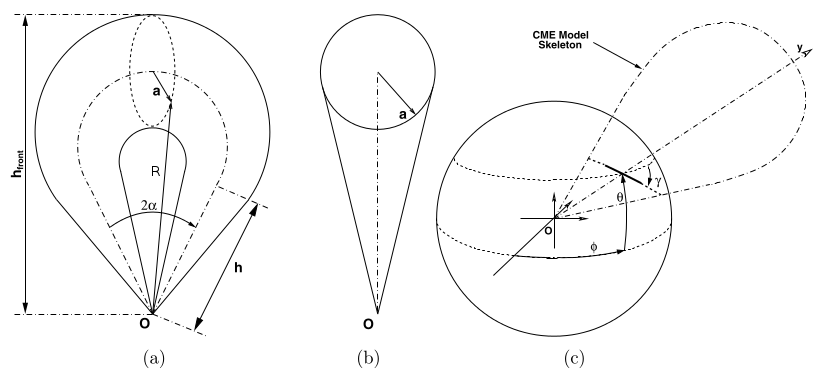}
\caption[Representations of the Graduated Cylindrical Shell (GCS) model]{Adapted from \citet{thr09}. Representations of the Graduated Cylindrical Shell (GCS) model (a) face-on and (b) edge-on. The dash-dotted line is the axis through the center of the shell. The solid line represents a planar cut through the cylindrical shell and the origin. O corresponds to the center of the Sun. (c) Positioning parameters. The loop represents the axis through the center of the shell, $\phi$ and $\theta$ are the longitude and latitude, respectively, and $\gamma$ is the tilt angle around the axis of symmetry of the model.}\label{GCS}
\end{figure}

This model is implemented by over-plotting the projection of the cylindrical shell structure onto both the SECCHI \index{coronagraph!SECCHI} A and B images. The observer then adjusts the six parameters of the model to get the best fit to the flux rope cavity. The GCS model is a sophisticated method of modeling the three-dimensional structure of the CME\index{CME!structure}. This model correctly handles the effects of projection for each image's point of view. The geometry of the GCS model is a good proxy for the flux rope \index{flux rope} like magnetic structure of CMEs. However, the technique used in this model has some limitations. Since the model defines a circular cross-section we can  only  fit the overall structure of the CME and cannot model any distortion in interplanetary space.

\subsection{Analysing SECCHI data}

\begin{figure}
\includegraphics[width = \columnwidth]{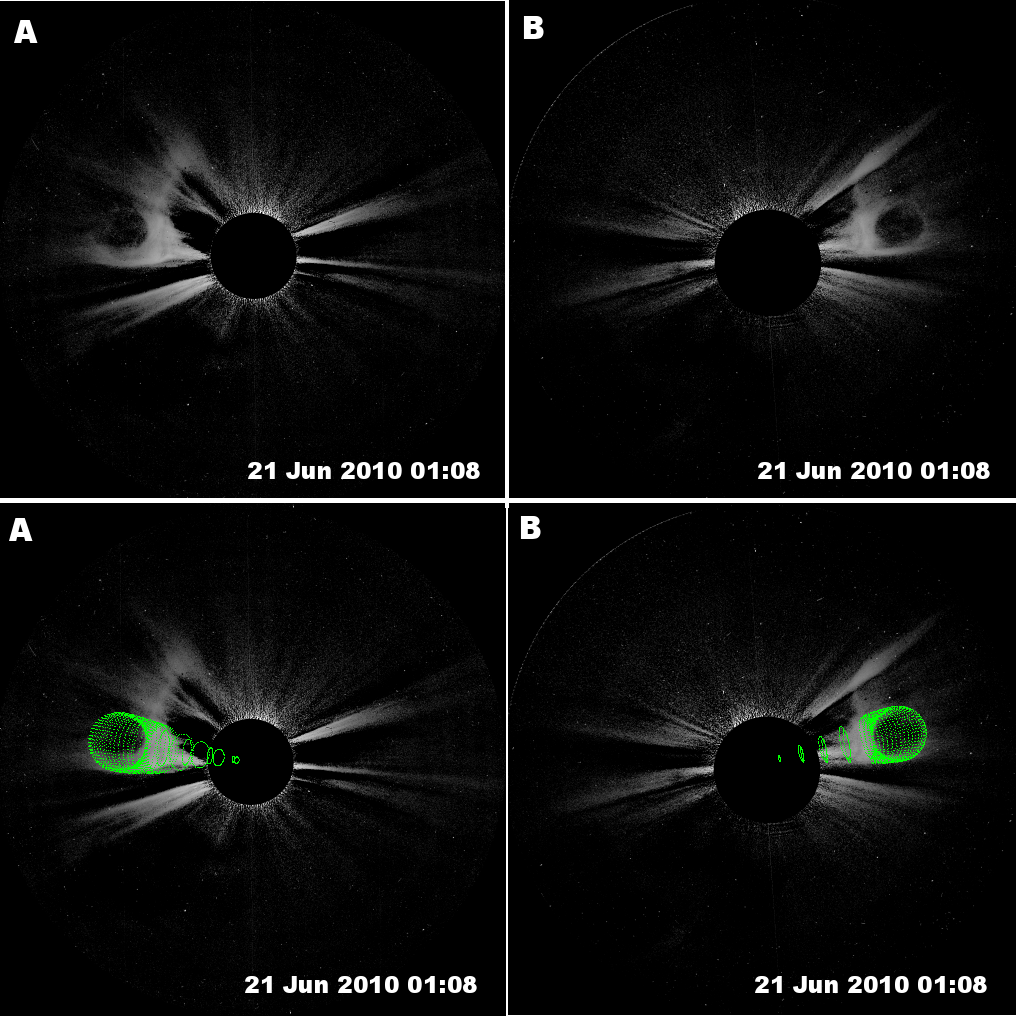}
\caption[Flux rope fitting to the CME]{A screenshot of a CME \index{CME} on 21 June 2010 that illustrates the flux rope fitting procedure. The left panels are from STEREO A and the right panels are from STEREO B. The upper panels show the white light CMEs \index{CME} data and the lower panels have the flux rope \index{flux rope} (displayed as a green wiremesh) structure superposed.} \label{crd}
\end{figure}

We have identified 9 well observed CMEs \index{CME} using the data from the SECCHI A and B coronagraphs \index{coronagraph!SECCHI} aboard the STEREO spacecraft. We used  the method explained in the section \ref{GCSM} to  fit a three-dimensional geometrical flux rope configuration to the images in SECCHI A and B coronagraphs simultaneously at each timestamp. A representative screenshot is shown in Figure \ref{crd}. This figure shows the flux rope fitting to the CME images at 01:08 UT on 21 June 2012. We have taken  care to ensure that the cross section of the flux rope \index{flux rope} is fitted only to the dark cavity visible in the coronagraph \index{coronagraph} images. The fitting of the GCS flux rope to the coronagraph images  yields a variety of geometrical parameters. Table \ref{tbl1} summarizes the most relevant ones for each of the events we have considered in this study. We followed CMEs only as far as it is possible to make a clear, unambiguous fit to the flux rope model.  The main results are summarized in Table \ref{tbl1}. Some salient features are:
 
\begin{itemize}
\item 
The quantity $\kappa$, which is the ratio of the flux rope \index{flux rope} minor radius to its major radius, remains approximately constant with time for a given CME. This conclusion holds for all the CMEs  we have studied, and is a clear demonstration of the fact that flux rope CMEs \index{CME} expand in a nearly self-similar \index{self-similar} manner.

\item
 The values of $\kappa$ for different CMEs \index{CME} in Table \ref{tbl1} are in the range  $0.44 \geq \kappa \geq 0.2$ 
\end{itemize}

\begin{longtable}{llcccccc}

\caption [Flux rope fits to STEREO COR2 data]{\label{tbl1} Flux rope \index{flux rope} fits to STEREO COR2 data }\\
\hline \hline
Date  & Time  & Longitude & Latitude & Tilt Angle &  Height & $\kappa \, = \, \frac{a}{R} $ \\
\hline 
\endfirsthead
\caption{Flux rope fits to STEREO COR2 data  }\\
\hline \hline
Date  & Time  & Longitude & Latitude & Tilt Angle &  Height & $\kappa \, = \, \frac{a}{R}$ \\ \hline
\endhead

\hline \multicolumn{7}{|r|}{{Continued on next page}} \\ \hline
\endfoot

\hline \hline
\endlastfoot
07/01/2010 & 07:07 & 135.28 & 10.06 & 3.35 & 7.50 &   \\
07/01/2010 & 08:08 & 134.16 & 09.50 & 3.35 & 8.43 &   \\
07/01/2010 & 09:08 & 133.05 & 09.50 & 3.91 & 9.50 &     \\
07/01/2010 & 10:08 & 131.93 & 10.62 & 3.35 & 10.21 &   {0.212$\pm$0.008}\\
07/01/2010 & 11:08 & 133.05 & 10.06 & 0.56 & 10.86 & \\
07/01/2010 & 12:08 & 133.05 & 10.62 & 2.80 & 12.14 &  \\ \hline \hline
01/02/2010 & 19:08 & 38.01 & -19.01 & 14.54 & 10.21 &  \\
01/02/2010 & 20:08 & 36.90 & -19.57 & 16.77 & 12.21 &  {0.315$\pm$0.006}\\
01/02/2010 & 21:08 & 35.78 & -19.01 & 17.89 & 14.57 & \\
01/02/2010 & 22:08 & 35.78 & -19.01 & 19.01 & 16.5 &  \\ \hline \hline
14/02/2010 & 03:08 & 210.18 & 13.98 & -34.66 & 8.14 &  \\
14/02/2010 & 04:08 & 210.18 & 12.30 & -30.75 & 9.79 &  {0.258$\pm$0.010}\\
14/02/2010 & 05:08 & 210.18 & 11.74 & -40.25 & 11.14 &  \\
14/02/2010 & 06:08 & 211.31 & 11.74 & -38.57 & 13.00 &  \\ \hline \hline
12/06/2010 & 15:08 & 338.76 & 35.22 & 72.67 & 9.21 &  \\
12/06/2010 & 16:08 & 336.52 & 34.10 & 72.67 & 11.29 &   {0.305$\pm$0.024}\\
12/06/2010 & 17:08 & 336.52 & 29.07 & 77.14 & 13.00 &   \\
12/06/2010 & 18:08 & 336.52 & 29.07 & 77.14 & 15.71 &   \\ \hline \hline
20/06/2010 & 22:08 & 310.81 & 11.18 & 0.56 & 7.14 &   \\
20/06/2010 & 23:08 & 310.81 & 11.18 & 0.56 & 8.14 &  \\
21/06/2010 & 00:08 & 310.81 & 11.18 & 0.56 & 9.07 &  {0.196$\pm$0.013}\\
21/06/2010 & 01:08 & 310.81 & 11.18 & 0.56 & 10.43 &   \\
21/06/2010 & 02:08 & 310.81 & 12.30 & 0.56 & 11.64 &   \\ \hline \hline
01/03/2010 & 05:08 & 24.60 & -16.77 & 3.35 & 10.57 &  \\
01/03/2010 & 06:08 & 24.60 & -16.77 & 3.35 & 12.50 &  {0.353$\pm$0.010}  \\
01/03/2010 & 07:08 & 24.60 & -16.77 & 2.80 & 13.93 &   \\
01/03/2010 & 08:08 & 23.48 & -15.65 & -3.91 & 15.93 &   \\ \hline \hline
26/03/2010 & 13:08 & 22.36 & -1.12 & 46.96 & 8.79 &   \\
26/03/2010 & 14:08 & 22.36 & -1.12 & 51.99 & 10.14 &   \\
26/03/2010 & 15:08 & 22.36 & -1.12 & 54.78 & 11.43 &   {0.216$\pm$0.011}\\
26/03/2010 & 16:08 & 22.36 & -0.56 & 55.90 & 13.21 &   \\
26/03/2010 & 17:08 & 22.36 & -1.12 & 87.20 & 14.71 &   \\ \hline \hline
13/04/2010 & 12:08 & 164.34 & 36.33 & -12.30 & 6.50 &  \\
13/04/2010 & 13:08 & 164.34 & 34.66 & -11.18 & 9.07 &  {0.438$\pm$0.032} \\
13/04/2010 & 14:08 & 164.34 & 34.10 & -14.54 & 12.07 &   \\
13/04/2010 & 15:08 & 164.34 & 33.54 & -13.98 & 15.98 &   \\ \hline \hline
29/01/2008 & 06:22 & 54.78 & 3.91 & -0.56 & 11.21 &   \\
29/01/2008 & 06:52 & 55.90 & 3.91 & -0.56 & 11.71 &   \\
29/01/2008 & 07:22 & 55.90 & 3.91 & -0.56 & 12.35 &  {0.203$\pm$0.008}\\
29/01/2008 & 07:52 & 55.90 & 3.35 & -0.56 & 13.29 &   \\
29/01/2008 & 08:22 & 55.90 & 3.91 & -0.56 & 13.86 &   \\
29/01/2008 & 09:22 & 55.90 & 4.47 & -1.12 & 15.71 &  \\
\end{longtable}

For a given CME, the geometrical flux rope fitting procedure we use allows it to have  different values of $\kappa$ at different timestamps. The approximate self-similarity observed in CME \index{CME!evolution} evolution as a result of the flux rope fitting  is thus physical. Several of the flux rope CMEs studied by \citet{kilpua2012} using the GCS method  also evolve in a self-similar \index{self-similar} manner, with the value of aspect ratio, $0.39 \ge \kappa \ge 0.23$.  Studies in the HI (Heliospheric Imager) field of view  \citep{robin2013}also reveal self-similar expansion, with $0.60 \ge \kappa \ge 0.25$. \citet{sv09} found that the subset of CMEs from \citet{sv07} that were subject to a net driving force show a constant value of $\kappa$, which  they used  to derive the axial current enclosed by these flux rope \index{flux rope} CMEs.  It is to be noted that \citet{sv09} used LASCO data, which did not have the advantage of two viewpoints that the current study does. \citet{sv09} selected CMEs \index{CME!propagation} which seemed to propagate mostly in the plane of the sky. They  interpreted the circular cross-section visible in LASCO images as the cross-section of these flux ropes. It is to be noted that there are CMEs whose expansion is not self-similar (e.g., \citealp{cheng2014}). This deviation from self-similarity might either  be an illusion arising out of CME rotation (e.g., \citealp{angelos2011}) or a genuine effect.  We now turn our attention to the implications of the observed self-similar propagation in the context of a flux rope model where the evolution is governed entirely by Lorentz self-forces\index{Lorentz self-forces}.

\section{Lorentz self-forces in flux ropes} \label{LSF}
\subsection{Self-similar expansion}

\begin{figure}
\centering
\includegraphics[width = 0.8\columnwidth]{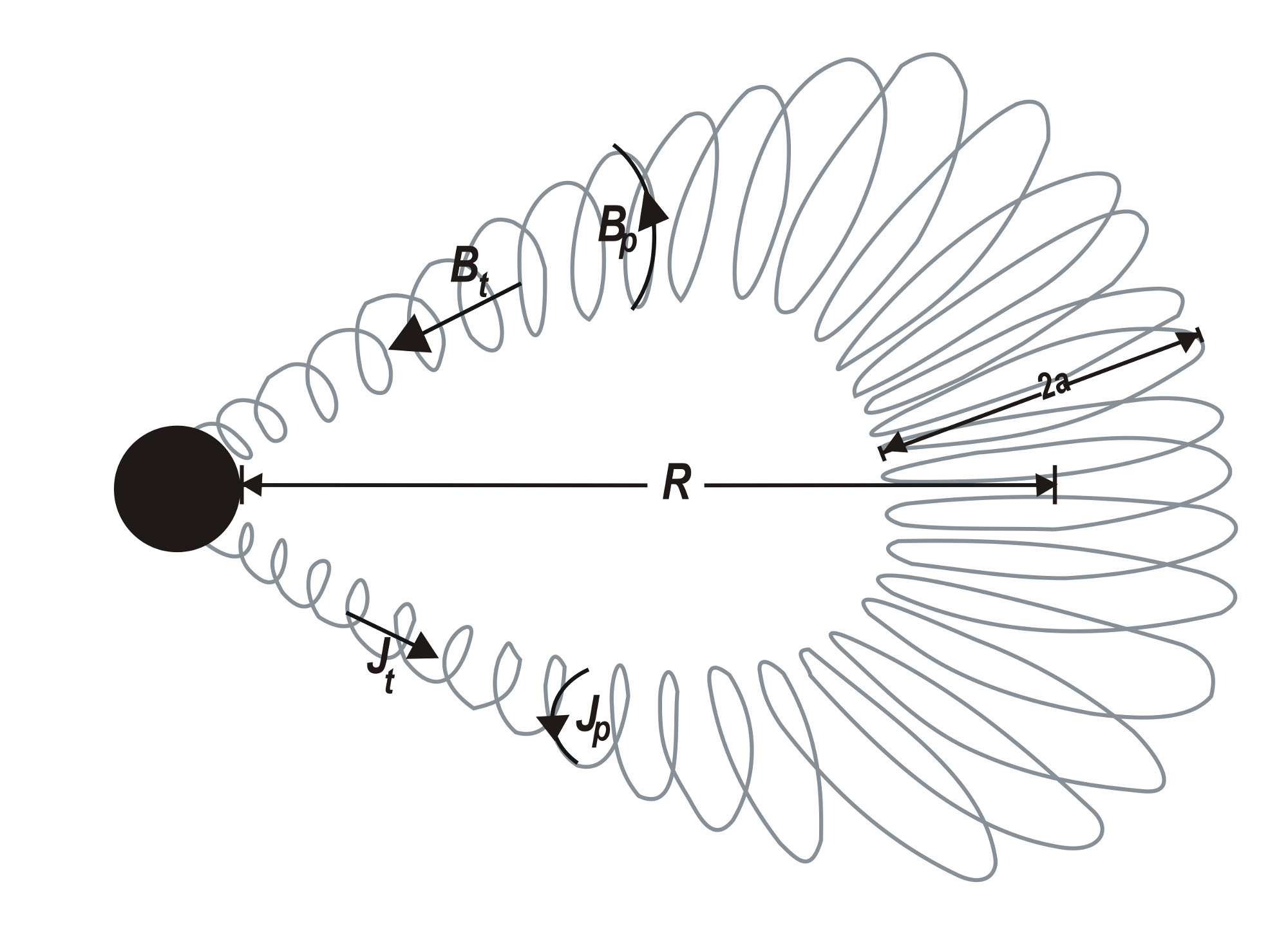}
\caption[A schematic of the fluxrope magnetic field]{A schematic of the fluxrope \index{flux rope} magnetic field. The fluxrope minor radius is $a$ and its major radius is $R$. The directions of the toroidal and poloidal current densities and magnetic fields are indicated.}
\label{flrp}
\end{figure}

The subject of Lorentz self-forces \index{Lorentz self-forces} in flux ropes has a long history, which starts  from \citet{sha66} on through treatments like \citet{anz, garchen, chn96, kmrst96, chkr, sv09, olmetal}. In a broad sense all these studies  appeal to variants of ${\mathbf J} \times {\mathbf B}$ forces, which  arise from  currents and magnetic fields carried by the flux rope structure.  A qualitative sketch of a fluxrope configuration with its poloidal and toroidal componenets of  ${\mathbf J} $ and $ {\mathbf B}$ are shown in figure \ref{flrp}. The assumption of self-similar \index{self-similar} flux rope evolution \index{CME!evolution} is built into several popular theoretical treatments of Lorentz self-force driving (e.g., \citealp{kmrst96, sv09, olmedo2010}). In the treatment of \citet{kmrst96}, self-similar evolution is a consequence of assuming that axial magnetic flux and helicity are both conserved. However, they do not use a specific value of the self-similarity parameter 
\begin{equation}
\kappa \equiv \frac{a}{R}
\label{eqkappa}
\end{equation}
 in their treatment. Here $a$ is the minor radius and $R$ is the major radius of the flux rope CME.

 We will now examine some other treatments   involving Lorentz self-forces \index{Lorentz self-forces}  (e.g., \citealp{chn96, chkr}) and do not explicitly apeal to self-similar \index{self-similar} expansion. It is often difficult to specify unique magnetic field and current configurations for a non-force-free \index{non-force-free} flux rope structure (see \citealp{chn12} for some examples), hence several authors have used the self-inductance of a slender, axisymmetric, circular flux rope as a starting point. This quantity (in cgs units) is given by \citep{sha66,lanlif}

\begin{equation}
L = 4\, \pi\, R \biggl [ {\rm ln}\biggl (\frac{8\,R}{a}\biggr ) - 1 \biggr ]\, ,
\label{eqinductance}
\end{equation}

where $R$ is the major radius of the flux rope \index{flux rope} and $a$ is its minor radius.

 The magnetic energy associated with a current loop such as this carrying an axial current $I$ is 

\begin{eqnarray}
\nonumber
U_{m} & = & (1/2) L I^{2} \\
  & = & 2 \, \pi\, R \, I^{2} \, \biggl [ {\rm ln}\biggl (\frac{8\,R}{a}\biggr ) - 1 \biggr ]
\end{eqnarray} 

Using this and considering the self similar \index{self-similar} expansion of flux rope, $\kappa \equiv \frac{a}{R}$  as constant, the Lorentz self-force \index{Lorentz self-forces} acting along the major radius is then derived as 

\begin{equation}
f_{R} = \frac{1}{c^{2}}\, \frac{\partial}{\partial R} U_{m} = \frac{2\, \pi\,I^{2}}{c^{2}}\,\biggl [ {\rm ln}\biggl (\frac{8\,R}{a}\biggr ) - 1 \biggr ]\, ,
\label{eqselfforce1}
\end{equation}

where $c$ is the speed of light.

Thereafter, the force per unit arc length acting along the major radial direction is calculated as 
\begin{equation} 
f_{\rm L} \,  = \, (1/2 \pi R) f_{R}
\end{equation}

  It maybe noted that the last step in equation~(\ref{eqselfforce1}) can be arrived at only if the quantity $\kappa \equiv a/R$ is assumed to be constant. In other words, any treatment that uses equation~(\ref{eqselfforce1}) implicitly assumes that the flux rope evolves in a self-similar manner.  However, we note that some treatments (e.g., \citealp{chn96, chkr}) use equation~(\ref{eqselfforce1}) (and therefore implicitly assume self-similar \index{self-similar} expansion) and yet have separate differential equations for the evolution \index{CME!evolution} of the flux rope major \index{flux rope} radius ($R$) and its minor radius ($a$).

In our study we use the observed values of the self-similarity \index{self-similar} parameter $\kappa$ to determine the relation between the local pitch angle \index{pitch angle $\gamma$} of the magnetic field configuration inside the flux rope \index{flux rope} and the misalignment angle \index{misalignment angle $\chi$} between the current density and the magnetic field.

\subsection{How misaligned are ${\mathbf J}$ and ${\mathbf B}$?}

Lorentz self-force \index{Lorentz self-forces} driving necessarily involves non-force-free \index{non-force-free} configurations of the current density (${\mathbf J}$) and the magnetic field (${\mathbf B}$). We therefore evaluate the angle between the current density ${\mathbf J}$ and the magnetic field ${\mathbf B}$. We decompose the current and magnetic field into poloidal and toroidal components:

{\begin{eqnarray}
\nonumber
 \mathbf{J}& = & J_p \mathbf{i_p} + J_t \mathbf{i_t} \equiv c_1 B_t \mathbf{i_p} + c_2 B_p \mathbf{i_t} \\
 \mathbf{B}& = & B_p \mathbf{i_p} + B_t \mathbf{i_t} \, ,
\label{eqdecompose}
\end{eqnarray}}
where ${\mathbf i_p}$ and ${\mathbf i_t}$ are unit vectors in the poloidal and toroidal directions respectively (see fig \ref{flrp}). The constant $c_1$ and $c_2$ can be defined as

\begin{equation}
c_{1} \equiv J_{p}/B_{t}\,\,\,\, {\rm and} \,\,\,\, c_{2} \equiv J_{t}/B_{p}, 
\label{eqc1c2def}
\end{equation}

The magnetic field pitch angle \index{pitch angle $\gamma$} $\gamma$ is given by the expression,

\begin{equation}
\gamma \equiv \tan^{-1} \frac{B_p}{B_t}\, .
\label{eqpitchangle}
\end{equation}

the angle $\chi$ \index{misalignment angle $\chi$} between the current density ${\mathbf J}$ and the magnetic field ${\mathbf B}$ can be written as

{\begin{eqnarray}
\nonumber
\sin \chi &  =  & \frac{| \mathbf{J} \times \mathbf{B}|}{|\mathbf{J} ||\mathbf{B}|} = \\
\nonumber 
&  =  & \frac{c_1 B_t^2 - c_2 B_p^2}{ \sqrt{c_1^2 B_t^2 + c_2^2 B_p^2} \sqrt{B_t^2 + B_p^2} } \\
&  =  & \frac{1 -  \frac{c_{2}}{c_{1}}\, \tan ^2 \gamma}{\biggl [  (1+ \frac{c_2^2}{c_1^2} \tan ^2 \gamma) \, (1 + \tan ^2 \gamma) \biggr ]^{1/2}} \, ,
\label{eqsinchi}
\end{eqnarray}}

where we have used Equation (\ref{eqdecompose}) for ${\mathbf J}$ and ${\mathbf B}$.   We note that Eq~(\ref{eqsinchi}) is independent of a specific model for the current density ${\mathbf J}$ and the magnetic field ${\mathbf B}$ inside the flux rope. It holds for any flux rope structure (fig \ref{flrp}), and does not make any assumptions about whether or not it is force-free\index{force-free}.  We now use two different methods to calculate the local pitch angle \index{pitch angle $\gamma$} (Eq. \ref{eqpitchangle}) for the flux rope \index{flux rope} magnetic field, which we describe herewith.

\subsubsection{Method 1}

The most popular concept to describe flux ropes is the force-free \index{force-free}Lundquist solution \citep{lundquist1950}.  A natural starting point would be to assume that Lorentz self-forces \index{Lorentz self-forces} arise from a situation where the flux rope structure deviates very little from the force-free state.  In Eq (51) of their paper, \citet{kmrst96} give the following expression for \index{misalignment angle $\chi$} $\sin \chi$:

\begin{equation}
 \sin \chi = \frac{\pi}{x_0} \left(\frac{a}{\pi R} \right) = \frac{\kappa}{x_0} \, ,
\label{eqrksinchi}
\end{equation}

where $x_{0} = 2.405$ is the first zero crossing of the Bessel function $J_{0}$ and we have used Eq~(\ref{eqkappa}). Apart from the assumptions regarding conservation of axial magnetic flux and helicity, this expression for $\sin \chi$ from \citet{kmrst96} relies crucially on the assumption that the flux rope \index{flux rope} deviates very little from the force-free \index{force-free}Lundquist solution \citep{lundquist1950}. It can be derived from Eqs~(16) and (50) of \citet{kmrst96} and using $ | {\mathbf J} | \,  | {\mathbf B} | = \alpha B^{2}$, which expresses the fact that the flux rope is nearly force-free.

 Equating  the equations for $\sin \chi$,  (\ref{eqsinchi}) and (\ref{eqrksinchi}) we get , 

\begin{equation}
\frac{1 - \frac{c_{2}}{c_{1}}\, \tan ^2 \gamma}{\biggl [(1 + \tan ^2 \gamma) \, (1+ \frac{c_2^2}{c_1^2} \tan ^2 \gamma) \biggr ]^{1/2}} =  \frac{\kappa}{x_0} \, .
\label{eqfirst}
\end{equation}

For the Lundquist force-free \index{force-free}solution (e.g., Eq 1, \citealp{kmrst96}; Eq 19, \citealp{linetal98}), since ${\mathbf J} = \alpha \, {\mathbf B}$, the ratio $c_{2}/c_{1}$ (Eq~\ref{eqc1c2def}) is given by

\begin{equation}
\frac{c_{2}}{c_{1}} \equiv \frac{J_{t}}{B_{p}} \frac{B_{t}}{J_{p}} = \biggl ( \frac{J_{0}(x_{0} y)}{J_{1}(x_{0} y)} \biggr )^{2} \, 
\label{eqc2c1forcefree}
\end{equation}

where $y$ is the fractional minor radius of the flux rope. In other words, $y < 1$ defines the interior of the flux rope \index{flux rope} and $y > 1$ its exterior. The quantity $J_{0}$ denotes the Bessel function of zeroth order while the quantity $J_{1}$ represents the Bessel function of first order.
Using equation~(\ref{eqc2c1forcefree}) for $c_{2}/c_{1}$ and the observationally determined values of the similarity parameter $\kappa$ (Table \ref{tbl1}), we can use Eq~(\ref{eqfirst}) to determine the pitch angle \index{pitch angle $\gamma$} $\gamma$ of the magnetic field configuration of a flux rope which deviates only slightly from a force-free \index{force-free}configuration. }

\begin{figure}
\centering
\includegraphics[width = 0.8\columnwidth]{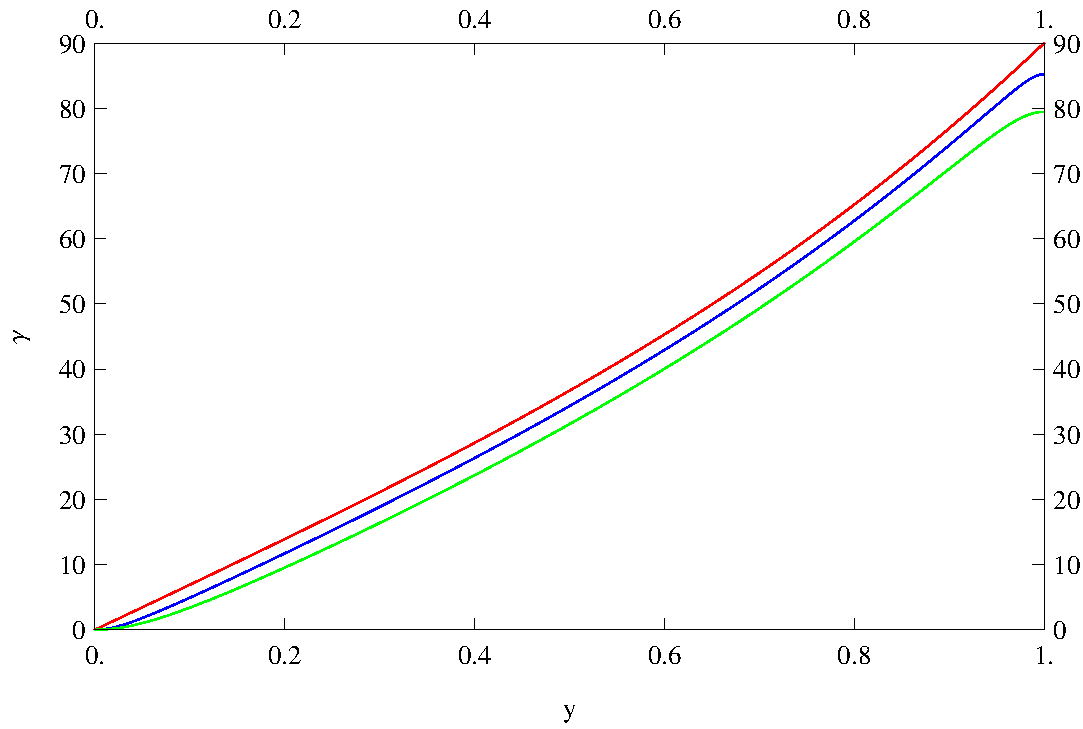}
\caption[A plot of the local magnetic field pitch angle] {A plot of the local magnetic field pitch angle  $\gamma$ as a function of fractional minor radius $y$. The red line denotes the force free \index{force-free}model (Eq~\ref{eqforcefree}) while the blue and green lines are obtained using method 1 (Eq~\ref{eqfirst}). The blue line uses $\kappa = 0.2$ and the green line employs $\kappa = 0.44$} 
\label{Ga}
\end{figure}

The poloidal and toroidal components of the magnetic field for the ideal force-free \index{force-free} Lundquist solution is given by   (e.g., Eq 1, \citealp{kmrst96};  Eq 19, \citealp{linetal98})

\begin{eqnarray}
\nonumber 
B_p & = & B_0 J_1(x_0 \, y)     \\
B_t & = & B_0 J_0(x_0 \, y)
\end{eqnarray}

Using this  the pitch angle \index{pitch angle $\gamma$} $\gamma$ for  the ideal force-free Lundquist solution is given by, 

\begin{equation}
\tan \gamma  \equiv \frac{B_{p}}{B_{t}} = \frac{J_1(x_0 \, y)}{J_0(x_0 \, y)}\, ,
\label{eqforcefree}
\end{equation}

{ The pitch angle \index{pitch angle $\gamma$} calculated using equations~(\ref{eqfirst}) and (\ref{eqc2c1forcefree}) are shown in the figure \ref{Ga}. In order to calculate the pitch angle we used some of the observed values of the similarity parameter $\kappa$ for all the CMEs \index{CME} in our list (table \ref{tbl1}).  The blue line in figure \ref{Ga} uses the smallest value of $\kappa$ observed in our sample ($\kappa = 0.2$), while the green line uses the largest observed value of $\kappa$ (= 0.44). For comparison, the pitch angle computed using the ideal force-free configuration (Eq \ref{eqforcefree}) is also overplotted in red. It is clear from the figure \ref{Ga} that the local magnetic field pitch angles for self-similarly  \index{self-similar} expanding flux ropes \index{flux rope} do not agree with that for an ideal force-free configuration. The larger the value of $\kappa$, the more is the disagreement. In other words, the magnetic field configurations in the observed (self-similarly expanding) flux ropes deviate considerably from a force-free one. This is despite the fact that the observed values of the self-similarity parameter ($\kappa$, Table \ref{tbl1}) correspond to misalignment angles \index{misalignment angle $\chi$} $\chi$ (equation~\ref{eqrksinchi}) of only 5$^{\circ}$ to 10$^{\circ}$. The nearly force-free \index{force-free} assumption is thus not consistent, and it is worth examining if such self-similarly expanding flux ropes can be better described by a non-force free model.}

\subsubsection{Method 2}

In this method we consider a prescription for a non-force-free \index{non-force-free} flux rope \index{flux rope} configuration given by \citet{ber13}. This prescription is a perturbative expansion on a force-free configuration, correct to order $\kappa \equiv a/R$, which incorporates the effect of large-scale curvature (see Fig \ref{flrp}). In this prescription, the pitch angle \index{pitch angle $\gamma$} is defined as

\begin{equation}
\tan \gamma \, \equiv \, \frac{B_p}{B_t} \, = \, \frac{J_1\left( A(y, \phi)\right)}{J_0\left( A(y, \phi)\right)} \, ,
\label{eqberdi1}
\end{equation}
where $\phi$ is the polar angle coordinate in the plane perpendicular to the toroidal axis and the quantity $A$ is defined by

\begin{equation}
A (y, \phi) = x_{o}\, y \left[  1+ \kappa y \, (\cos \phi - | \sin \phi |) \right ] \, .
\label{eqberdi2}
\end{equation}

\begin{figure}
\centering
\includegraphics[width = 0.8\columnwidth]{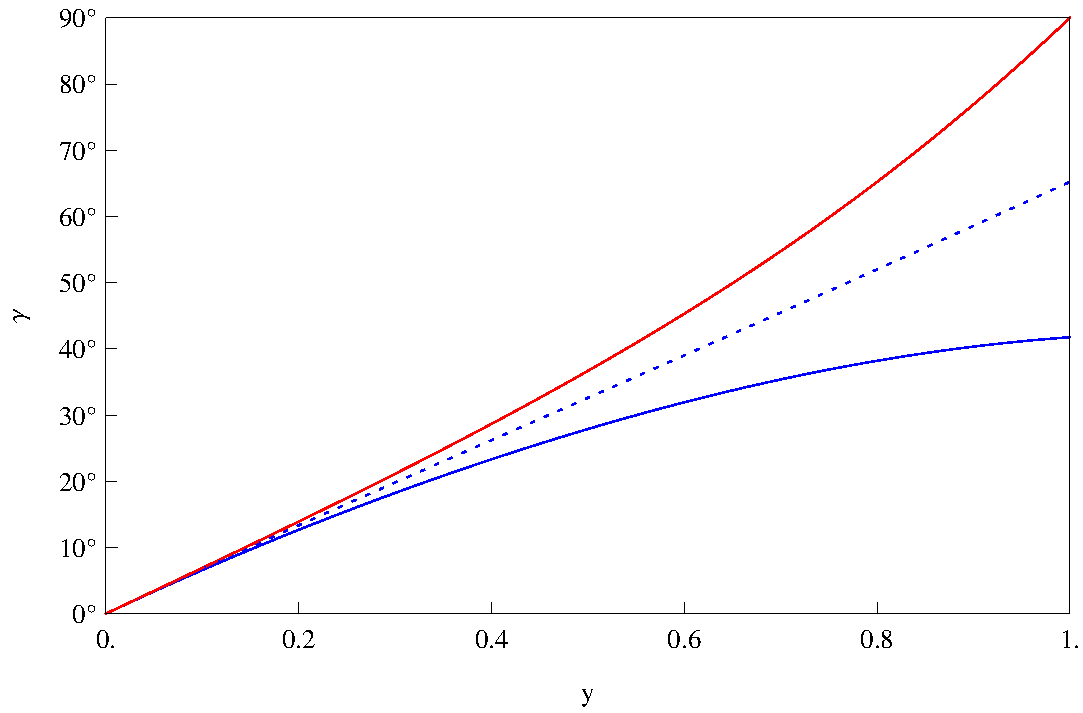}
\caption[A plot of $\gamma$ as a function of fractional minor radius $y$  using method 2]{A plot of $\gamma$ as a function of fractional minor radius $y$  using method 2 (Eq~\ref{eqberdi1}). The red solid line is for $\phi=\pi/4$, the blue dotted line is for $\phi=\pi/2$ and $\kappa=0.2$ while the blue solid line is for $\phi=\pi/2$ and $\kappa=0.44$.}
\label{TGXk}
\end{figure}

The quantity $A (y, \phi)$ expresses the effect of the curvature of the major radius. For a straight flux rope, an observer looking through the cross section will see only one circle, because the circles defining the cross section overlap each other. For a bent flux rope, on the other hand, the observer will see a few circles displaced from each other with the center of circles lying in the curvature of fluxrope. The more the flux rope \index{flux rope} curvature, the farther the centers of these circles are displaced from each other. From a direct  comparison of equations~(\ref{eqberdi1}) and (\ref{eqforcefree}), it is evident that the non-force free \index{non-force-free} expression is identical to the force-free expression for the  values of $\phi = \pi/4$ and $7 \pi/4$. 

We can calculate the local magnetic field pitch angle \index{pitch angle $\gamma$} ($ \gamma$) for this non-force-free \index{non-force-free} configuration using Eqs~(\ref{eqberdi1}) and (\ref{eqberdi2}). This magnetic field pitch angle ($ \gamma$) is a function of the observed similarity parameter $\kappa$ (table \ref{tbl1}) and the fractional minor radius $y$. The local magnetic field pitch angle $\gamma$ is depicted as a function of the fractional minor radius $y$ for a few representative values of $\kappa$ and $\phi$ in figure \ref{TGXk}. {  The red solid line is for $\phi=\pi/4$, the blue dotted line is for $\phi=\pi/2$ and $\kappa=0.2$ while the blue solid line is for $\phi=\pi/2$ and $\kappa=0.44$. } Since $\phi = \pi/4$ corresponds to the force-free \index{force-free} case, it is independent of $\kappa$. 

Using the values of $\gamma$ obtained from this method as shown in figure \ref{TGXk}, we can compute the misalignment angle \index{misalignment angle $\chi$} $\chi$ between ${\mathbf J}$ and ${\mathbf B}$ using equation~(\ref{eqsinchi}). {  However, since we are considering a non-force free \index{non-force-free} configuration, it is not appropriate to use equation~(\ref{eqc2c1forcefree}) for the ratio $c_{2}/c_{1}$, which is valid only for a force-free \index{force-free} configuration. Instead, we recognize that the poloidal magnetic field $B_p$ is generated by a toroidal current $I_{t}$, while the toroidal magnetic field $B_t$ is generated by a poloidal current $I_{p}$:

\begin{eqnarray}
\nonumber
B_p & = & \frac{2I_t}{ca} \\ 
B_t & = & \frac{2I_p}{cR} 
\label{eqJB}
\end{eqnarray}

where $c$ denotes the speed of light. Furthermore, the total currents are related to their respective densities by (e.g., \citealp{chn89})

\begin{eqnarray}
\nonumber
I_t \, = & 2 \pi \int_0^a r J_t dr \, & = \, \pi a^2 J_t  \\
I_p \, = & 2 \pi R \int_0^a J_p dr \, & = \, 2 \pi R a J_p \, ,
\label{eqIJ}
\end{eqnarray}
where, for the sake of concreteness, we have assumed that the current density is uniform throughout the body of the flux rope. We note that other current distributions are possible. Equations (\ref{eqc1c2def}), (\ref{eqJB}) and (\ref{eqIJ}) yield

\begin{equation}
\frac{c_{2}}{c_{1}} = 2 \, .
\label{eqc1c2}
\end{equation}

\begin{figure}
\centering
\includegraphics[width = 0.8\columnwidth]{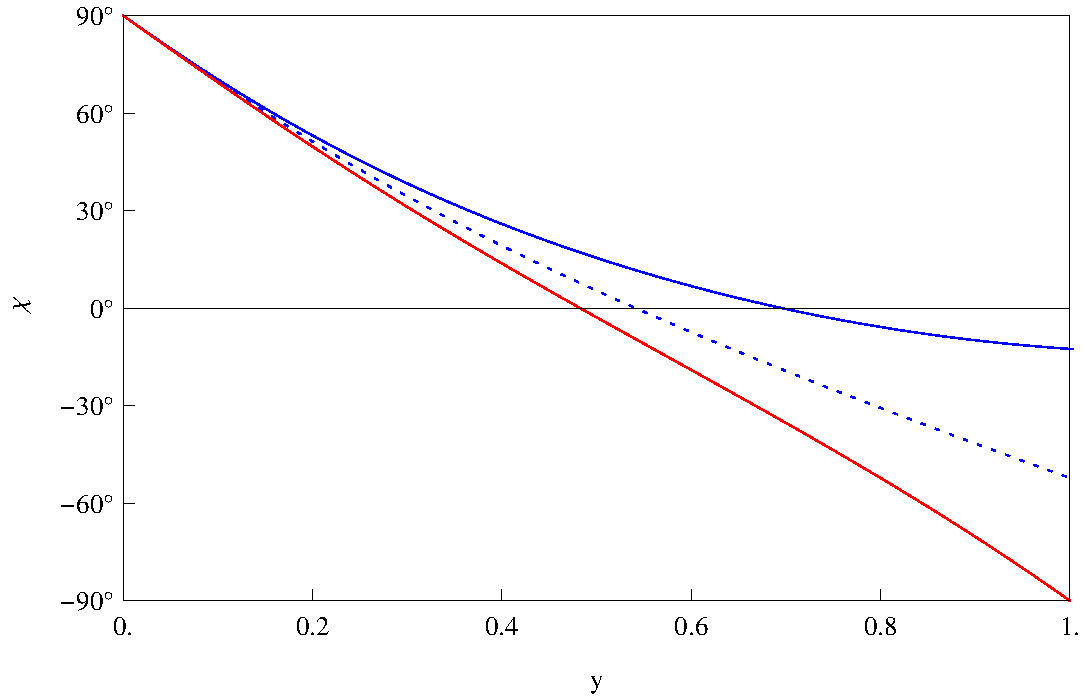}
\caption[A plot of the angle ($\chi$) between ${\mathbf J}$ and ${\mathbf B}$ as a function of fractional minor radius $y$ ]{A plot of the angle ($\chi$) between ${\mathbf J}$ and ${\mathbf B}$ as a function of fractional minor radius $y$ using method 2. The linestyles are the same as that used in figure~\ref{TGXk}.}
\label{CX}
\end{figure}

Using the values of $\tan \gamma$ from equations (\ref{eqberdi1}) and (\ref{eqberdi2}) in equation~(\ref{eqsinchi}) with the quantity $c_{2}/c_{1}$ given by equation~(\ref{eqc1c2}), we get the values of $\chi$ \index{misalignment angle $\chi$} depicted in figure~\ref{CX}.  {  The red solid line is for $\phi=\pi/4$, the blue dotted line is for $\phi=\pi/2$ and $\kappa=0.2$ while the blue solid line is for $\phi=\pi/2$ and $\kappa=0.44$. } Since $\phi = \pi/4$ corresponds to the force-free \index{force-free}case, it is independent of $\kappa$.  
Clearly, the angle between ${\mathbf J}$ and ${\mathbf B}$ can be substantial, which means that the flux rope deviates considerably from a force-free state.}

\section{Discussion}

Our findings imply that, in the coronagraph \index{coronagraph} field of view, the current (${\mathbf J}$) and the magnetic field (${\mathbf B}$) within flux rope \index{flux rope} CMEs \index{CME!propagation} that propagate in a self-similar \index{self-similar} manner can be substantially misaligned. This is the first conclusive evidence of the non-force-free \index{non-force-free} nature of flux rope CMEs. This misaligned  ${\mathbf J}$ and ${\mathbf B}$ forms the basis for Lorentz self-force \index{Lorentz self-forces} driving. The magnitude of the Lorentz self force ($|{\mathbf J}|\,|{\mathbf B}|\,\sin \chi$) depends upon the magnitudes of the current density ($|{\mathbf J}|$) and magnetic field ($|{\mathbf B}|$) as well as the the angle ($\chi$) \index{misalignment angle $\chi$} they subtend on each other. The results from this work regarding $\chi$ can constrain the magnitudes of the current and magnetic field in flux ropes needed to explain an observationally mandated driving force (e.g., \citealp{sv07}). Furthermore, findings from \citet{sv07} imply that there is excess magnetic energy that is available within the flux rope structure which can be expended in translating and expanding the CME, in (often) driving a shock ahead of it, and in heating the plasma that is inside it. It is not yet clear how the available magnetic energy is partitioned among these different avenues.

\begin{figure}
\centering
\includegraphics[width = 0.8 \columnwidth]{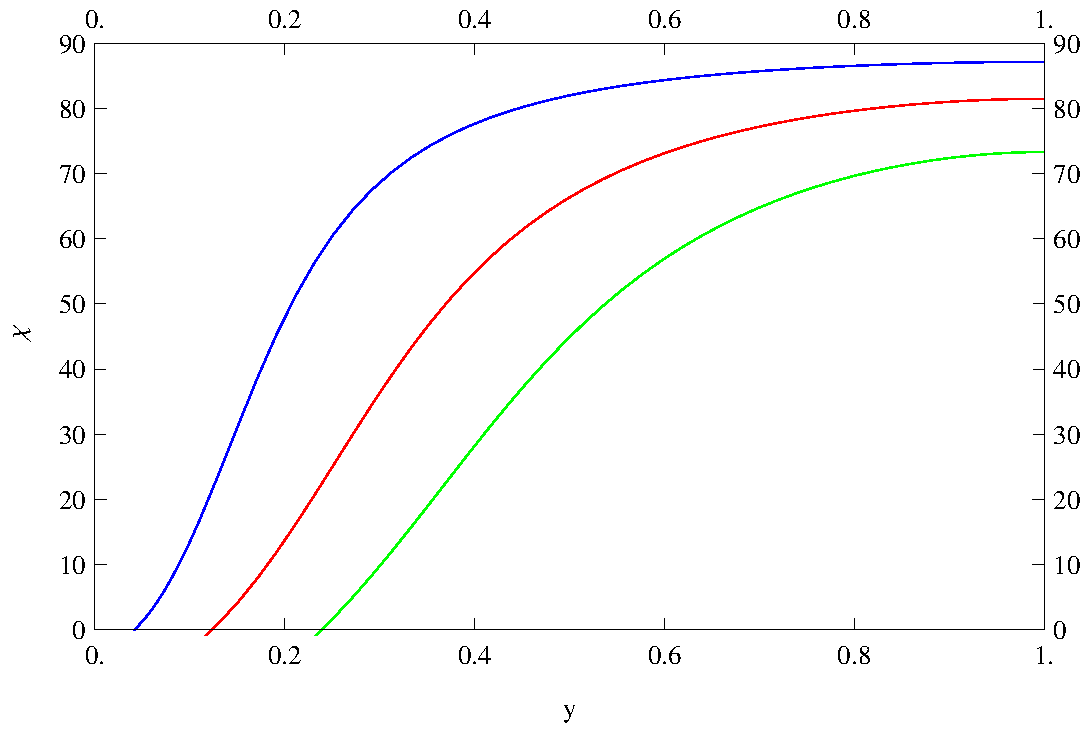}
\caption[The misalignment angle is plotted as a function of the fraction radius ($y$) inside the flux rope.]{The misalignment  angle ($\chi$) between the current density (${\mathbf J}$) and the magnetic field (${\mathbf B}$) implied by inferences of the magnetic field pitch angle  ($\gamma$) near the Earth. The misalignment angle is plotted as a function of the fraction radius ($y$) inside the flux rope. The blue curve is plotted for $\tan \gamma = 0.05$, the red one for $\tan \gamma = 0.15$ and the green one for $\tan \gamma = 0.3$}
\label{NearEarth}
\end{figure}

It is worth comparing our estimates for the magnetic field pitch angle \index{pitch angle $\gamma$} (figures \ref{Ga} and \ref{TGXk}) with values for this quantity near the Earth. 
Observations of near-Earth magnetic clouds (which are generally modeled as force-free \index{force-free} flux ropes\index{flux rope}) suggest that $0.05 \le \tan \gamma \le 0.3$ \citep{lar91, lea04, guli05}. Along with plausible guesses for the number of field line turns, these values for $\tan \gamma$ \index{pitch angle $\gamma$} have been used to infer total field line lengths in near-Earth magnetic clouds \citep{kah11}, which in turn are  used to address questions related to whether or not a force-free flux rope configuration is a good model for these structures, and if their legs are still connected to the Sun. { In keeping with the general expectation that flux ropes observed near the Earth are force-free structures, we use method 1 (which relies on the nearly force-free assumption) to check if these observed magnetic field pitch angles \index{pitch angle $\gamma$} are consistent with small values of the misalignment angle \index{misalignment angle $\chi$} $\chi$, as they are often assumed to be. We use equations~(\ref{eqsinchi}) and (\ref{eqc2c1forcefree}) to calculate the values of $\chi$ implied by the observed range $0.05 \le \tan \gamma \le 0.3$. The results are shown in figure~(\ref{NearEarth}). This figure shows the misalignment angle ($\chi$) between the the current density and the magnetic field inside the flux rope corresponding to $\tan \gamma = 0.05$ (blue line), $\tan \gamma = 0.15$ (red line) and $\tan \gamma = 0.30$ (green line).} This shows that, even near the Earth, the angle ($\chi$) between ${\mathbf J}$ and ${\mathbf B}$ is often quite substantial, and the force-free assumption is probably not valid.

\section{Summary} \label{sum}
Several CMEs \index{CME} are driven in the coronagraph \index{coronagraph} field of view; i.e., from a few to a few tens of $R_{\odot}$. We have fitted the \citet{thr09} 3D flux rope (GCS) model to nine well observed CMEs in the SECCHI/STEREO \index{coronagraph!SECCHI} field of view. One of the main conclusions from this exercise is that the flux rope CMEs propagate in a nearly self-similiar \index{self-similar}manner, which means that the ratio ($\kappa$) of the flux rope minor to major radius remains approximately constant as it propagates outwards. This conclusion is consistent with those from similar exercises using COR2 data (\citealp{kilpua2012}) and HI data (\citealp{robin2013}).

It is generally accepted that Lorentz self-forces \index{Lorentz self-forces} are responsible for the evolution \index{CME!evolution} of CMEs in both expansion as well as translation. The  Lorentz self-forces are assumed to arise from misaligned current density (${\mathbf J}$) and magnetic field (${\mathbf B}$). Eventhough the Lorentz self-forces involvement in the dynamics of CMEs are generally accepted, we really dont know much about the degree of misalignment with the current density (${\mathbf J}$) and magnetic field (${\mathbf B}$). We have derived a general relation (Eq~\ref{eqsinchi}) between the local pitch angle \index{pitch angle $\gamma$} ($\gamma$) of the flux rope magnetic field and the misalignment angle \index{misalignment angle $\chi$} ($\chi$) between ${\mathbf J}$ and ${\mathbf B}$, which will give the nature of the misalignment angle ($\chi$) inside the flux rope. This clearly means that the flux rope configurations are not force-free \index{non-force-free} while they are driven. 

We have used the observed values of the self-similarity \index{self-similar} parameter ($\kappa$) to calculate the local pitch angle \index{pitch angle $\gamma$} of the flux rope magnetic field ($\gamma$) using two different prescriptions. 
{ In the first one, we have assumed that the flux rope \index{flux rope} deviates only slightly from a force-free \index{force-free}equilibrium following the prescription of \citet{kmrst96}. Even though this prescription predicts that the misalignment angle \index{misalignment angle $\chi$} ($\chi$) between ${\mathbf J}$ and ${\mathbf B}$ is only 5$^{\circ}$ to 10$^{\circ}$, the local magnetic field pitch angle calculated deviates appreciably from that calculated using the purely force-free assumption (figure \ref{Ga}). This implies that the nearly force-free assumption is not well justified.}

We therefore adopt a second method that employs an explicit expression for magnetic fields in a non-force-free flux rope configuration. This is a first order perturbation (in the quantity $\kappa$) to a force-free flux rope \citep{ber13}. This method yields values of the local magnetic field pitch angle as a function of radial position inside the flux rope as well as the azimuthal angle (figure \ref{TGXk}). Since the second method does not assume a priori that the flux rope is nearly force-free, we contend that results using this method (figures \ref{TGXk} and \ref{CX}) are more reliable. 
The values for the angle ($\chi$) between ${\mathbf J}$ and ${\mathbf B}$ deduced from method 2 (figure \ref{CX}) are substantial; they range from $-50^{\circ}$ to $90^{\circ}$. These values may be contrasted with the rather small values for $\chi$ (around $3^{\circ}$) that are required for flux rope prominences to be supported against gravity (\citealp{rustkumar94}). 


\chapter[Conclusion and Future Work]{Conclusions and Future Work}
\label{conclu}

\noindent\makebox[\linewidth]{\rule{\textwidth}{3pt}} 
{\textit {In this chapter we summarize the main conclusions from this thesis. We also give a flavour of future work arising from the work done in this thesis.}  }\\
\noindent\makebox[\linewidth]{\rule{\textwidth}{3pt}} 

\section{Conclusions}

The broad goal of this thesis was to understand the near-Earth structure of Earth directed CMEs and the driving force acting on the CMEs.  Earth-directed Coronal mass ejections (CMEs) emanating from the Sun are the primary drivers of space weather disturbances. To understand the near-Earth structure of CMEs we used cosmic rays as a proxy. The effect of CMEs near the Earth is often manifested as transient decreases in galactic cosmic ray intensity, which are called Forbush decreases (FDs). We used FD events observed by the GRAPES-3 muon telescope to study the turbulence levels in the sheath region between the CME and the shock. To study the kinematics of CMEs we used the observation of CMEs in the STEREO/SECCHI A and B coronagraphs. This study helps us understand how misaligned are the current density and magnetic field in flux-rope CMEs; this is the basis of Lorentz self-force driving. 

\subsection{Forbush decreases observed in GRAPES-3}

The very large-area tracking muon telescope operating as a part of the GRAPES-3 experiment is a unique instrument capable of studying the muon variation rates in 225 solid angle bins. Since there can be a directional spread due to the influence of the terrestrial, solar, and interplanetary magnetic fields a regrouping of bins are made out of 169 after eliminating the outer bins as shown in the figure \ref{G225}. Thus the muon telescope is capable of looking in nine different directions, which corresponds to nine different cut-off rigiditites (Table \ref{cuto}). The data is summed over a time interval of one hour for each of the nine bins, which improves the signal-to-noise ratio. We used a low pass filter which removes all frequencies higher than 1 $day^{-1}$ in the data. This filter helps to remove the oscillations due to diurnal variations. The Forbush decrease events are better evident with the filtered data. 

\subsection{Forbush decrease models}

The relative contributions of shocks and coronal mass ejections (CMEs) in causing Forbush decreases is a matter of debate. We investigated this issue using multirigidity FD data from the GRAPES-3 muon telescope.  We consider two different models - the CME-only cumulative diffusion model and the shock-only model. We have used observationally derived parameters for the CME and shock (Table \ref{TCME} ) in constraining our models. The only free parameter in our study was the magnetic turbulence level ($\sigma \, =\,  \left(\frac {\langle B_{tur}^2 \rangle}{B_0^2}\right)^{0.5}$ ) in the sheath region. We studied the short listed events using the short-listing criteria described in section \S\ref{ESC}. Figure~\ref{Call} shows the results of the CME-only cumulative diffusion model fits to multi-rigidity data for each of the short-listed events. Table~\ref{Sig} summarizes the values of these turbulence levels that we have used for each of the FD events in the final short-list.  We compare these values with the estimate of 6--15 \% density turbulence level in the quiescent solar wind \citep{spa02}. We found that a good model fit using the CME-only cumulative diffusion model requires a turbulence level only a little higher than the quiescent solar wind,  whereas a good fit with the shock-only model demands a turbulence level which is an order of magnitude higher than that of the quiet solar wind. We conclude that a scenario where the FD is caused by the cumulative diffusion of galactic cosmic rays into the CME as it travels from the Sun to the Earth (through the turbulent sheath region) is a plausible one.

\subsection{Relation of FD profiles with Interplanetary magnetic field enhancements}

Forbush decreases observed at the Earth are related with the interplanetary magnetic field enhancements. The magnetic field enhancement mainly comprises the sheath region, which is the region between the shock and CME (Figure \ref{timi} \& \ref{tur}). The magnetic field turbulence level also gets enhanced in this region (Figure \ref{tur}). We find that the FD profile closely resembles the corresponding magnetic field compression (Figure \ref{23may}). The FD profile looks like a lagged copy of the magnetic field compression. We studied this resemblance using the quantities $B_{total}$, $B_z$,  $B_y $ and  $B_x$ (Table \ref{T1}). The observed lag in the correlation can be due to cross-field diffusion of high energy protons through the turbulent magnetic fields in the sheath region. We have verified that the lag between the IP magnetic field enhancement and the FD corresponds to few tens to few hundreds of cross-field diffusion times (Table \ref{T2}, Table \ref{T3}). This provides quantitative support to the idea of FDs arising out of galactic cosmic ray protons diffusing across the turbulent sheath region into the CME.

\subsection{Self-similar expansion of solar coronal mass ejections}

Our observations of flux rope CMEs in the coronograph field of view show that they are expanding  in a self-similar manner; in other words, the ratio of the flux rope minor radius to its major radius ($\kappa$) remains almost constant. We have used the observed values of the self-similarity parameter ($\kappa$) to calculate the local pitch angle of the flux rope magnetic field ($\gamma$) using two different prescriptions (\citealp{kmrst96} and \citealp{ber13}). Both these methods reveals that the magnetic field-current configuration in these flux ropes are substantially non force-free (Figure \ref{Ga} \& \ref{TGXk}). Lorentz self-forces, which are generally thought to be responsible for CME propagation, arise from misaligned current density (J) and magnetic field (B).  Our study found that for the self-similar expanding CMEs in coronagraph field of view (Figure \ref{CX}),  the current density (${\mathbf J}$) and the magnetic field (${\mathbf B}$) can be substantially misaligned. This is the first conclusive evidence of the non force-free nature of flux rope CMEs, which forms the basis for Lorentz self-force driving.

\section{Future Work}

\subsection{Forbush decreases}
We would like to study the Forbush decreases in greater detail. In the work carried out in this thesis we studied the Forbush decreases and their associated CMEs during the years 2001-2004. We would like to extend our studies to many more interesting events from 2004 onwards. We used two different approaches to study Forbush decreases in this thesis. In chapter \ref{model} we studied the cumulative effect of diffusion of cosmic rays into the  CME through the turbulent magnetic field in the sheath region; in this approach the cross-field diffusion was envisaged to happen across an idealized thin boundary. In chapter \ref{corrIP} we studied local diffusion of the cosmic rays through the entire magnetic field compression profile. In future work we would like to combine both these approaches so as to obtain a comprehensive physical understanding of the entire Forbush decrease. 

We have observed that the time evolution of Forbush decreases observed in galactic cosmic rays by the GRAPES-3 muon telescope bears a remarkable similarity to that of the observed compression in the near-Earth interplanetary magnetic field. Motivated by this, we can carry out modelling of the observed Forbush decreases that will concentrate on particle orbits in the observed magnetic field compressions in the presence of MHD turbulence. This treatment will result in a clear understanding of the MHD turbulence level in the sheath region ahead of the CME, which is a vital indicator of the strength of the geomagnetic storm that can be triggered by the event. It will also provide one of the few observational constraints on the magnitude of the diffusion coefficients of charged particles in the presence of turbulent magnetic fields.

\subsection{Forbush decrease precursors}
We would like to look at Forbush decrease precursors: something that will contribute substantially to advance warnings of the strength of geomagnetic storms. Forbush decrease precursors are pre-increase/pre-decrease observed in cosmic ray intensity before the onset of Forbush decreases.  Many studies suggest that the pre-increases/pre-decreases of the cosmic ray intensity (known as precursors) which usually precede a Forbush decrease could serve as a useful tool for studying space weather effects. Anomalies in the cosmic ray intensity distribution such as pre-increases or pre-decreases along with the anisotropy are often observed. These changes are observed from one to 24 hours before the arrival of the shock and can be used to forecast the intensity of the impending geomagnetic storms. 

We need to examine the cosmic ray  data and identify the loss-cone precursors associated with the Forbush decreases so as to understand the physics behind the precursors of the Forbush decrease. We believe that the unique multi-rigidity measurements from GRAPES-3 can contribute substantially to this field. 

\subsection{CME kinematics}
In chapter \ref{fluxrope} we found that the CMEs expanding in a self-similar manner are not force-free. The misalignment angle between the between the current density $\mathbf J $ and the magnetic field $\mathbf B$ inside the CMEs provides evidence for Lorentz self-force driving. In our future work we would to examine the magnetic field structure of CMEs observed in-situ by near-Earth spacecraft and calculate the pitch angle ($\gamma$) of the flux rope magnetic field. This will help us to identify the misalignment angle in near-Earth CMEs and establish the force-free nature (or lack thereof) of CMEs observed near the Earth.

In section \S~\ref{VP} we constructed the two stage velocity profile using a constant acceleration for the first stage and using a constant drag coefficient $C_D$ for the second. In future work we would like to gain a better understanding of the Lorentz self-force driving of CMEs. \citet{slb12} describes the drag coefficient in terms of the viscosity of the solar wind. This model will help us to use a more physical drag coefficient rather than a constant $C_D$ through out the heliosphere. Combining Lorentz self-force driving and the viscous drag will allows us to obtain a more physical CME velocity profile, which can predict the CME transit time more accurately.


{\small 
\textit{
\cleardoublepage
\phantomsection 
\addcontentsline{toc}{chapter}{\indexname} 
\begin{singlespace}
\printindex
\end{singlespace}
}}
\end{document}